\documentstyle[preprint,pra,aps]{revtex}
\input{psfig}
\begin{document}
\draft
\title{\bf Violations of fundamental symmetries in atoms and tests of 
unification theories of elementary particles}
\author{J.S.M. Ginges and V.V. Flambaum}
\address{School of Physics, University of New South Wales, 
Sydney 2052, Australia}
\date{\today}
\maketitle

\tightenlines

\begin{abstract}
High-precision measurements of violations of fundamental symmetries in atoms 
are a very effective means of testing the standard model of elementary 
particles and searching for new physics beyond it. Such studies complement 
measurements at high energies. 
We review the recent progress in atomic parity nonconservation and atomic electric 
dipole moments (time reversal symmetry violation), 
with a particular focus on the atomic theory required 
to interpret the measurements. 
\end{abstract}
\vspace{1cm}
\pacs{PACS: 32.80.Ys,11.30.Er,12.15.Ji,31.15.Ar}

%*************************************************************************
\tableofcontents
%*************************************************************************
\newpage

\section{Introduction}
\label{introduction}

The Glashow-Weinberg-Salam $SU(3)\times SU(2)\times U(1)$ standard model 
of elementary particles \cite{standardmodel} has enjoyed 30 years of undisputed success. 
It has been tested in physical processes covering a range in momentum 
transfer exceeding ten orders of magnitude. 
It correctly predicted the existence of new particles such as the 
neutral Z boson. 
However, the standard model fails to provide a deep explanation for 
the physics that it describes. 
For example, why are there three generations of fermions? 
What determines their masses and the masses of gauge bosons?
What is the origin of CP violation?
The Higgs boson (which gives masses to the particles in the standard model) 
has not yet been found.
The standard model is unable to explain Big Bang baryogenesis which is believed 
to arise as a consequence of CP violation. 

It is widely believed that the standard model is a low-energy manifestation 
of a more complete theory (perhaps one that unifies the four forces).
Many well-motivated extensions to the standard model have been proposed, 
such as supersymmetric, technicolor, and left-right symmetric models, 
and these give predictions for physical phenomena that differ from 
those of the standard model.% even at low energies.

Some searches for new physics beyond the standard model are performed at 
high-energy and medium-energy particle colliders where new processes or particles 
would be seen directly. However, a very sensitive probe can be carried out at 
low energies through precision studies of quantities that can be described 
by the standard model. The new physics is manifested indirectly through 
a deviation of the measured values from the standard model predictions.
The atomic physics tests that are the subject of this review lie in this 
second category. These tests exploit the fact that low-energy phenomena 
are especially sensitive to new physics that is manifested in the 
violations of fundamental symmetries, in particular 
P (parity) and T (time-reversal), that occur in the weak interaction.
The deviations from the standard model, or the effects themselves, may be very small. 
To this end, exquisitely precise measurements and calculations are required.

More than twenty years ago atomic experiments played an important role in the 
verification of the standard model.
While the first evidence for neutral weak currents (existence of the neutral Z boson) was 
discovered in neutrino scattering \cite{neutralcurrents}, the fact that neutral currents 
violate parity was first established in atomic experiments \cite{BZ} and only later 
observed in high-energy electron scattering \cite{prescott1978}. 
Now atomic physics plays a major role in the search for possible physics beyond 
the standard model. 
Precision atomic and high-energy experiments have different sensitivities to models 
of new physics and so they provide complementary tests. 
In fact the energies probed in atomic measurements exceed those currently accessible 
at high-energy facilities. For example, the most precise measurement of parity 
nonconservation (PNC) in the cesium atom sets a lower bound on an extra Z boson 
popular in many extensions 
of the standard model that is tighter than the bound set directly at the Tevatron 
(see Section \ref{section3iib}).   
Also, the null measurements of electric dipole moments (EDMs) in atoms 
(an EDM is a $P$- and $T$-violating quantity) place severe restrictions on 
new sources of $CP$-violation which arise naturally in models beyond the standard model 
such as supersymmetry. (Assuming CPT invariance, $CP$-violation is accompanied by $T$-violation.) 
Such limits on new physics have not been set by the detection of CP-violation in the 
neutral K \cite{kmeson} and B \cite{bmeson} mesons 
(see, e.g., Ref. \cite{cpreview} for a review of CP violation in these systems).

Let us note that while new physics would bring a relatively small correction to a 
very small signal in atomic parity violation, in atomic EDMs the standard 
model value is suppressed and is many orders of magnitude {\it below} 
the value expected from new theories. 
Therefore, detection of an EDM would be unambiguous evidence of new physics.

This review is motivated by the great progress that has been made recently in both 
the measurements and calculations of violations of fundamental symmetries in atoms. 
This includes the discovery of the nuclear anapole moment (an electromagnetic 
multipole that violates parity) \cite{wieman}, the measurement of the parity violating 
electron-nucleon interaction in cesium to 0.35\% accuracy \cite{wieman}, the 
improvement in the accuracy (to 0.5\%) of the atomic theory required to interpret the 
cesium measurement \cite{DFG02}, and greatly improved limits on the atomic \cite{hgedm} and 
electron \cite{tledm02} electric dipole moments.

The aim of this review is to describe the theory of parity and time-reversal
violation in atoms and explain how atomic experiments
are used to test the standard model of elementary particles and search for 
new physics beyond it.
We track the recent progress in the field. 
In particular, we clarify the situation in atomic parity violation in cesium:   
it is now firmly established that the cesium measurement \cite{wieman} is in 
excellent agreement with the standard model; see Section \ref{section3iib}. 
%The origin of the apparent $2.5\sigma$ deviation is explained in Section \ref{section3iib}.  

The structure of the review is the following. 
Broadly, it is divided into two parts. The first part, Section \ref{section2} to 
Section \ref{section6}, 
is devoted to parity violation in atoms.
The second part, Section \ref{section7} to 
Section \ref{section7iv}, is concerned with atomic electric dipole moments.

In Section \ref{section2} the sources of parity violation, and the standard model predictions, 
are described. In Section \ref{section3iia} a summary of the measurements of parity violation 
in atoms is given, with particular emphasis on the measurements with cesium. Also the 
atomic calculations are summarized. In Section \ref{section3} we present a detailed description 
of the methods for high-precision atomic structure calculations applicable to atoms with a 
single valence electron. 
The methods are applied to parity violation in cesium in Section 
\ref{section3iib} and the value for the weak nuclear charge is extracted and compared with 
the standard model prediction.
A discussion of the new physics constraints is also 
presented. In Section \ref{section3iic} a brief description for the method of atomic 
structure calculations for atoms with more than one valence electron is given, and 
the thallium PNC work is discussed. A brief discussion of the prospects for measuring PNC along 
a chain of isotopes is also presented. Then in Section \ref{section6} work on the anapole moment 
is reviewed.      

A description of electric dipole moments in atoms is given in Section \ref{section7}, 
with a summary of all the measurements and a discussion of the $P,T$-violating sources at 
different energy scales. 
Then in Section \ref{section7ii} a review of $P,T$-violating nuclear moments is given. 
In Section \ref{section7iv} a summary of the best limits on $P,T$-violating parameters can be 
found.  

Concluding remarks are presented in Section \ref{conclusion}.

For a general introduction to atomic $P$-violation and $P,T$-violation we refer 
the reader to the excellent books by Khriplovich \cite{khriplovichpnc} and 
Khriplovich and Lamoreaux \cite{cpoddbook}.

\section{Manifestations and sources of parity violation in atoms}
\label{section2}

Parity nonconservation (PNC) in atoms arises largely due to the exchange of $Z^{0}$-bosons 
between atomic electrons and the nucleus. 
The weak electron-nucleus interaction violating parity, but conserving 
time-reversal, is given by the following product of axial vector (A) 
and vector (V) currents: 
\begin{equation}
\label{eq:genpv}
\hat{h}=\frac{G}{\sqrt{2}}\sum_{N}\Big[ 
C_{1N}\bar{e}\gamma _{\mu}\gamma _{5}e\bar{N}\gamma ^{\mu}N 
+C_{2N}\bar{e}\gamma _{\mu}e\bar{N}\gamma ^{\mu}\gamma _{5}N\Big] \ . 
\end{equation}
Here $G=1.027\times 10^{-5}/m_{p}^{2}$ is the Fermi weak constant,  
$N$ is a nucleon wave function, and the sum runs over all protons 
$p$ and neutrons $n$ in the nucleus. 
The Dirac matrices are defined as
\begin{eqnarray}
\label{eq:diracmatrices}
{\gamma _{0}=
\left(
\begin{array}{cc}
I & 0 \\
0 & -I 
\end{array}
\right)} \ ,
&
\qquad {\gamma _{i}=
\left(
\begin{array}{cc}
0 & \sigma _{i} \\
-\sigma _{i} & 0 
\end{array}
\right)} \ ,
&
\qquad {\gamma _{5}=
\left(
\begin{array}{cc}
0 & -I \\
-I & 0 
\end{array}
\right)} \ ,
\end{eqnarray}
and $\mbox{\boldmath$\sigma$}=2{\bf s}$ are the Pauli spin matrices. 
The coefficients $C_{1N}$ and 
$C_{2N}$ give different weights to the contributions of protons and neutrons to 
the parity violating interaction. 
To lowest order in the electroweak interaction,
\begin{eqnarray}
\nonumber
C_{1p}&=&1/2\Big(1-4\sin ^{2}\theta _{W}\Big)\approx 0.04 \ , 
\quad C_{1n}=-1/2 \ , \\ 
\label{eq:smcoeff}
C_{2p}&=&-C_{2n}=1/2\Big( 1-4\sin^{2}\theta _{W}\Big)g_{A}\approx 0.05 \ ,
\end{eqnarray} 
where $g_{A}\approx 1.26$. 
The Weinberg angle $\theta _{W}$ is a free parameter; 
experimentally it is $\sin ^{2}\theta _{W}\approx 0.23$.
The suppression of the coefficients $C_{1p}$ and $C_{2N}$ 
due to the small factor $(1-4\sin ^{2}\theta)$ makes 
$|C_{1n}|$ about 10 times larger than $C_{1p}$ and $|C_{2N}|$. 

There is a contribution to atomic parity violation arising due to 
$Z^{0}$ exchange between electrons. However, this effect is negligibly small 
for heavy atoms \cite{bouchiats1,SF1978,BJS1990}. It is suppressed 
by a factor $(1-4\sin ^{2}\theta)K/(Q_{W}R(Z))$ compared to the dominant 
electron-nucleon parity violating interaction, where $K$ is a 
numerical factor that decreases with $Z$ and $R(Z)$ is a relativistic factor 
that increases with $Z$ \cite{SF1978}. 
For $^{133}$Cs $6S-7S$, $K\approx 2$ and $R(Z)=2.8$ and 
so the suppression factor is $\approx 0.04\%$ of the dominant amplitude 
\cite{SF1978}. This number was confirmed in \cite{BJS1990}. 
We will consider this interaction no further. 

\subsubsection{The nuclear spin-independent electron-nucleon interaction; 
the nuclear weak charge}

Approximating the nucleons as non-relativistic, 
the time-like component of the interaction $(A_{e}, V_{N})$ 
is given by the nuclear spin-independent Hamiltonian (see, e.g., \cite{khriplovichpnc})
\begin{equation}
\label{eq:pncnp}
\hat{h}_{W}=\frac{G}{\sqrt{2}}\gamma _{5}\Big[ 
ZC_{1p}\rho _{p} (r)+NC_{1n}\rho _{n}(r)\Big] \ ,
\end{equation}
$Z$ and $N$ are the number of protons and neutrons. 
This is an effective single-electron operator. The proton and neutron 
densities are normalized to unity, $\int \rho _{n,p}d^{3}r=1$.
Assuming that these densities coincide, $\rho _{p}=\rho _{n}=\rho$, this 
interaction reduces to 
\begin{equation}
\label{pncham}
\hat{h}_{W}=\frac{G}{2\sqrt{2}}Q_{W}\rho (r)\gamma _{5} \ ,
\end{equation}
where $Q_{W}$ is the nuclear weak charge.
%The nuclear weak charge $Q_{W}$ is given by the sum of the 
%weak charges of the up (u) and down (d) quarks which constitute the nucleus,
%\begin{equation}
%Q_{W}=[2Q_{W}({\rm u})+Q_{W}({\rm d})]Z+[Q_{W}({\rm u})+2Q_{W}({\rm d})]N
%\end{equation}
%To lowest order in the electroweak interaction, $Q_{W}({\rm u})$ 
%and $Q_{W}({\rm d})$ are
%\begin{equation}
%Q_{W}({\rm u})=(1-8/3 \sin ^{2}\theta _{W})
%\qquad
%Q_{W}({\rm d})=-(1-4/3 \sin ^{2}\theta _{W}) \ .
%\end{equation}
The nuclear weak charge $Q_{W}$ is very close to the neutron number. 
To lowest order in the electroweak interaction, it is 
\begin{equation}
\label{eq:Q_{W}tree}
Q_{W}=-N+Z(1-4\sin ^{2}\theta _{W}) \approx -N \ .
\end{equation}
This value for $Q_{W}$ is modified by radiative corrections.
The prediction of the standard electroweak model for the value of 
the nuclear weak charge $Q_{W}$ in cesium is \cite{PDG}
\begin{equation}
\label{eq:Q_{W}SM}
Q_{W}^{SM}(^{133}_{55}{\rm Cs})=-73.10\pm 0.03 \ .
\end{equation}

The nuclear weak charge $Q_{W}$ is protected from strong-interaction effects 
by conservation of the nuclear vector current. 
The clean extraction of the weak couplings of the quarks from atomic measurements 
makes this a powerful method of testing the standard model and searching 
for new physics beyond it.

The nuclear spin independent effects arising from 
the nuclear weak charge give the largest contribution to 
parity violation in heavy atoms compared to other mechanisms. 
However, note that the weak interaction (\ref{pncham}) 
does not always ``work''. This interaction 
can only mix states with the same electron angular momentum (it is a scalar). 
Nuclear spin-dependent mechanisms (see below), which produce much 
smaller effects in atoms, can change electron angular momentum and so can 
contribute exclusively to certain transitions in atoms and dominate 
parity violation in molecules.

\subsubsection{Nuclear spin-dependent contributions to atomic parity violation; 
the nuclear anapole moment}
\label{sssec:NSDPNC}

Using the non-relativistic approximation for the nucleons, 
the nuclear spin-dependent interaction due to neutral weak currents is 
(see, e.g., \cite{khriplovichpnc})
\begin{equation}
\label{eq:NSDnc}
\hat{h}_{NC}=\frac{G}{\sqrt{2}}\sum_{N}C_{2N}\mbox{\boldmath$\alpha$}\cdot 
\mbox{\boldmath$\sigma$}_{N}\rho (r) \ ,
\end{equation}
%\rho should be valence nucleon distribution
where $\alpha _{i}=\gamma _{0}\gamma _{i}$. 
This term arises from the space-like component of the $(V_{e}, A_{N})$ coupling.
Averaging this interaction over the nuclear state with angular momentum $I$ in the 
single-particle approximation gives
\begin{equation}
\label{eq:NSDVeAn}
\hat{h}^{I}_{NC}=-\frac{G}{\sqrt{2}}\kappa_{2}\frac{K -1/2}{I(I+1)}
\mbox{\boldmath$\alpha$}\cdot {\bf I}\rho (r) \ , 
\end{equation}
where $K =(I+1/2)(-1)^{I+1/2-l}$ and $\kappa_{2}=-C_{2}$.
There are two reasons for the suppression of this contribution to 
parity violating effects in atoms. 
First, unlike the spin-independent effects [Eq. (\ref{pncham})], 
the nucleons do not contribute coherently; 
in the nuclear shell model only the unpaired nucleon which carries 
nuclear spin $I$ makes a contribution. 
Second, the factor $C_{2}\propto (1-4\sin ^{2}\theta _{W})$ 
is small in the standard model.
 
% ??? collective anapole moment in quadrupole deformed nuclei ???

There is another contribution to nuclear spin-dependent PNC in 
atoms arising from neutral currents: the ``usual'' weak interaction 
due to the nuclear weak charge, $\hat{h}_{W}$, perturbed by the 
hyperfine interaction \cite{FK85}. 
In the single-particle approximation this interaction can be written as 
\cite{FK85,BPplb1991}
\begin{equation}
\label{eq:NSDQW}
\hat{h}_{Q}^{I}=\frac{G}{\sqrt{2}}\kappa_{Q}\frac{\mbox{\boldmath$\alpha$}
\cdot {\bf I}}{I}\rho (r) \ ,
\end{equation}
with 
\begin{equation}
\label{eq:qwa2/3}
\kappa _{Q}=-\frac{1}{3}Q_{W}\frac{\alpha \mu _{N}}{m_{p}R_{N}}
=2.5\times 10^{-4}A^{2/3}\mu _{N} \ ,
\end{equation}
$R_{N}=r_{0}A^{1/3}$ is the nuclear radius, $r_{0}=1.2~{\rm fm}$, 
and $\mu _{N}$ is the magnetic moment of the nucleus in nuclear magnetons. 
For $^{133}$Cs, $\mu _{N}=2.58$ and $\kappa _{Q}=0.017$.

However, the neutral currents are not the dominant source of parity violating 
spin-dependent effects in heavy atoms.
It is the nuclear anapole moment $\kappa_{a}$ that gives the largest effects 
\cite{FK1980}. 
This moment arises due to parity violation inside the nucleus,    
and manifests itself in atoms through the usual electromagnetic interaction 
with atomic electrons.
The Hamiltonian describing the interaction between the nuclear anapole 
moment and an electron is\footnote{
In fact, the distribution of the anapole magnetic vector potential is different from 
the nuclear density. However, the corrections produced by this difference are 
small; see Section \ref{section6}.}
\begin{equation}
\label{eq:NSDa}
\hat{h}_{a}=\frac{G}{\sqrt{2}}\kappa_{a}\frac{K}{I(I+1)}
\mbox{\boldmath$\alpha$}\cdot {\bf I}\rho (r) \ .
\end{equation}
The anapole moment $\kappa _{a}$ increases with atomic number, 
$\kappa_{a}\propto A^{2/3}$. This is the reason it leads to 
larger parity violating effects in heavy atoms compared to 
other nuclear spin-dependent mechanisms. 
In heavy atoms $\kappa _{a}\sim \alpha A^{2/3}\sim 0.1-1$ \cite{FK1980,FKS1984}.
(Note that the interaction (\ref{eq:NSDQW},\ref{eq:qwa2/3}) also increases 
as $A^{2/3}$, however the numerical coefficient is very small.) 

The spin-dependent contributions 
[Eqs. (\ref{eq:NSDVeAn},\ref{eq:NSDQW},\ref{eq:NSDa})] 
have the same form and produce the same effects in atoms. 
We will continue our discussion of the nuclear anapole moment and of 
nuclear spin-dependent effects in atoms in Section \ref{section6}.

\subsection{Simple calculation of the weak interaction in atoms induced
by the nuclear weak charge; the $Z^{3}$ enhancement}

In 1974 the Bouchiats showed that parity violating effects in atoms 
increase with the nuclear charge $Z$ faster than $Z^{3}$ 
\cite{bouchiats1,bouchiats2}.
This result was the incentive for studies of parity violation in 
heavy atoms.

Let us briefly point out where the factor of $Z^{3}$ originates. 
Taking the non-relativistic limit of the electron wave functions 
and considering the approximation of infinite boson exchange mass, 
the Hamiltonian (\ref{pncham}) reduces to
\begin{equation}
\hat{h}_{W}=\frac{G}{4\sqrt{2}m}\Big( 
\mbox{\boldmath$\sigma$}\cdot {\bf p} \delta^{3}({\bf r}) 
+\delta ^{3}({\bf r})\mbox{\boldmath$\sigma$}\cdot {\bf p}\Big)Q_{W} \ ,
\end{equation}
where $m$, $\mbox{\boldmath$\sigma$}$, ${\bf p}$ are the electron mass, 
spin, and momentum.   
The weak Hamiltonian $\hat{h}_{W}$ mixes electron states of opposite parity 
and the same angular momentum (it is a scalar). 
It is a local operator, so we need only consider the mixing of $s$ and $p_{1/2}$ states. 
The matrix element $\langle p_{1/2}|\hat{h}_{W}|s\rangle$, 
with non-relativistic single-particle $s$ and $p_{1/2}$ electron states, 
is proportional to $Z^{2}Q_{W}$. One factor of $Z$ here comes from the 
probability for the valence electron to be at the nucleus, 
and the other from the operator ${\bf p}$ which, near the nucleus 
(unscreened by atomic electrons), is proportional to $Z$.
The nuclear weak charge $|Q_{W}|\approx N\sim Z$.
(See \cite{bouchiats1,bouchiats2,khriplovichpnc} for more details.) 
It should be remembered that relativistic effects are important, 
since Dirac wave functions diverge at $r=0$, 
$\psi _{j}\propto r^{\gamma-1}$, $\gamma =\sqrt{(j+1/2)^2-Z^{2}\alpha ^{2}}$.
Taking into account the relativistic nature of the wave functions 
brings in a relativistic factor $R(Z)$ which increases with the 
nuclear charge $Z$. The factor $R\approx 10$ when $Z=80$.

As a consequence, the parity nonconserving effects in atoms increase as
\begin{equation}
\langle p_{1/2}|\hat{h}_{W}|s\rangle \propto R(Z)Z^{2}Q_{W}\ , 
\end{equation}
that is, faster than $Z^{3}$.

\section{Measurements and calculations of parity violation in atoms}
\label{section3iia}

An account of the dramatic story of the search for parity violation 
in atoms can be found in the book \cite{khriplovichpnc}. 
Below we will briefly discuss how parity violation in atoms is manifested, 
which experiments have yielded non-zero signals of parity violation, 
what quantity is measured in the atomic experiments, and what is required to 
interpret the measurements.

Parity violation in atoms produces a spin helix, 
and this helix interacts differently with right- and 
left-polarized light (see, e.g., Ref. \cite{khriplovichpnc}). 
The polarization plane of linearly polarized light will therefore 
be rotated in passing through an atomic vapour.

The weak interaction mixes states of opposite parity (parity violation), 
e.g., $|p\rangle +\beta |s\rangle$. 
Therefore, an M1 transition in atoms will have a component originating from 
an E1 transition between states of the same nominal parity, $E_{PNC}$, 
e.g. $p_{1/2}-p_{3/2}$. 
The rotation angle per absorption length in such a transition is 
proportional to the ratio ${\rm Im}(E_{PNC})/M1$. While it may appear that it is more 
rewarding to study M1 transitions that are highly forbidden, where 
there is a larger rotation angle, the ordinary M1 transitions are in 
fact more convenient for experimental investigation since the angle per unit 
length $\approx {\rm Im}{E_{PNC}}~{M1}$ 
(see, e.g., \cite{khriplovichpnc}). 

In measurements of parity violation in highly-forbidden M1 transitions, 
an electric field $\varepsilon$ is applied to open up the forbidden transition. 
The M1 transition then contains a Stark-induced E1 component $E_{\rm Stark}$
which the parity violating amplitude interferes with. 
In such experiments the ratio ${\rm Im}(E_{PNC})/\beta$ is measured, 
where $\beta$ is the vector transition polarizability, 
$E_{\rm Stark}\sim \beta \varepsilon$.

Atomic many-body theory is required to 
calculate the parity nonconserving E1 transition amplitude $E_{PNC}$. 
This is expressed in terms of the fundamental $P$-odd parameters like the 
nuclear weak charge $Q_{W}$.  
Interpretation of the measurements in terms of the $P$-odd parameters also 
requires a determination of $M1$ or $\beta$.

\subsection{Summary of measurements}

Zel'dovich was the first to propose optical rotation experiments in atoms 
\cite{zeldovich1959}. 
Unfortunately, he only considered hydrogen where PNC effects are small. 
Optical rotation experiments in Tl, Pb, and Bi were proposed by 
Khriplovich \cite{khriplovich1974}, Sandars \cite{sandars1975prop}, 
and Sorede and Fortson \cite{SF1975}. 
These proposals followed those by the Bouchiats to measure PNC in 
highly forbidden transitions in Cs and Tl \cite{bouchiats2,bouchiats1}.

The first signal of parity violation in atoms was seen in 1978 at 
Novosibirsk in an optical rotation experiment with bismuth \cite{BZ}. 
Now atomic PNC has been measured in bismuth, lead, thallium, and cesium. 
%what about Fr, Yb, Dy, Sm, Ba^{+} - are any of these complete?
PNC effects were measured by optical rotation in the following atoms and transitions:  
in $^{209}$Bi in the transition
$6s^{2}6p^{3}~ ^{4}S_{3/2}-6s^{2}6p^{3}~ ^{2}D_{5/2}$ 
by the Novosibirsk \cite{BZ}, Moscow \cite{birich84}, and 
Oxford \cite{taylor87,warrington93} groups 
and in the transition $6s^{2}6p^{3}~ ^{4}S_{3/2}-6s^{2}6p^{3}~ ^{2}D_{3/2}$ 
by the Seattle \cite{hollister81} and Oxford \cite{macpherson87,macpherson91} groups; 
in $6s^{2}6p^{2}~ ^{3}P_{0}-6s^{2}6p^{2}~ ^{3}P_{1}$ in $^{208}$Pb 
at Seattle \cite{emmons84,meekhof93} and Oxford \cite{phipp96}; and 
in the transition $6s^{2}6p~ ^{2}P_{1/2}-6s^{2}6p~ ^{2}P_{3/2}$ in natural Tl 
($70.5\%$ $^{205}$Tl and $29.5\%$ $^{203}$Tl) at Oxford \cite{wolfenden91,edwards95} and 
Seattle \cite{vetter95}. 
The highest accuracy that has been reached in each case is: 
$9\%$ for $^{209}$Bi $^{4}S_{3/2}-^{2}D_{5/2}$ \cite{warrington93}, 
$2\%$ for $^{209}$Bi $^{4}S_{3/2}-^{2}D_{3/2}$ \cite{macpherson91}, 
$1\%$ for $^{208}$Pb \cite{meekhof93}, and 
$1\%$ for Tl \cite{vetter95}
% also cite oxford experiment?
.
%are other isotopes pure?

The Stark-PNC interference method was used to measure PNC in the 
highly-forbidden M1 transitions:  
$6s~ ^{2}S_{1/2}-7s~ ^{2}S_{1/2}$ in $^{133}$Cs 
at Paris \cite{bouchiat82,bouchiat84,bouchiat85,guena2003} and 
Boulder \cite{gilbert85,noecker88,wieman} and 
$6s^{2}6p~ ^{2}P_{1/2}-6s^{2}7p~ ^{2}P_{1/2}$ in $^{203,205}$Tl at 
Berkeley \cite{conti81,drell84}. 
In the most precise Tl Stark-PNC experiment \cite{drell84} an accuracy of 
20\% was reached. 
In 1997, PNC in Cs was measured with an accuracy of 
0.35\% \cite{wieman} -- an accuracy unprecedented in measurements of PNC in atoms. 

Results of atomic PNC measurements accurate to sub-$5\%$ are listed in 
Table \ref{tab:pncexp<5}. 

Several PNC experiments in rare-earth atoms have been prompted by the 
possibility of enhancement of the PNC effects due to the presence of anomously 
close levels of opposite parity \cite{DFK86}. 
Another attractive feature of rare earth atoms is their abundance of stable isotopes. 
Taking ratios of measurements of PNC in different isotopes of the same element 
removes from the interpretation the dependence on atomic theory \cite{DFK86}; 
see Section \ref{section3iic}.
Null measurements of PNC have been reported for M1 transitions in the 
ground state configuration $4f^{6}6s^{2}$ of samarium at Oxford 
\cite{wolfenden93,lucas98} 
and for the $4f^{10}6s^{2}~J=8 - 4f^{10}5d6s~J=10$ transition in dysprosium 
at Berkeley \cite{nguyen97}. 
The upper limits were smaller than expected by theory. 
%add reference?

For a recent review of measurements of atomic PNC, we refer the reader to \cite{budker1999}; 
for a review of the early measurements, see, e.g., \cite{fortsonlewis}. 
For comprehensive reviews, please see the book \cite{khriplovichpnc} and the more recent 
review \cite{bouchiatsreview}. 

\subsection{Summary of calculations}

The interpretation of the PNC measurements is limited by atomic structure 
calculations. 
%Calculations of PNC amplitudes have been performed for 
%bismuth in the transition ${^{4}S}_{3/2}-{^{2}D}_{3/2}$ 
%\cite{.......} 
%and the transition ${^{4}S}_{3/2}-{^{2}D}_{5/2}$ 
%\cite{.........}; 
%for lead ${^{3}P}_{0}-{^{3}P}_{1}$ 
%\cite{.......}; 
%for thallium in the transition $6P_{1/2}-6P_{3/2}$ 
%\cite{NSK1976,HW1976,NC1977,HKW1977,DFSS1987b,hartley1990b,hartleytl,kozlovtl} 
%and the transition $6P_{1/2}-7P_{1/2}$ 
%%also Martensson-Pendrill ???
%\cite{bouchiats1,SFK1976,NC1977,das1982,calc10,johnson8586,DFSS1987b,parpia1988,hartley1990b,hartleytl}; 
%and for cesium $6S-7S$ 
%\cite{bouchiats1,bouchiats2,calc2,calc3,calc4,calc5,calc6,calc7,calc8,M-P1985jp,calc10,calc11,johnson8586,calc13,DFSS1987b,calc15,parpia1988,DFS1989pnc,hartley1990a,hartley1990b,BJS1990,safronova2000,kozlovcs,DFG02}
The theoretical uncertainty for thallium is at the level of 2.5-3\% 
for the transition $6P_{1/2}-6P_{3/2}$ \cite{DFSS1987b,kozlovtl}, 
and is worse for the transition $6P_{1/2}-7P_{1/2}$ at 6\% \cite{DFSS1987b} 
and for lead (8\%) \cite{DFSS88} and bismuth (12\% for the $876~{\rm nm}$ transition 
${^{4}S}_{3/2}-{^{2}D}_{3/2}$
and about 70\% for the $648~{\rm nm}$ transition ${^{4}S}_{3/2}-{^{2}D}_{5/2}$) 
\cite{DFSS88,DFS1989pnc}. 
The sizeable error in the calculation for the Bi $648~{\rm nm}$ transition arises 
because there is a strong cancellation of the zeroth order contribution by the 
first-order correlation corrections, with the amplitude then being comprised 
largely of the contributions of 
higher-order correlations \cite{DFSS88,DFS1989pnc}. 
Cesium is the simplest atom of interest in PNC experiments, 
it has one electron above compact, closed shells. 
The precision of the atomic calculations for Cs is $0.5\%$ 
\cite{DFG02} 
(see also calculations accurate to better than 1\%, 
\cite{DFS1989pnc,BJS1990,kozlovcs}).
For references to earlier calculations for the above atoms and transitions, 
see, e.g., the book \cite{khriplovichpnc}.

In Table \ref{tab:pnccalcs} we present the values of the most precise calculations for the 
PNC amplitudes corresponding to those atoms and transitions in which high-precision measurements 
($<5\%$ error) have been performed (Table \ref{tab:pncexp<5}). 

\subsection{Cesium}

Because of the extraordinary precision that has been achieved in 
measurements of cesium, and the clean interpretation of the measurements 
(compared to other heavy atoms), in this review we concentrate mainly 
on parity violation in cesium. The high precision of the nuclear weak charge 
extracted from cesium has made this system important in low-energy tests 
of the standard model and has made it one of the most sensitive probes of new physics. 
Measurements of parity violation in cesium have also opened up a new window 
from which parity violation within the nucleus (the nuclear anapole moment; 
see Section \ref{section6}) can be studied. 

Below we list the measurements and calculations for cesium that have 
been performed over the years, culminating in a 0.35\% measurement 
and 0.5\% calculation.

\subsubsection{Measurements}

Measurements of parity violation in the highly forbidden 
$6S-7S$ transition in Cs were first suggested and considered in 
detail in the landmark works of the Bouchiats 
\cite{bouchiats1,bouchiats2}. Measurements 
have been performed independently by the Paris group 
\cite{bouchiat82,bouchiat84,bouchiat85,guena2003} 
and the Boulder group \cite{gilbert85,noecker88,wieman}.
The results of the Cs PNC experiments are summarized in Table \ref{tab:csexps}.

The Paris result in the first row is the average \cite{bouchiat85} of their 
(revised) results for the measurements of PNC in the transitions 
$6S_{F=4}-7S_{F=4}$ \cite{bouchiat82} and $6S_{F=3}-7S_{F=4}$ \cite{bouchiat84}. 
(The nuclear angular momentum of $^{133}$Cs $I=7/2$ and the electron 
angular momentum $J=1/2$, so the total angular momentum of the atom is 
$F=3,4$). The Paris group have very recently performed a new measurement of PNC in Cs 
(last row) using a novel approach, chiral optical gain \cite{guena2003}.

Each of the Boulder results \cite{gilbert85,noecker88,wieman} cited in the 
table is an average of PNC in the hyperfine transitions 
$6S_{F=4}-7S_{F=3}$ and $6S_{F=3}-7S_{F=4}$. 
The accuracy of the latest result is 0.35\%, several times 
more precise than the best measurements of parity violation 
in other atoms.  

The PNC nuclear spin-independent component, 
arising from the nuclear weak charge, makes the same contribution 
to all hyperfine transitions. So averaging the PNC amplitudes 
over the hyperfine transitions gives the contribution from the 
nuclear weak charge.

PNC in atoms dependent on the nuclear spin 
was detected for the first (and only) time in Ref. \cite{wieman} 
where it appeared as a difference in the PNC amplitude in 
different hyperfine transitions. 
The dominant mechanism for nuclear spin dependent effects in atoms, 
the nuclear anapole moment, is the subject of Section \ref{section6}.

\subsubsection{Calculations}

Numerous calculations of the Cs $6S-7S$ $E_{PNC}$ amplitude have 
been performed over the years. These calculations are summarized 
in Table \ref{tab:cscalcs}. 
%Semi-empirical methods were used in the works 
%\cite{bouchiats1,bouchiats2,calc2,calc3,calc6,johnson8586,calc13}.
%In 
%\cite{calc4,calc5,calc7,calc8,M-P1985jp,calc10,calc11,johnson8586,DFSS1987b,calc15,DFS1989pnc,hartley1990b,BJS1990,safronova2000,kozlovcs,DFG02} 
%{\it ab initio} many-body methods were used. 
%In \cite{parpia1988,hartley1990a} a combination of many-body and 
%semi-empirical methods were employed.
The many-body calculations \cite{DFS1989pnc,BJS1990}, accurate to 
1\%, performed more than ten years ago represented a significant step forward 
%more modest word? 
for atomic many-body theory and parity violation in atoms. 
At the time, these calculations were unmatched by the PNC measurements 
which were accurate to 2\%. 
The method of calculation used in Ref. \cite{DFS1989pnc} is the 
subject of Sections \ref{section3},\ref{section3iib}. 
The method used in Ref. \cite{BJS1990} is based 
on the popular coupled-cluster method, and we refer the interested reader 
to this work for details.

In the last ten years a series of new measurements have been performed for 
quantities used to test the accuracy of the atomic calculations \cite{DFS1989pnc,BJS1990}, 
such as electric dipole transition amplitudes (see \cite{BW}). 
The new measurements are in agreement with the calculations, resolving a previous 
discrepancy between theory and experiment. 
This inspired Bennett and Wieman \cite{BW} to claim that the atomic theory is 
accurate to 0.4\% rather than 1\% claimed by theorists. 
Since then, a number of previously unaccounted for contributions to the PNC 
amplitude have been discovered, the Breit interaction and more recently the 
strong-field radiative corrections, that enter above the 0.4\% level, but 
below 1\% (see Section \ref{section3iib}).  

%The recent calculation \cite{kozlovcs} is a simplified version of 
%the method used in Ref. \cite{DFS1989pnc}.

A re-calculation of the work \cite{DFS1989pnc}, 
with some further improvements, was performed recently, with 
a full analysis of the accuracy of the PNC amplitude.  
This work, Ref. \cite{DFG02}, represents the most accurate (0.5\%) 
calculation to date. It is described in 
detail in Section \ref{section3iib}.

The result of \cite{DFG02} differs from 
\cite{DFS1989pnc,BJS1990} by only $\sim 0.1\%$ if 
Breit, vacuum polarization, and neutron distribution corrections 
are excluded. One may interpret this as grounds for asserting that 
the many-body calculations \cite{DFS1989pnc,BJS1990,kozlovcs,DFG02} 
have an accuracy of $0.5\%$ in agreement with the conclusion of \cite{BW}.

\section{Method for high-precision atomic structure calculations in heavy alkali-metal atoms} 
\label{section3}

In this section we describe methods that can be used to obtain high accuracy 
in calculations involving many-electron atoms with a single valence electron. 
These are the methods that have been used to obtain the most precise calculation 
of parity nonconservation in Cs. They were originally developed in works 
\cite{DFSS1985,DFSS1988,DFS1989energy,DFSS1987b} 
and applied to the calculation of PNC in Cs in Ref. \cite{DFS1989pnc}. 
In \cite{DFS1989pnc} it was claimed that the atomic theory is accurate to 1\%. 
A complete re-calculation of PNC in Cs using this method, with a new analysis 
of the accuracy, indicates that the error is as small as 0.5\% \cite{DFG02}.
(We refer the reader to Section \ref{section3iib}, where this question of accuracy is 
discussed in general; please also see Section \ref{section3iib} for an in-depth 
discussion of PNC in Cs.)

In this section the method is applied to energies, electric dipole (E1) transition 
amplitudes, and hyperfine structure (hfs). A comparison of the calculated and 
experimental values gives an indication of the quality of the many-body wave functions.
Note that the above quantities are sensitive to the wave functions at different 
distances from the nucleus. Hyperfine structure, energies, and E1 amplitudes are 
dominated by the contribution of the wave functions at small, intermediate, and large 
distances from the nucleus.
We concentrate on calculations for Cs relevant to the $6S-7S$ PNC E1 amplitude 
(see Eq. (\ref{eq:pncsum}) and Section \ref{sssec:acc}).

A brief overview of the method is presented in Section \ref{overview}. 
For those not interested in the technical details of the 
atomic structure calculations, Sections \ref{ssec:RHF}-\ref{ssec:extfields} 
may be omitted without loss of consistency.

\subsection{Overview}
\label{overview}

The calculations begin in the relativistic Hartree-Fock (RHF) approximation. 
The $N-1$ self-consistent RHF orbitals of the core are found 
($N$ is the total number of electrons in the atom), 
and the external electron is solved in the potential of the core electrons 
(the $\hat{V}^{N-1}$ potential). RHF wave functions, energies, and 
Green's functions are obtained in this way.

Correlation corrections to the external electron orbitals are included 
in second (lowest) order in the residual interaction 
($\hat{V}_{\rm exact}-\hat{V}^{N-1}$), where $\hat{V}_{\rm exact}$ is the 
exact Coulomb interaction between the atomic electrons.
The correlations are included into the external electron orbitals 
by adding the correlation potential (the self-energy operator) to the RHF 
potential when solving for the external electron. 
Using the Feynman diagram technique, 
important higher-order diagrams are included into the self-energy in all 
orders: screening of the electron-electron interaction and 
the hole-particle interaction. The self-energy is then iterated 
using the correlation potential method.

Interactions of the atomic electrons with external fields are calculated 
using the time-dependent Hartree-Fock (TDHF) method; 
this method is equivalent to the random-phase approximation (RPA) with exchange. 
Using this approach we can take into account the polarization of the 
atomic core by external fields to all orders. 
Then the major correlation corrections are included as corrections to electron 
orbitals (Brueckner orbitals). Small correlation corrections (structural radiation, 
normalization) are taken into account using many-body perturbation theory.

\subsection{Zeroth-order approximation: relativistic Hartree-Fock method}
\label{ssec:RHF}

The full Hamiltonian we wish to solve is the many-electron Dirac equation\footnote{
In Section \ref{sssec:Breit} we discuss the inclusion of 
the Breit interaction into the Hamiltonian.}
\begin{equation}
\label{exacth}
\hat{H}=
\sum _{i=1}^{N}[\mbox{\boldmath$\alpha$}_{i}\cdot {\bf p}_i+(\beta -1)m -Ze^{2}/r_i] 
+ \sum_{i<j}\frac{e^2}{|{\bf r}_i-{\bf r}_j|} \ .
\end{equation}
Here $p$ is the electron momentum, $\mbox{\boldmath$\alpha$}$ and $\beta$ are 
Dirac matrices, $Ze$ is the nuclear charge and 
$N$ is the number of electrons in the atom ($N=55$ for cesium).
This equation cannot be solved exactly, 
so some approximation scheme must be used.
This is done by excluding the complicated Coulomb term and adding 
instead some averaged potential in which the electrons move. 
The Coulomb term, minus the averaged potential, can be added back 
into the equation perturbatively. 

\begin{figure}[b]
\centerline{\psfig{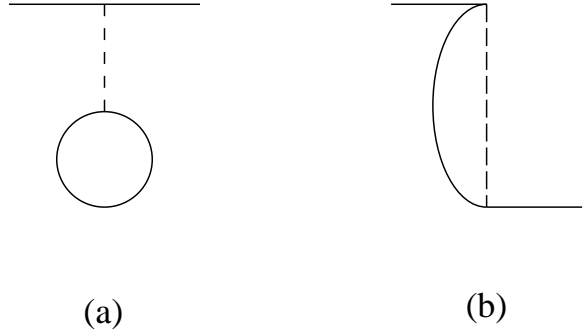}}
\caption{Hartree-Fock (a) direct and exchange (b) diagrams for energies. 
The solid and dashed lines are the electron and Coulomb lines, respectively.}
\label{fig:hf}
\end{figure}

It is well known that choosing the electrons to move in the self-consistent 
Hartree-Fock potential $\hat{V}^{N-1}$, in the zeroth order approximation, 
simplifies the calculations of higher-order terms 
(we will come to this in the next section).
The single-particle relativistic Hartree-Fock (RHF) Hamiltonian is 
\begin{equation}
\label{eq:RHF}
\hat{h}_{0}=\mbox{\boldmath$\alpha$}\cdot {\bf p}+(\beta -1)m -Ze^{2}/r+\hat{V}^{N-1} 
\ ,
\end{equation}
$\hat{H}_{0}=\sum _{i}\hat{h} _{0}^{(i)}$, where the Hartree-Fock potential
\[\hat{V}^{N-1}=\hat{V}_{\rm dir}+\hat{V}_{\rm exch} \ ,\]
is the sum of the direct and 
nonlocal exchange potentials created by the $(N-1)$ core 
electrons $n$,
\begin{eqnarray}
\hat{V}_{\rm dir}\psi (\bf r)&=&
e^2\sum _{n=1}^{N-1}\int \frac{\psi ^{\dag}_{n}
({\bf r}_{1})\psi _{n}({\bf r}_1)}{|{\bf r}-{\bf r}_1|}
d{\bf r}_{1}\psi ({\bf r})\\
\hat{V}_{\rm exch}\psi (\bf r)&=&
-e^2\sum _{n=1}^{N-1}\int \frac{\psi ^{\dag}_{n}
({\bf r}_{1})\psi ({\bf r}_1)}{|{\bf r}-{\bf r}_1|}
d{\bf r}_{1}\psi _n ({\bf r}) \ .
\end{eqnarray}
The direct and exchange Hartree-Fock potentials are presented diagrammatically 
in Fig. \ref{fig:hf}.
The Schr\"{o}dinger equation
\begin{equation}
\label{eq:schr}
\hat{h}_{0}\psi _i =\epsilon _i \psi _i \ ,
\end{equation}
where $\psi _i$, $\epsilon _i$ are single-particle wave functions and 
energies, is solved self-consistently for the $N-1$ core electrons. 
The Hartree-Fock potential is then kept ``frozen'' and the RHF equation 
(\ref{eq:RHF},\ref{eq:schr}) is solved for the states of the external electron. 
The Hamiltonian $\hat{H}_{0}$ thus generates a complete 
orthogonal set of single-particle orbitals for the core and valence 
electrons \cite{DFS1983}.

Because we are performing calculations for heavy atoms, 
and we are interested in interactions that take place in the vicinity of the nucleus 
(the weak and hyperfine interactions), 
the finite size of the nucleus needs to be taken into account. 
We use the standard formula for the charge distribution in the nucleus
\begin{equation}
\label{density}
\rho (r)=\frac{\rho _{0}}{1+{\rm exp}[(r-c)/a]},
\end{equation}
where $\rho _{0}$ is the normalization constant found from the condition 
$\int \rho (r) {\rm d}^{3}r=1$, $t=a(4\ln 3)$ is the skin-thickness, 
and $c$ is the half-density radius.
We take $t=2.5~{\rm fm}$ and $c=5.6710~{\rm fm}$ 
($\langle r^{2}\rangle ^{1/2}=4.804~{\rm fm}$) \cite{fricke}.
%In the range of small $r$, the potential of the nucleus is determined by the 
%numerical integration of the charge density.

Energy levels of cesium states relevant to the $6S$-$7S$ E1 PNC transition 
are presented in Table \ref{tab:energies}. 
It is seen that the RHF energies agree with experiment to 10\%.

In order to obtain more realistic wave functions, we need to take into account 
the effect of correlations between the external electron and the core. 
We describe the techniques used to calculate these correlations in the 
following sections.

\subsection{Correlation corrections and many-body perturbation theory}
\label{ssec:corr}

The subject of this section is the inclusion of electron-electron 
correlations into the single-particle electron orbitals using
many-body perturbation theory. 
We will see that high accuracy can 
be reached in the calculations by using the Feynman diagram technique 
as a means of including dominating classes of diagrams in all orders.

The correlation corrections can be most accurately calculated 
in the case of alkali-metal atoms (for example, cesium). 
This is because the external electron has very little overlap with 
the electrons of the tightly bound core, enabling the use of perturbation 
theory in the calculation of the residual interaction of the 
external electron with the core.  

\begin{figure}[b]
\centerline{\psfig{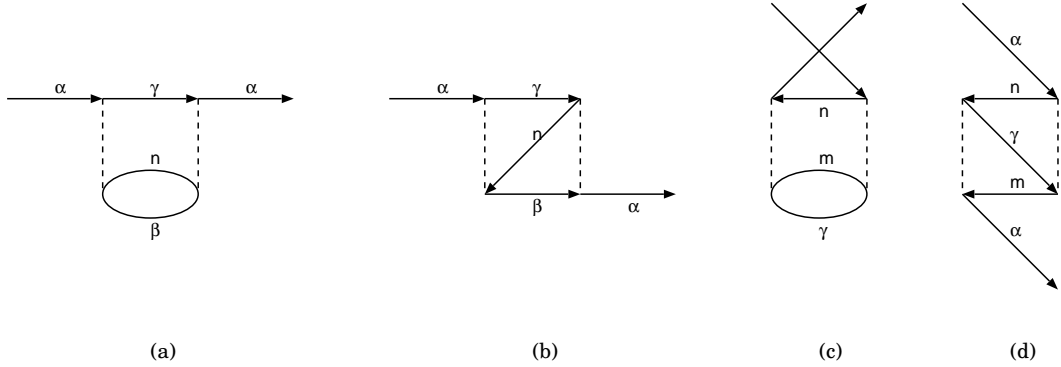}}
\caption{Second-order correlation diagrams for the valence 
electron ($\hat \Sigma$ operator). Dashed line is the Coulomb interaction 
between core and valence electrons. Loop is the polarization of the atomic
core which corresponds to the virtual creation of the excited electron and
a hole in the core shells. Here $\alpha $ is the state of the external 
electron; $n$, $m$ are core states; and $\beta$, $\gamma$ are states outside 
the core.}
\label{2ndordercorr}
\end{figure}

The exact Hamiltonian of an atom [Eq. (\ref{exacth})] 
can be divided into two parts: 
the first part is the sum of the single-particle Hamiltonians, 
and the second part represents the residual Coulomb interaction
\begin{eqnarray}
\hat{H}&=&\sum _{i=1}^N \hat{h}_{0}({\bf r}_i)+\hat{U} \ , \\
\hat{U}&=&\sum_{i<j}\frac{e^2}{|{\bf r}_i-{\bf r}_j|}-
\sum_{i=1}^{N}\hat{V}^{N-1}({\bf r}_i) \ . \label{eq:residual}
\end{eqnarray}
Correlation corrections to the single-particle orbitals are included 
perturbatively in the residual interaction $\hat{U}$.
By calculating the wave functions in the Hartree-Fock potential 
$\hat{V}^{N-1}$ for the zeroth-order approximation, the perturbation 
corrections are simplified. The first-order corrections (in the 
residual Coulomb interaction $\hat{U}$) to the ionization 
energy vanish, since the diagrams in first-order in the Coulomb interaction 
are nothing but the Hartree-Fock ones (Fig. \ref{fig:hf}). 
So two terms in Eq. (\ref{eq:residual}) cancel each other. 
The lowest-order corrections therefore correspond to those arising in 
second-order perturbation theory, $\hat{U}^{(2)}$. These corrections are determined 
by the four Goldstone diagrams in Fig. \ref{2ndordercorr} \cite{DFS1983}.
They can be calculated by direct summation over intermediate states \cite{DFS1983} 
or by the ``correlation potential'' method \cite{DFSS1985}. 
This latter method gives higher accuracy and, 
along with the Feynman diagram technique to be discussed in the following 
section, enables 
the inclusion of higher-order effects: electron-electron screening, 
the hole-particle interaction, and the nonlinear contributions of 
the correlation potential.

The correlation potential method corresponds to adding a nonlocal 
correlation potential $\hat{\Sigma}$ to the potential $\hat{V}^{N-1}$ 
in the RHF equation (\ref{eq:RHF}) and then solving for the states 
of the external electron. The correlation potential is defined such 
that its average value coincides with the correlation correction to 
the energy,
\begin{eqnarray}
\delta \epsilon_{\alpha}&=&\langle \alpha |\hat{\Sigma}|\alpha\rangle\\
\hat{\Sigma}\psi _{\alpha}&=&\int\hat{\Sigma}({\bf r}_1,{\bf r}_2,\epsilon_{\alpha})
\psi_{\alpha}({\bf r}_1)d^{3}r_{1} \ .
\end{eqnarray}
It is easy to write the correlation potential explicitly. For example, 
a part of the operator $\hat{\Sigma}({\bf r}_1,{\bf r}_2,\epsilon_{\alpha})$ 
corresponding to Fig. \ref{2ndordercorr}(a) is given by
\begin{equation}
\hat{\Sigma}^{a}({\bf r}_1,{\bf r}_2,\epsilon_{\alpha})=
e^4\sum_{n,\beta,\gamma}\int\int
d^{3}r_3d^{3}r_4\frac{\psi _{n}^{\dag}({\bf r}_{4})r_{24}^{-1}
\psi _{\beta}({\bf r}_{4})\psi _{\gamma}({\bf r}_{2})
\psi _{\beta}^{\dag}({\bf r}_{3})\psi ^{\dag}_{\gamma}({\bf r}_{1})
r_{13}^{-1}\psi _{n}({\bf r}_{3})}
{\epsilon_{\alpha}+\epsilon_{n}-\epsilon_{\gamma}-\epsilon_{\beta}} .
\end{equation}
Note that $\hat{\Sigma}$ is a single-electron and energy-dependent 
operator.
By solving the RHF equation for the states of the external electron in 
the field $\hat{V}^{N-1}+\hat{\Sigma}$, we obtain ``Brueckner'' 
orbitals and energies.\footnote{
Note that there is a slight distinction in the definition of these Brueckner orbitals 
and those defined in, e.g., \cite{lindgrenbook}.}
The largest correlation corrections are included in the Brueckner orbitals. 
 
See Table \ref{tab:energies} for Brueckner energies of the lower states of cesium 
calculated in the secon-order correlation potential. 
$\hat{\Sigma} ^{(2)}\equiv \hat{U}^{(2)}$ denotes the ``pure'' second-order correlation potential 
(without screening, etc.).
It is seen that the inclusion of these corrections improves the  
energies significantly, from the level of 10\% deviation from experiment 
for the RHF approximation to the level of 1\%.

\subsection{All-orders summation of dominating diagrams}
\label{ssec:sum}

We saw in the previous section that when we take into account 
second-order correlation corrections, the accuracy for energies 
is improved significantly beyond that for energies calculated in 
the RHF approximation. 
However, the corrections are overestimated. This overestimation is 
largely due to the neglect of screening in the electron-electron 
interaction.

In this section we describe the calculations of three series of 
higher-order diagrams: 
screening of the electron-electron interaction and
the hole-particle interaction, which are inserted into 
the correlation potential $\hat{\Sigma}$; 
and iterations of $\hat{\Sigma}$.
With the inclusion of these diagrams the accuracy for energies
is improved to the level of 0.1\% (see Table \ref{tab:energies}).

The screening of the electron-electron interaction is a collective 
phenomenon and is similar to Debye screening in a plasma; 
the corresponding chain of diagrams is enhanced by a factor approximately 
equal to the number of electrons in the 
external closed subshell (the $5p$ electrons in cesium) 
\cite{DFSS1988}. 
The importance of this effect can be understood by looking at 
a not dissimilar example in which screening effects are important, 
for instance, the screening of an external electric field in an atom. 
According to the Schiff theorem \cite{schiff}, 
a homogeneous electric field is screened by atomic electrons 
(and at the nucleus it is zero). (See \cite{DFSS1986} where 
a numerical calculation of an external electric field inside the atom has 
been performed.)

The hole-particle interaction is enhanced by the large zero-multipolarity 
diagonal matrix elements of the Coulomb interaction \cite{DFS1989energy}. 
The importance of this effect can be seen by noticing that the 
existence of the discrete spectrum excitations in noble gas atoms 
are due only to this interaction (see, e.g., \cite{amusia1975}).

The non-linear effects of the correlation potential are 
calculated by iterating the self-energy operator. 
These effects are 
enhanced by the small denominator, which is the energy for the 
excitation of an external electron 
(in comparison with the excitation energy of a core electron) 
\cite{DFS1989energy}.

All other diagrams of perturbation theory are proportional to 
powers of the small parameter $Q_{nd}/\Delta \epsilon_{\rm int}\sim 10^{-2}$, 
where $Q_{nd}$ is a nondiagonal Coulomb integral and 
$\Delta \epsilon_{\rm int}$ is a large energy denominator corresponding 
to the excitation of a core electron \cite{DFS1989energy}.

\subsubsection{Screening of the electron-electron interaction}
\label{sss:screening}

\begin{figure}[b]
\centerline{\psfig{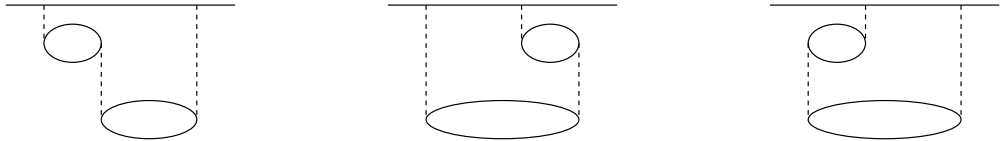}}
\caption{Lowest order screening corrections to the diagram in 
Fig. \ref{2ndordercorr}(a).}
\label{screeningeg}
\end{figure}

The main correction to the correlation potential comes from the 
inclusion of the screening of the Coulomb field by the 
core electrons. Some examples of the lowest-order screening 
corrections are presented in Fig. \ref{screeningeg}. 
When screening diagrams in the lowest (third) order of perturbation 
theory are taken into account, a correction is obtained of 
opposite sign and almost the same absolute value as the 
corresponding second-order diagram \cite{DFSS1988}. 
Due to these strong cancellations 
there is a need to sum the whole chain of screening diagrams. 
However, this task causes difficulties in standard perturbation theory 
as the screening diagrams in the correlation correction cannot be 
represented by a simple geometric progression due to the overlap 
of the energy denominators of different loops (such an overlap indicates a 
large number of excited electrons in the intermediate states, see e.g. 
Fig. \ref{screeningeg}(b,c)). This summation problem is solved by 
using the Feynman diagram technique.

\begin{figure}[b]
\centerline{\psfig{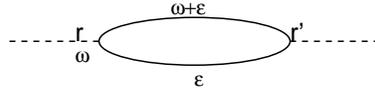}}
\caption{Polarization operator.}
\label{polarop}
\end{figure}

The correlation corrections to the energy in the Feynman diagram 
technique are presented in Fig. \ref{feyncorr}. The Feynman Green's 
function is of the form
\begin{equation}
\label{Greenfn}
\hat{G}_{{\bf r}_{1}{\bf r}_{2}}(\epsilon)=
\sum_{n}\frac{\psi_{n}({\bf r}_{1})\psi ^{\dagger}_{n}({\bf r}_{2})}
{\epsilon-\epsilon_{n}-i\delta} +    
\sum_{\gamma}\frac{\psi_{\gamma}({\bf r}_{1})
\psi ^{\dagger}_{\gamma}({\bf r}_{2})}
{\epsilon-\epsilon_{\gamma}+i\delta}\ ,\qquad \delta \rightarrow 0\ ,
\end{equation}
where $\psi_{n}$ is an occupied core electron state, $\psi_{\gamma}$ 
is a state outside the core. While the simplest way of calculating 
the Green's function is by direct summation over the discrete and 
continuous spectrum, there is another method in which higher numerical 
accuracy can be achieved. As is known, the radial Green's function $G_0$ 
for the equation without the nonlocal exchange interaction $V_{\rm exch}$ 
can be expressed in terms of the solutions $\chi _0$ and $\chi _\infty$ 
of the Schr\"{o}dinger or Dirac equation that are regular at 
$r\rightarrow 0$ and $r\rightarrow \infty$, respectively:
$G_{0}(r_1,r_2)\sim \chi_0(r_<)\chi_{\infty}(r_>)$, 
$r_<={\rm min}(r_1,r_2)$, 
$r_>={\rm max}(r_1,r_2)$. The exchange interaction is taken into account 
by solving the matrix equation 
$\hat{G}=\hat{G}_0+\hat{G}_0\hat{V}_{\rm exch}
\hat{G}$. The polarization operator (Fig. \ref{polarop}) is given by
\begin{equation}
\hat{\Pi}_{{\bf r}_{1}{\bf r}_{2}}(\omega)=
\int_{-\infty}^{\infty}\frac{d\epsilon}{2\pi}
\hat{G}_{{\bf r}_{1}{\bf r}_{2}}(\omega +\epsilon)
\hat{G}_{{\bf r}_{2}{\bf r}_{1}}(\epsilon).
\end{equation}  
This integration is carried out analytically, giving
\begin{eqnarray}
\label{Pi}
\hat{\Pi}_{{\bf r}_{1}{\bf r}_{2}}(\omega)
&=&\sum_{n,\gamma}\frac{2i(\epsilon _n-\epsilon _{\gamma})}
{(\epsilon _n-\epsilon_{\gamma})^2-{\omega}^2}
\psi _n^{\dag}({\bf r}_1)\psi_{\gamma}({\bf r}_1)
\psi _{\gamma}^{\dag}({\bf r}_2)\psi_n({\bf r}_2) \nonumber \\
&=&i\sum_{n}\psi_{n}^{\dag}({\bf r}_1)[\hat{G}(\epsilon _n+\omega)+\hat{G}
(\epsilon _n-\omega)]\psi_n ({\bf r}_2).
\end{eqnarray}
Using formulae (\ref{Greenfn}) and (\ref{Pi}), it is easy to perform 
analytical integration over $\omega$ in the calculation of the 
diagrams in Fig. \ref{feyncorr}.
After integration, diagram 
\ref{feyncorr}(a) transforms to \ref{2ndordercorr}(a,c) and diagram 
\ref{feyncorr}(b) transforms to \ref{2ndordercorr}(b,d).

\begin{figure}[b]
\centerline{\psfig{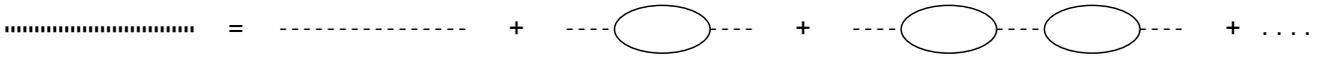}}
\caption{Screening diagram chain for effective polarization operator.}
\label{screening}
\end{figure}

\begin{figure}[b]
\centerline{\psfig{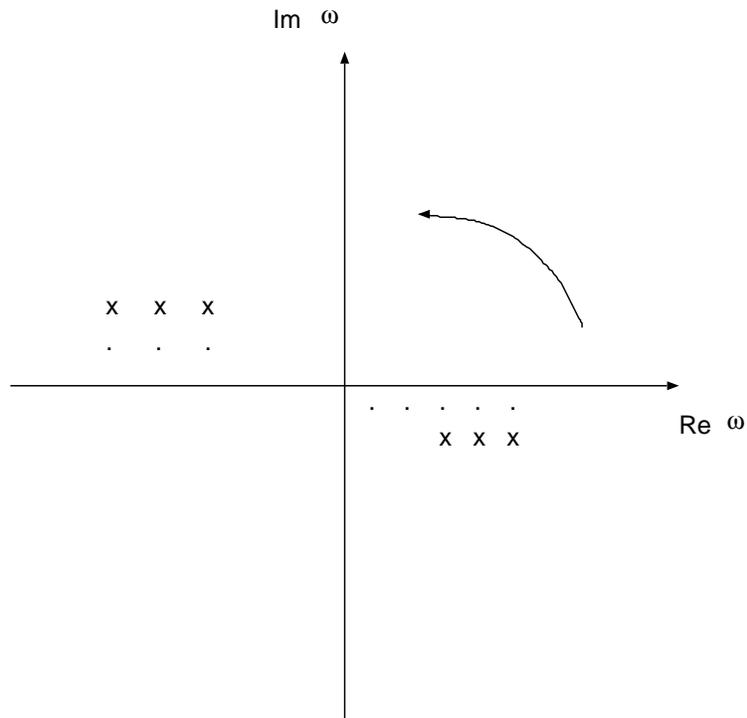}}
\caption{Rotation of the integration contour over $\omega$.  The 
points indicate the positions of Green function poles; the crosses 
denote the positions of the poles of the polarization operator.}
\label{complane}
\end{figure}

Electron-electron screening, to all orders in the Coulomb interaction, 
corresponds to the diagram chain presented in Fig. \ref{screening}. 
To calculate this we perform summation of the polarization operators 
before carrying out the integration over $\omega$. 
The whole sum of 
screening diagrams in Fig. \ref{screening} can be represented by
\begin{equation}
\label{screenedPi}
\hat{\pi}(\omega)=\hat{\Pi}(\omega)[1+i\hat{Q}\hat{\Pi}(\omega)]^{-1}.
\end{equation}
The integration over $\omega$ is performed numerically. 
The integration contour is rotated $90^{\rm o}$ from the real axis to 
the complex $\omega$ plane parallel to the imaginary axis 
(see Fig. \ref{complane}) -- this aids the numerical convergence by 
keeping the poles far from the integration contour.

The all-order electron-electron screening reduces the second-order 
correlation corrections to the energies of $S$ and $P$ states of $^{133}$Cs 
by 40\%.

\subsubsection{The hole-particle interaction}
\label{sss:holepart}

\begin{figure}[b]
\centerline{\psfig{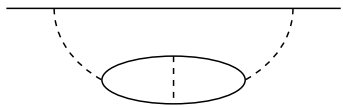}}
\caption{Insertion of the hole-particle interaction into the 
second order correlation correction.}
\label{hp}
\end{figure}

The hole-particle interaction is presented diagrammatically in Fig. \ref{hp}.
This diagram describes the alteration of the core potential 
due to the excitation of the electron from the core to the virtual 
intermediate state. 
This electron now moves in the potential created 
by the $N-2$ electrons, and no longer contributes to the 
Hartree-Fock potential. Denoting $\hat{V}_0$ as the zero multipolarity direct
potential of the outgoing electron, the potential which describes 
the excited and core states simultaneously is \cite{DFS1989energy}
\begin{equation}
\label{corexpotential}
\hat{V}=\hat{V}^{N-1}-(1-\hat{P})\hat{V}_{0}(1-\hat{P}) \ ,
\end{equation}
where $\hat{P}$ is the projection operator on the core orbitals,
\begin{equation} 
\hat{P}=\sum_{n=1}^{N-1}|n\rangle\langle n| \ .
\end{equation}
The projection operator $\hat{P}$ is introduced into the potential 
to make the excited states orthogonal to the core states.
It is easily seen that for the occupied orbitals 
$\langle \hat{V}\rangle=\langle \hat{V}^{N-1}\rangle$, while for the 
excited orbitals $\langle \hat{V}\rangle =\langle \hat{V}^{N-1}\rangle 
-\langle \hat{V}_{0}\rangle$. Strictly one should also make 
subtractions for higher multipolarities and for the exchange 
interaction as well, however these contributions are relatively small 
and are therefore safe to ignore \cite{DFS1989energy}. 

To obtain high accuracy, the hole-particle interaction in the 
polarization operator needs to be taken into account in all orders 
(see Fig. \ref{hpchain}). This is achieved by calculating  
the Green's function in the potential (\ref{corexpotential}) and then 
using it in the expression for the polarization operator 
(\ref{Pi}). The screened polarization operator, with hole-particle 
interaction included, is found by using 
the Green's function in Eq. (\ref{screenedPi}).

\begin{figure}[b]
\centerline{\psfig{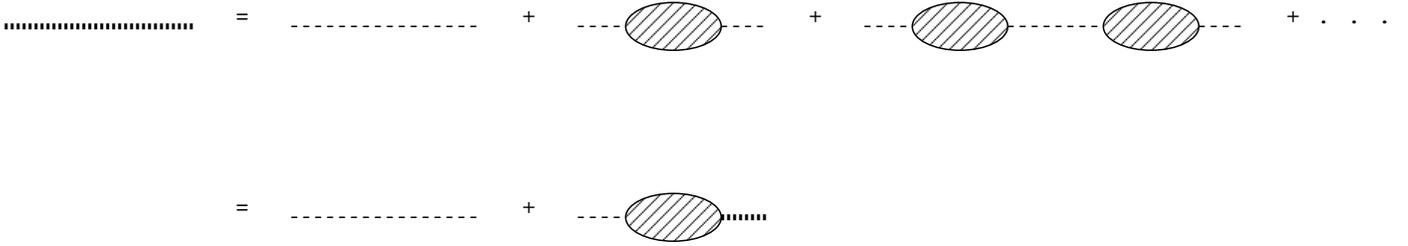}}
\caption{Renormalization of the Coulomb line due to the hole-particle 
interaction and screening.}
\label{hpscreening}
\end{figure}

The Coulomb interaction, with screening and the hole-particle 
interaction included in all orders, is calculated from the matrix equation 
\cite{DFS1989energy}
\begin{equation}
\label{eq:renormq}
\tilde{Q}=\hat{Q}-i\hat{Q}\hat{\pi}\hat{Q}.
\end{equation}  
This is depicted diagrammatically in Fig. \ref{hpscreening}.
 
\begin{figure}[b]
\centerline{\psfig{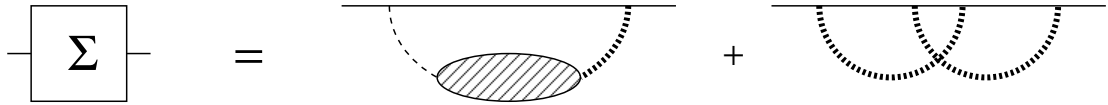}}
\caption{The electron self-energy operator with screening and hole-particle 
interaction included.}
\label{hpscreense}
\end{figure}

The infinite series of diagrams representing the screening and 
hole-particle interaction can now be included into the correlation 
potential. This is done by introducing the renormalized 
Coulomb interaction (Fig. \ref{hpscreening}) and the 
polarization operator (Fig. \ref{hpchain}) into the second-order 
diagrams according to Fig. \ref{hpscreense}.

The screened second-order correlation corrections to the energies of $S$ and $P$ 
states of cesium are increased by 30\% when the hole-particle interaction is 
taken into account in all orders.

\subsubsection{Chaining of the self-energy}
\label{sss:sechain}

\begin{figure}[b]
\centerline{\psfig{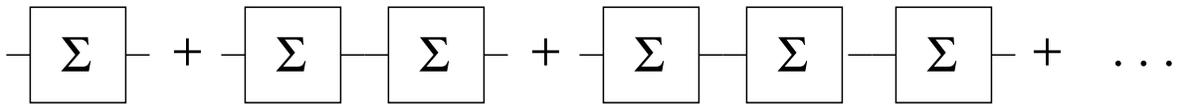}}
\caption{Chaining of the self-energy operator.}
\label{sechain}
\end{figure}

The accuracy of the calculations can be further improved by 
taking into account the nonlinear contributions of the 
correlation potential $\hat{\Sigma}$ (see Fig. \ref{sechain}). 
The chaining of the correlation potential (Fig. \ref{hpscreense}) 
to all orders is calculated by adding $\hat{\Sigma}$ to the Hartree-Fock 
potential, $\hat{V}^{N-1}$, and solving the equation
\begin{equation}
(\hat{h}_{0}+\hat{\Sigma}-\epsilon)\psi =0
\end{equation}
iteratively for the states of the external electron.
The inclusion of $\hat{\Sigma}$ into the Schr\"{o}dinger equation 
is what we call the ``correlation potential method'' and the 
resulting orbitals and energies ``Brueckner'' orbitals and 
``Brueckner'' energies (see Section \ref{ssec:corr}). 

Iterations of the correlation potential $\hat{\Sigma}$ increase the 
contributions of $\hat{\Sigma}$ (with screening and hole-particle interaction) 
to the energies of $S$ and $P$ states of cesium by about 10\%. 
 
The final results for the energies are listed in Table \ref{tab:energies}. 
The inclusion of the three series of higher-order diagrams improves 
the accuracy of the calculations of the energies to the level of $0.1\%$. 

\subsection{Other low-order correlation diagrams}
\label{ssec:othercorr}

\begin{figure}[b]
\centerline{\psfig{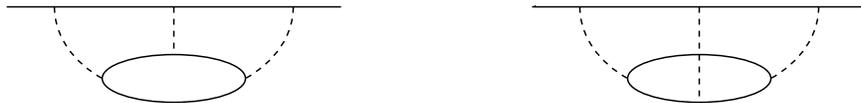}}
\caption{Third-order diagrams of the interaction of a hole and 
particle from the loop with an external electron.}
\label{other3order}
\end{figure}

Third-order diagrams for the interaction of a hole and particle 
in the polarization loop with an external electron are depicted in 
Fig. \ref{other3order}. These are not taken into account 
in the method described above. However, these diagrams are of opposite 
sign and cancel each other almost exactly \cite{DFS1989energy}: the small and almost 
constant potential of a distant external electron practically does not 
influence the wave functions of the core and excited electrons in the loop; 
it shifts the energies of the core and excited electrons by the same 
amount. This cancellation was proved in the work \cite{JIS1987} by direct 
calculation.

\begin{figure}[b]
\centerline{\psfig{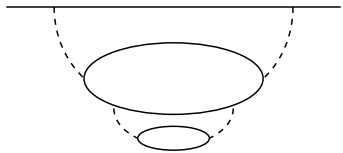}}
\caption{Correlation corrections to occupied orbitals of closed 
shells.}
\label{corrocc}
\end{figure}

Also, correlation corrections to the external electron energy arising 
from the inclusion of the self-energy into orbitals belonging to 
closed electron shells, depicted in Fig. \ref{corrocc}, are small 
and can be safely omitted 
\cite{DFSS1985}. 
%what about virtual excited states???

\subsection{Empirical fitting of the energies}
\label{ssec:fitting}

The calculations of the external electron wave functions can be 
refined by placing coefficients before the self-energy operator, 
$\hat{\Sigma}\rightarrow f\hat{\Sigma}$, 
such that the energies are reproduced exactly.
This can be considered as a way of including higher-order diagrams 
not explicitly included in the calculations. 
Comparison of quantities calculated with and without fitting can be 
used to test the stability of the wave functions and to estimate 
the contribution of unaccounted diagrams.

\subsection{Asymptotic form of the correlation potential}
\label{ssec:asymptotic}

At large distances, the correlation potential $\hat{\Sigma}$ approaches the 
local polarization potential \cite{DFSS1987a}, 
\begin{equation}
\hat{\Sigma}_{r\rightarrow \infty}\approx 
-\alpha e^{2}/2r^{4} \ ,
\end{equation}
where $\alpha$ is the polarizability of the core. This explains the 
universal behaviour of the correlation corrections to the energies of 
states of the external electron.

\subsection{Interaction with external fields}
\label{ssec:extfields}

In this section we will describe the procedures used to achieve 
high accuracy in the calculations of the interactions between 
atomic electrons and external fields (in particular, we will present 
calculations for E1 transition amplitudes and 
hyperfine structure (hfs) constants, arising from the interaction of 
the atomic electrons with the electric field of the photon and 
with the magnetic field of the nucleus, respectively).

We need to calculate the effect of the external field on the wave functions 
of the core electrons (core polarization) and then take into account the effect 
of this polarization and the external field on the valence electron.
This is achieved by using the time-dependent Hartree-Fock (TDHF) method. 
We describe this method in Section \ref{sssec:tdhf} 
and apply it to the calculation of E1 transition amplitudes and 
hfs constants in Sections \ref{sssec:e1} and \ref{sssec:hfs}, respectively.

\begin{figure}[b]
\centerline{\psfig{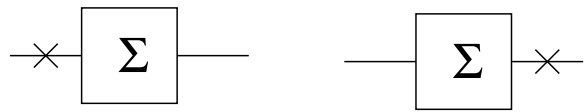}}
\caption{External field (denoted by a cross) in the external electron lines.}
\label{tdhfbrueckner}
\end{figure}

\begin{figure}[b]
\centerline{\psfig{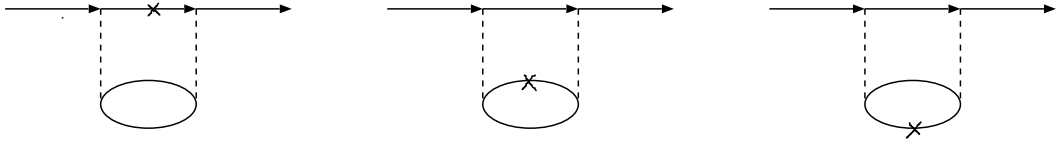}}
\caption{Structural radiation in diagram \ref{2ndordercorr}(a) of the 
self-energy operator.}
\label{structrad}
\end{figure}

The dominant correlation corrections correspond to those diagrams in which the 
interactions occur in the external lines of the self-energy operator 
(``Brueckner-type corrections''). These diagrams are presented in Fig. \ref{tdhfbrueckner}; 
they are enhanced by the small energy denominator $\epsilon _{\rm ext}$ corresponding to the 
excitation of the external electron in the intermediate states.
The Brueckner-type corrections are calculated in a similar way 
to the correlation corrections to energy (see Sections 
\ref{sssec:tdhf}, \ref{sssec:e1}, \ref{sssec:hfs}).
In Section \ref{sssec:srnorm} we describe the calculations 
of the remaining second-order corrections, i.e., ``structural radiation'' and 
normalization of states. 
Structural radiation diagrams are presented in Fig. \ref{structrad}; 
in these diagrams the external fields 
occur in the internal lines, and so their contributions are small
due to the large energy denominators $\epsilon _{\rm int}$ corresponding 
to the excitation of the core electrons. 
The relative suppression of the contributions of structural radiation compared to Brueckner-type 
corrections is $\epsilon _{\rm ext}/\epsilon _{\rm int}\sim 1/10$.

This section is largely based on the works \cite{DFSS1987b,DFSS1987a}.

\subsubsection{Time-dependent Hartree-Fock method}
\label{sssec:tdhf}

The polarization effects are taken into account by using 
the time-dependent Hartree-Fock method (see, e.g., 
\cite{DFSS1987b,DFSS1987a} and references therein). 
We will consider how the RHF equations are modified in the presence 
of a time-dependent field
\begin{equation}
\label{hint}
\hat{h}_{\rm ext}=(\hat{f}e^{-i\omega t}+ \hat{f}^{\dag}e^{i\omega t}).
\end{equation}  
We can assume that the time-dependent single-particle orbitals are 
then given by
\begin{equation}
\label{eq:solnvar}
\tilde{\psi}_k=(\psi _k +\chi _ke^{-i\omega t}+\eta _ke^{i\omega t}),
\end{equation}
with corresponding eigenvalues (``quasi-energies'')
\begin{equation}
\tilde{\epsilon}_{k}=\epsilon _{k}+\delta \epsilon^{(1)} _{k}e^{-i\omega t} +
\delta \epsilon^{(2)} _{k}e^{i\omega t};
\end{equation}
$\chi _k$, $\eta _k$ and $\delta \epsilon^{(1)}_{k}$, 
$\delta \epsilon^{(2)}_{k}$ are corrections to the RHF wave functions 
$\psi _{k}$ and energies $\epsilon _{k}$, respectively, induced by 
$\hat{h}_{\rm ext}$.
The Schr\"{o}dinger equation
\begin{equation}
(\hat{h}_{0}(\tilde\psi)+\hat{h}_{\rm ext})
\tilde{\psi} _k=i\frac{\partial}{\partial t}\tilde{\psi} _k,
\end{equation}
can now be used to obtain equations for the corrections $\chi _k$ and 
$\eta _k$, 
\begin{eqnarray}
\label{eq:chieta}
(\hat{h}_0-\epsilon _k -\omega )\chi _k&=&-(\hat{f}+
\delta \hat{V}_{\rm ext})\psi _k +\delta\epsilon_{k}\psi_{k} 
\nonumber \\
(\hat{h}_0-\epsilon _k +\omega )\eta _k&=&-(\hat{f}^{\dag}+
\delta \hat{V}^{\dag}_{\rm ext})\psi _k +\delta\epsilon_{k}\psi_{k},
\end{eqnarray}
where we take into account terms up to first order; 
the energy shift 
$\delta\epsilon_{k}=\delta\epsilon_{k}^{(1)}=\delta\epsilon_{k}^{(2)}=
\langle \psi _{k}|\hat{f}+\delta \hat{V}_{\rm ext}|\psi _{k}\rangle$. 
The RHF Hamiltonian $\hat{h}(\tilde{\psi})$ corresponds to $\hat{h}_{0}$ 
with the Hartree-Fock potential $\hat{V}^{N-1}$ calculated with 
the new core wave functions $\tilde{\psi}$, and $\delta \hat{V}_{\rm ext}$ 
is the difference between the potential found in the external field and 
the RHF potential,
\begin{eqnarray}
\label{eq:potcor}
\delta \hat{V}_{\rm ext}=&&\hat{V}^{(N-1)}(\tilde{\psi})-
\hat{V}^{(N-1)}(\psi) \nonumber \\
\delta \hat{V}_{\rm ext}\psi ({\bf r}) 
=&&e^2\sum_{n=1}^{N-1}\int \frac{{\rm d}^{3}{\bf r}_1}
{|{\bf r}-{\bf r}_1|}
\Big[ \Big( \eta _n ^{\dag}({\bf r}_1)\psi _n ({\bf r}_1)+\psi _n ^{\dag}({\bf r}_1)\chi_n({\bf r}_1)\Big) 
\psi ({\bf r})- \\
&&-\Big( \eta _n ^{\dag}({\bf r}_1)\psi _n({\bf r})+\psi _n ^{\dag}({\bf r}_1)\chi _n ({\bf r})\Big) 
\psi ({\bf r}_1)\Big]
\nonumber.
\end{eqnarray}
Eqs. (\ref{eq:chieta}), (\ref{eq:potcor}) should be solved self-consistently 
for the $(N-1)$ core electrons. The wave function of the external electron 
is then found in the field of the frozen core (the $V^{N-1}$ approximation). 
So in the same way as in the RHF case, here we can find a complete set of 
orthonormal TDHF orbitals $\tilde \psi _{k}$ with quasi-energy 
$\tilde{\epsilon}_{k}$.

The external electron transition amplitude $M_{\beta \alpha}$ 
from state $|\alpha\rangle$ to state $|\beta\rangle$ induced by the 
field $\hat{h}_{\rm ext}$ can be found by comparison of
the amplitude obtained from Eqs. (\ref{eq:solnvar}) and (\ref{eq:chieta}),
\begin{equation}
\label{eq:mba}
\langle \psi _{\beta}|\tilde{\psi}_{\alpha}\rangle
=\langle \psi _{\beta}|\chi _{\alpha}e^{-i\omega t}\rangle =
\frac{\langle \psi _{\beta}|\hat{f}+\delta \hat{V}_{\rm ext}|\psi _{\alpha}\rangle }
{\epsilon_{\alpha}-\epsilon_{\beta}+\omega}e^{-i\omega t},
\end{equation} 
with conventional time-dependent perturbation theory,
\begin{equation}
\label{eq:mbapert}
\tilde{\psi}_{\alpha}=\psi_{\alpha}+\frac{M_{\beta \alpha}}
{\epsilon_{\alpha}-\epsilon_{\beta}+\omega}\psi _{\beta}e^{-i\omega t},
\end{equation}
where only the resonant term ($\omega \approx \epsilon_{\beta}-\epsilon_{\alpha}$) 
is considered.
Comparing Eqs. (\ref{eq:mbapert}) and (\ref{eq:mba}) gives 
\cite{DFSS1987b,DFSS1987a}
\begin{equation}
\label{eq:transition}
M_{\beta\alpha}=
\langle \psi _\beta |\hat{f}+\delta \hat{V}_{\rm ext}|\psi _{\alpha}\rangle \ .
\end{equation}
This formula corresponds to the well-known random-phase approximation (RPA) with 
exchange 
(see, e.g., \cite{amusia1975}.
When the orbitals $\alpha$ and $\beta$ are calculated in the potential 
$\hat{V}^{N-1}$, the transition amplitude corresponds to the RHF value. 
%when calculated in the potential $\hat{V}^{N-1}+\hat{\Sigma}$, 
%Brueckner-type corrections are also included.

\begin{figure}[b]
\centerline{\psfig{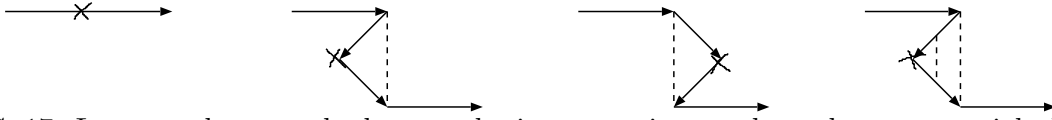}}
\caption{Lowest order many-body perturbation corrections to the 
exchange potential; the cross denotes the external field.}
\label{tdhfseries}
\end{figure}

Using the TDHF procedure described above, core polarization is included in 
all orders of perturbation theory. 
This is equilavent to summation of the diagram series presented 
in Fig. \ref{tdhfseries}.  

The Brueckner-type correlation corrections are calculated in the same way as the 
corrections to energies. Brueckner, instead of RHF, orbitals are used for 
$\alpha$ and $\beta$ in Eq. (\ref{eq:transition}). 
This is equivalent to calculating the diagrams presented in Fig. \ref{tdhfbrueckner}. 
However, it must be noted that by using this technique we neglect the diagrams 
in which the interaction takes place in the internal lines 
(see Fig. \ref{structrad}), although these diagrams give only small 
corrections; 
we will deal with these contributions in Section \ref{sssec:srnorm}. 
Also in Section \ref{sssec:srnorm} we look at another second-order correction: 
the normalization of the many-body states.

\subsubsection{E1 transition amplitudes}
\label{sssec:e1}

The Hamiltonian of the electron interaction with the electric field 
\begin{equation}
{\bf E}(t)={\bf E}_{0}(e^{-i\omega t}+e^{i\omega t})
\end{equation} 
of an electromagnetic wave
depends on the choice of gauge. In ``length'' form 
$\hat{f}_{l}=e{\bf r}\cdot {\bf E}_{0}$ and in ``velocity'' form 
$\hat{f}_{v}=-ie(\mbox{\boldmath$\alpha$}\cdot {\bf E}_{0}/\omega)$, 
where $\hat{h}_{E1}=\hat{f}e^{-i\omega t}+\hat{f}^{\dagger}e^{i\omega t}$.

It is known that in TDHF calculations the amplitude (\ref{eq:transition}) 
is gauge invariant (see, e.g., \cite{amusia1975,kobe1979,M-P1985jp}.
Comparing results obtained from the two forms for the amplitudes 
(\ref{eq:transition}) provides a test of the numerical calculation.
In \cite{DFSS1987a}, the length and velocity forms were shown to give 
the same results for the E1 transition amplitudes when the correlation 
corrections (correlation potential, frequency shift in the velocity 
form operator, structural radiation, and normalization of states) 
are taken into account. 
Calculations using the dipole operator 
in length form are more stable than those obtained in velocity form 
(see, e.g., \cite{GK1979,sandars1980,DFSS1987a}). 
For this reason we consider the calculations of E1 transition amplitudes 
in length form.

As seen from Eq. (\ref{eq:transition}), inclusion of the core polarization 
into the E1 transition amplitude is reduced to the addition of the 
operator $\delta\hat{V}_{E1}$.
Because the E1 operator can only mix 
opposite parity states, there is no energy shift, and so when calculating 
the wave function corrections $\chi _{k}$ and $\eta _{k}$ from Eq. 
(\ref{eq:chieta}), $\delta \epsilon _k=0$. 

The results for E1 transition amplitudes calculated in length form 
between the lower states of cesium are presented in Table \ref{tab:e1}. 

Another test of the accuracy and self-consistency of the TDHF 
equations is the value of an external electric field at the nucleus. 
According to the Schiff theorem \cite{schiff}, an external static 
electric field ($\omega =0$ in TDHF) at the nucleus in a neutral atom 
is shielded completely by atomic electrons. 
For an atom with charge $Z_{i}$ the total static electric field ${\bf E}_{\rm tot}$ 
at the nucleus is \cite{DFSS1986}
\begin{equation}
{\bf E}_{\rm tot}(0)={\bf E}_0 +\langle {\bf E}_e (0)\rangle
={\bf E}_0 Z_{i}/Z \ ,
\end{equation}  
where $E_0$ is the external field and $E_e$ is the induced electron field.
TDHF equations reproduce this result correctly. 
The oscillations of the electric field inside the atom are quite complex. 
We refer the reader to \cite{DFSS1986} for a plot of the electric field 
inside Tl$^{+}$.\footnote{Note that in the paper \cite{DFSS1986} the figures 
and the captions have been switched.}

\subsubsection{Hyperfine structure constants}
\label{sssec:hfs}

The hyperfine interaction between a relativistic electron and 
a point nucleus is given by
\begin{equation}
\hat{h}_{hf}=e\mbox{\boldmath$\alpha$}\cdot {\bf A} \ ,
\end{equation}
where
\begin{equation}
{\bf A}=\frac{\mbox{\boldmath$\mu$}\times {\bf r}}{r^{3}}
\end{equation}
is the vector potential created by the nuclear magnetic moment $\mbox{\boldmath$\mu$}$ 
and $\mbox{\boldmath$\alpha$}$ is the Dirac matrix.
When performing the calculation of the hyperfine structure of 
heavy atoms, it is important to take into account the finite size of the 
nucleus. Using a simple model in which the nucleus represents a 
uniformly magnetized ball, the hyperfine interaction is given by
\begin{equation}
\hat{h}_{hf}=e\mbox{\boldmath$\mu$}\cdot {\bf f}({\bf r}) \ , \qquad \quad 
{\bf f}({\bf r})=\left \{ \begin{array}{ll}
{{\bf r}\times \mbox{\boldmath$\alpha$}}/{r^{3}_{m}} & , \quad r< r_m\\
{{\bf r}\times \mbox{\boldmath$\alpha$}}/{r^{3}} & , \quad r\geq r_m
\end{array} \right.
\end{equation}
where we take the magnetic radius $r_{m}=1.1A^{1/3}$fm ($A$ is the mass number of the nucleus).
Note that while the distribution of currents in the nucleus (produced by 
unpaired nucleons) is very complex, the hfs is only weakly dependent 
on its form 
(see, e.g., \cite{DFS1984}). 

Corrections to the RHF wave functions induced by the hyperfine interaction 
are calculated using Eq. (\ref{eq:chieta}). There is no time-variation 
in the hyperfine interaction, and so we set $\omega=0$.
The hfs equations are then
\begin{eqnarray}
(\hat{h}_0-\epsilon)\delta \psi &=&-(\hat{h}_{hf}+
\delta \hat{V}_{hf})\psi +\delta\epsilon \psi\\
\delta\epsilon &=&\langle\psi |\hat{h}_{hf}+
\delta \hat{V}_{hf}|\psi \rangle \ .
\end{eqnarray}
The hyperfine interaction does not alter the direct contribution to the 
core potential, so
$\delta \hat{V}_{hf}=\hat{V}_{\rm exch}(\tilde{\psi})-
\hat{V}_{\rm exch}(\psi)$. 
We use the $\hat{V}^{N-1}$ approximation to calculate the complete 
set of orbitals.
The external electron correction $\delta \epsilon$ determines the atomic 
hyperfine structure.

Hyperfine structure results for cesium are presented in Table \ref{tab:hfs}.
The correlations increase the density of the external electron in the 
nuclear vicinity by about $30\%$. An accurate inclusion of the correlations is 
therefore very important when considering interactions singular on the nucleus, 
for example the PNC weak interaction.

\subsubsection{Structural radiation and normalization of states}
\label{sssec:srnorm}

``Structural radiation'' is the term we use for correlation corrections 
with the external field in the internal lines. 
(We will make a further distinction when we are dealing with 
the PNC E1 amplitude $E_{PNC}$, since here there are two fields. 
We call correlation corrections with the weak interaction in the internal 
lines ``weak correlation potential''; see Section \ref{ssec:cscalcs}.)
Examples of diagrams which represent structural radiation are presented in 
Fig. \ref{structrad}.
In second-order perturbation theory in the residual interaction there is 
another contribution which arises due to the change of the 
normalization of the single-particle wave functions 
due to admixture with many-particle states. 

For the E1 transition amplitudes in length form, 
the following approximate formula is used to calculate the 
structural radiation \cite{DFSS1987a},
\begin{equation}
\label{eq:strappl}
{\bf M}_{\rm str}=-\frac{1}{2}\langle \beta|{\bf D}
\frac{\partial \hat{\Sigma}}{\partial \epsilon} +
\frac{\partial \hat{\Sigma}}{\partial \epsilon}{\bf D}|\alpha \rangle \ ,
\end{equation}
where ${\bf D}\equiv {\bf d}+\delta {\bf V}$.
The derivation can be found in Ref. \cite{DFSS1987a}.
We refer the interested reader to Refs. \cite{DFSS1987a,DFSS1987b} 
for the structural radiation in the velocity gauge.
The normalization contribution for E1 transitions has the form 
\cite{DFSS1987a,DFSS1987b}
\begin{equation}
\label{eq:norm}
{\bf M} _{\rm norm}=\frac{1}{2}\langle \beta|{\bf D}|\alpha\rangle
\Big( \langle \beta |\frac
{\partial \hat{\Sigma}}{\partial \epsilon}|\beta\rangle +
 \langle \alpha |
\frac{\partial \hat{\Sigma}}{\partial \epsilon}|\alpha\rangle \Big) 
\ . 
\end{equation}

The structural radiation and normalization contributions 
to the hyperfine structure and E1 transition amplitudes of low-lying states 
of cesium are small. Their combined contribution, for both hfs and E1 amplitudes, 
usually lies in the range $0.1-1.0\%$.

\section{High-precision calculation of parity violation in cesium and 
extraction of the nuclear weak charge}
\label{section3iib}

In this section the method described in the preceding section is applied to 
the $6S-7S$ parity violating amplitude in cesium and the value for the nuclear 
weak charge is extracted from the measurement of Wood {\it et al.} \cite{wieman}. 
%We include other significant corrections 
%to the PNC amplitude (e.g. strong-field radiative corrections) 
%and extract the value for the nuclear weak charge from the measurement 
%of Wood {\it et al.} \cite{wieman}. 
%The results of the atomic calculations 
%presented in the following sections were obtained in the work \cite{DFG02}.
The value for this amplitude has been the source of much confusion recently. 
It has jumped around erratically in the last few years, and finally it has stabilized.
The origin of this instability is described below.
%The source of much confusion recently is the value for the $6S-7S$ parity violating amplitude 
%in cesium calculated from atomic theory. The value for this amplitude has been 
%jumping around erratically these last few years, but has now stabilized. 
%The origin of this instability is described below. 

%The interpretation of these results gives us reliable information about parity 
%and time-reversal violating interactions and provides a vital window on
%new physics beyond the standard model of particle physics, such as
%supersymmetry, grand unification, and extended Higgs structures.
%For example, atomic and neutron electric dipole moment measurements 
%have closed the Weinberg $CP$-violation model; atomic parity violation 
%experiments have given the best limits on the extra $Z$-boson in the 
%$SO(10)$ Grand Unification Model, and also on leptoquarks and composite 
%fermions (pre-quarks) predicted in other models; a Generic Technicolor model 
%has been found to be in contradiction with atomic experiments.

In 1999, experimentalists Bennett and Wieman argued that the standard model is in 
contradiction with atomic experiments by 2.5 $\sigma$ \cite{BW}. 
This conclusion was mainly based on their analysis of the accuracy of atomic structure 
calculations which were published in Ref. \cite{DFS1989pnc} in 1989 
and in Ref. \cite{BJS1990} in 1990,1992. The point is that the new measurements of 
electromagnetic amplitudes in atoms have demonstrated that the accuracy of the atomic 
calculations is much better than it seemed to be ten years ago. 
Indeed, all disagreements between theory and experiments were resolved in favour 
of theory. Based on this, Bennett and Wieman reduced the theoretical error from 
the 1\% claimed by theorists \cite{DFS1989pnc,BJS1990} to 0.4\% 
and came to the conclusion that there may be new physics beyond the standard model. 

Particle physicists put forward new physics possibilities, 
such as an extra Z boson in the weak interaction, leptoquarks, and composite fermions; 
see, e.g., Refs. \cite{newphysics}.

However, the 1\% error placed on the atomic calculations \cite{DFS1989pnc,BJS1990} was based 
not only on the comparison of measured and calculated quantities such as E1 transition 
amplitudes. It was based also on an allowance for unnaccounted contributions. 
It was soon after discovered that the Breit contribution is 
larger than expected (-0.6\%) \cite{derevianko00} (Section \ref{sssec:Breit}). 
Then it was suggested that strong-field 
radiative corrections could make a contribution as large as Breit \cite{sushkovbreitrad}. 
The Uehling contribution has been found to give a contribution of 0.4\% \cite{MS02,JBS01} 
(Section \ref{sssec:radcorr}). 
And just recently it has been established that self-energy and vertex contributions are 
$\approx -0.8\%$ \cite{KF2002,MST2002} (Section \ref{sssec:radcorr}). 
A recent comprehensive calculation of atomic PNC is accurate to 0.5\% \cite{DFG02}. 
The results of this calculation, and the discussion of accuracy, are presented in the 
following sections.  
The final value for parity violation in cesium is in 
agreement with the standard model and tightly constrains possible new physics 
(Section \ref{ssec:finalvalue}). 

%This example shows the important role played by atomic theory.
%The conclusions derived from experiments may depend crucially on one's
%ability to perform very accurate calculations.

\subsection{High-precision calculations of parity violation in cesium}
\label{ssec:cscalcs}

The weak interaction $\hat{H}_{W}=\sum _{i}^{N}\hat{h}_{W}^{i}$ 
[Eq. (\ref{pncham})]\footnote{
It is seen from Eqs. (\ref{eq:smcoeff},\ref{eq:pncnp}), by inserting the coefficients 
$C_{1N}$, that the density $\rho (r)$ is essentially the (poorly understood) neutron density 
in the nucleus. In the calculations, $\rho (r)$ will be taken equal to the charge density, 
Eq. (\ref{density}), and then in Section \ref{sssec:neutron} we will consider 
the effect on the PNC E1 amplitude as a result of correcting for $\rho (r)$.}
mixes atomic wave functions of opposite parity 
and leads to small opposite-parity admixtures in atomic states $\Psi$,   
$\tilde{\Psi}=\Psi +\delta \Psi$.  
This gives rise to E1 transitions between states of the same 
nominal parity. The parity violating $6S-7S$ E1 transition amplitude 
in Cs is
\begin{equation}
\label{eq:csadmixed}
E_{PNC}=\langle \widetilde{7S}|\hat{H}_{E1}|\widetilde{6S}\rangle 
       =\langle \delta (7S)|\hat{H}_{E1}|6S\rangle + 
\langle 7S|\hat{H}_{E1}|\delta (6S)\rangle \ .
\end{equation}
%
%Nuclear spin-independent parity nonconserving E1 transition amplitudes arise 
%due to the simultaneous interaction of atomic electrons with the nuclear weak charge 
%and with the photon field. 

Calculations of PNC E1 amplitudes can be performed using the following approaches:
from a mixed-states approach, in which there is a small opposite-parity 
admixture in each state [Eq. (\ref{eq:csadmixed})];
or from a sum-over-states approach, in which the amplitude [Eq. (\ref{eq:csadmixed})] is broken down 
into contributions arising from opposite-parity admixtures and a direct summation over 
the intermediate states is performed [Eq. (\ref{sum})]. 

%\subsubsection{Sum-over-states calculation}
%\label{sssec:sospnc}

In the sum-over-states approach, the Cs $6S-7S$ PNC E1 transition amplitude 
is written in terms of a sum over intermediate, many-particle states $NP_{1/2}$
\begin{equation}
\label{sum}
E_{PNC}=
\sum_{N}\Big[ 
\frac{\langle 7S|\hat{H}_{E1}|NP_{1/2} \rangle 
\langle NP_{1/2}|\hat{H}_{W}|6S \rangle}{E_{6S} -E_{NP_{1/2}}}
+\frac{\langle 7S|\hat{H}_{W}|NP_{1/2} \rangle
\langle NP_{1/2}|\hat{H}_{E1}|6S \rangle}{E_{7S}-E_{NP_{1/2}}} \Big] \ .
\end{equation}
If one neglects configuration mixing, this sum can be represented in terms of 
single-particle states; in this case, the sum also runs over core 
states (corresponding to many-particle states with a single core 
excitation), 
\begin{equation}
E_{PNC}=
\sum_{n}\Big[ 
\frac{\langle 7s|\hat{h}_{E1}|np_{1/2} \rangle 
\langle np_{1/2}|\hat{h}_{W}|6s\rangle}{\epsilon_{6s}-\epsilon_{np_{1/2}}}
+\frac{\langle 7s|\hat{h}_{W}|np_{1/2} \rangle 
\langle np_{1/2}|\hat{h}_{E1}|6s\rangle}{\epsilon_{7s}-\epsilon_{np_{1/2}}} 
\Big] \ .
\end{equation}
There are three dominating contributions to this sum:
\begin{eqnarray}
E_{PNC}&=& 
\label{sumcs}
\frac{\langle 7s|\hat{h}_{E1}|6p_{1/2}\rangle 
\langle 6p_{1/2}|\hat{h}_{W}|6s\rangle}{\epsilon_{6s}-\epsilon_{6p_{1/2}}} +
\frac{\langle 7s|\hat{h}_{W}|6p_{1/2}\rangle 
\langle 6p_{1/2}|\hat{h}_{E1}|6s\rangle}{\epsilon_{7s}-\epsilon_{6p_{1/2}}}\nonumber \\ 
&& \qquad  +\frac{\langle 7s|\hat{h}_{E1}|7p_{1/2}\rangle 
\langle 7p_{1/2}|\hat{h}_{W}|6s\rangle}{\epsilon_{6s}-\epsilon_{7p_{1/2}}}+... 
%\nonumber 
\label{eq:pncsum}
\\
&=& -1.908+1.493+1.352+...=0.937 +...  \ .\nonumber 
\end{eqnarray} 
The numbers are from the work \cite{BJS1990} where the sum-over-states method was used; 
here we just demonstrate that these terms dominate.
An advantage of the sum-over-states approach is that experimental values for the energies 
and E1 transition amplitudes
can be explicitly included into the sum. This was the procedure for some of the early 
calculations of PNC in Cs (see, e.g., \cite{calc13}).
%which other works?!!!!!!! 

In Refs. \cite{DFS1989pnc,BJS1990,kozlovcs,DFG02} PNC calculations were performed in 
the mixed-states approach, and in Ref. \cite{BJS1990} a calculation was carried out in the 
sum-over-states approach also. Here we refer to the most precise calculations, 
$\leq 1\%$ accuracy.

\subsubsection{Mixed-states calculation}
\label{sssec:mixedpnc}

In the TDHF method (Section \ref{sssec:tdhf}, Eq. (\ref{eq:solnvar})), 
a single-electron wave function in external weak and E1 fields is
\begin{equation}
\label{eq:wf}
	\psi = \psi_0 + \delta \psi + Xe^{-i\omega t}+Ye^{i\omega t}+
	\delta Xe^{-i\omega t}+ \delta Ye^{i\omega t},
\end{equation}
where $\psi _{0}$ is the unperturbed state, 
$\delta \psi$ is the correction due to the weak interaction 
acting alone, $X$ and $Y$ are corrections due to the photon field 
acting alone, and $\delta X$ and $\delta Y$ are corrections due 
to both fields acting simultaneously.
These corrections are found by solving self-consistently the system
of the TDHF equations for the core states
\begin{eqnarray}
	(\hat{h}_{0}-\epsilon)\delta \psi &=&
	-(\hat{h}_{W}+\delta \hat{V}_{W})\psi, \label{eq:WE1:1}\\
	(\hat{h}_{0}-\epsilon -\omega)X &=&
	-(\hat{h}_{E1}+\delta \hat{V}_{E1})\psi, \label{eq:WE1:2}\\
	(\hat{h}_{0}-\epsilon +\omega)Y&=&
	-(\hat{h}_{E1}^{\dagger}+\delta \hat{V}_{E1}^{\dagger})\psi, 
	\label{eq:WE1:3}\\
	(\hat{h}_{0}-\epsilon -\omega)\delta X &=&
	-\delta \hat{V}_{E1}\delta \psi 
	-\delta \hat{V}_{W}X 
	-\delta \hat{V}_{E1W}\psi, \label{eq:WE1:4}\\
	(\hat{h}_{0}-\epsilon +\omega)\delta Y&=&
	-\delta \hat{V}_{E1}^{\dagger}\delta \psi
	-\delta \hat{V}_{W}Y
	-\delta \hat{V}_{E1W}^{\dagger} \psi,\label{eq:WE1:5}
\end{eqnarray} 
where $\delta \hat{V}_W$ and $\delta \hat{V}_{E1}$ are corrections to the
core potential due to the weak and E1 interactions, respectively, 
and $\delta \hat{V}_{E1W}$ is the correction to the core potential 
due to the simultaneous action of the weak field and the electric field 
of the photon.

The TDHF contribution to $E_{PNC}$ between the states $6S$ and $7S$ 
is given by
\begin{equation}
\label{eq:TDHF}
E_{PNC}^{TDHF}=\langle \psi _{7s}|\hat{h}_{E1}+\delta \hat{V}_{E1}|
\delta \psi _{6s}\rangle +
\langle \psi _{7s}|\hat{h}_{W}+\delta \hat{V}_{W}|
X _{6s}\rangle +
\langle \psi _{7s}|\delta \hat{V}_{E1W}|\psi _{6s}\rangle  \ . 
\end{equation}
The corrections $\delta \psi_{6s}$ and $X_{6s}$ are found by solving 
the equations~(\ref{eq:WE1:1}-\ref{eq:WE1:2}) in the field of the 
frozen core (of course, the amplitude (\ref{eq:TDHF}) can instead 
be expressed in terms of corrections to $\psi _{7s}$). 

\begin{figure}[b]
\psfig{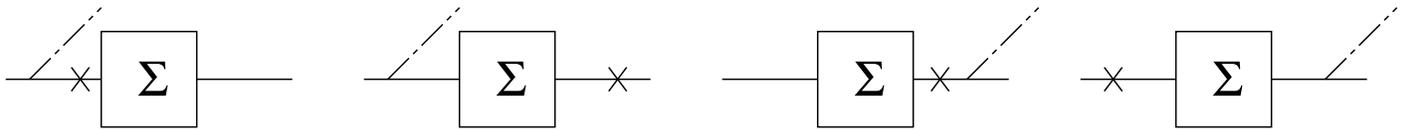}
\caption{Brueckner-type correlation corrections to the PNC E1 transition 
amplitude in first order in the weak interaction; 
the crosses denote the weak interaction and the 
dashed lines denote the electromagnetic interaction.}
\label{fig:pncdom}
\end{figure}

\begin{figure}[b]
\psfig{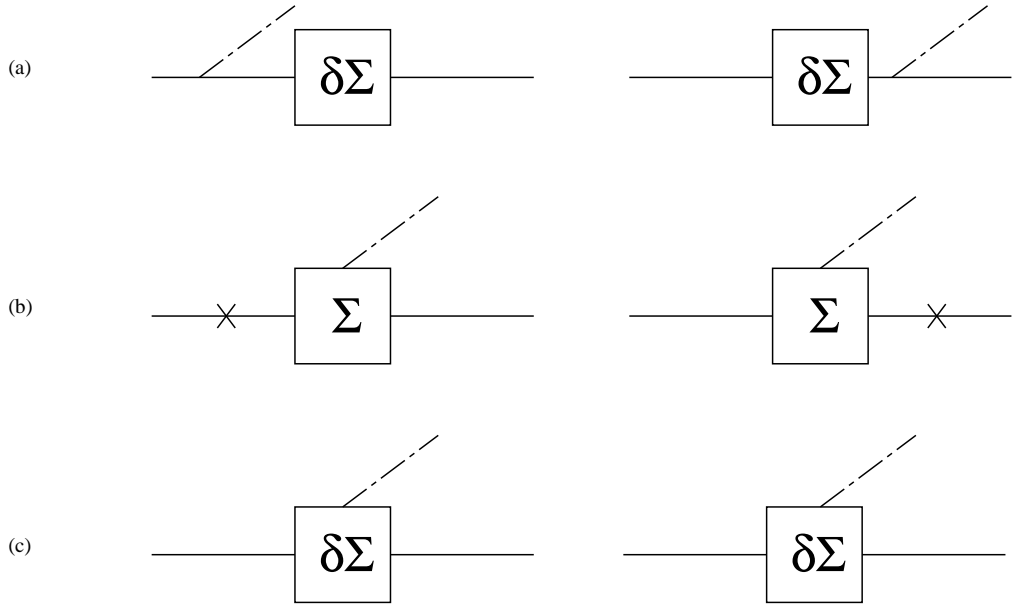}
\caption{External field inside the correlation potential.
In diagrams (a) the weak interaction is inside the 
correlation potential ($\delta \Sigma$ denotes the change in $\Sigma $ due to 
the weak interaction); this is known as the weak correlation potential.
Diagrams (b,c) represent the structural radiation 
(photon field inside the correlation potential).
In diagram (b) the weak interaction occurs in the external lines;
in diagram (c) the weak interaction is included in the 
electromagnetic vertex.}
\label{fig:pncint1}
\end{figure}

Now we need to include the correlation corrections to the 
PNC E1 amplitude. In the previous sections 
(Sections \ref{ssec:corr},\ref{ssec:sum},\ref{sssec:srnorm}) we have discussed 
two types of corrections: the dominant Brueckner-type corrections, 
represented by diagrams in which the external field appears in 
the external electron line (see Fig. \ref{fig:pncdom}); 
and structural radiation, in which the external field acts on 
an internal electron line.
In the case of PNC E1 amplitudes, in order to distinguish between 
structural radiation diagrams with different fields, we refer 
to diagrams with the weak interaction attached to the internal electron 
line as ``weak correlation potential'' diagrams.
Structural radiation and the weak correlation potential diagrams are 
presented in Fig. \ref{fig:pncint1}.

We will consider first the dominating Brueckner-type corrections to 
the E1 PNC amplitude.
Remember that the correlation potential is energy-dependent, 
$\hat{\Sigma}=\hat{\Sigma}(\epsilon)$. 
This means that the $\hat{\Sigma}$ operators for the $6s$ and $7s$ 
states are different.
We should consider the proper energy-dependence at least in first-order 
in $\hat{\Sigma}$ (higher-order corrections are small and the proper 
energy-dependence is not important for them).
The first-order in $\hat{\Sigma}$ correction to $E_{PNC}$ is 
presented diagrammatically in Fig. \ref{fig:pncdom}.  
We can write this as 
\begin{equation}
\label{eq:pnc-cor}
\langle \psi _{7s}|\hat{\Sigma}_s(\epsilon_{7s})|\delta X_{6s}\rangle 
+\langle \delta\psi _{7s}|\hat{\Sigma}_p(\epsilon_{7s})|X_{6s}\rangle  
+\langle \delta Y_{7s}|\hat{\Sigma}_s(\epsilon_{6s})|\psi_{6s}\rangle
+\langle Y_{7s}|\hat{\Sigma}_p(\epsilon_{6s})|\delta \psi_{6s}\rangle \ .
\end{equation}

The non-linear in $\hat{\Sigma}$ contribution to the Brueckner-type 
correction is found using the correlation potential method 
(Section \ref{ssec:corr}): the all-orders in $\hat{\Sigma}$ 
contribution is calculated and from this the first-order contribution,
found in the same method, is subtracted.
The all-orders term is calculated using external electron orbitals, 
and corrections to these orbitals induced by the weak interaction 
and the photon field, found in the potential $\hat{V}^{N-1}+\hat{\Sigma}$. 
The PNC E1 amplitude is then calculated, using these new orbitals, 
in the same way as in the usual time-dependent Hartree-Fock method.
The all-orders contribution to $E_{PNC}$ is
\begin{equation}
E_{PNC}^{\rm all-orders}=
E_{PNC}(\hat{V}^{N-1}+\hat{\Sigma})-
E_{PNC}(\hat{V}^{N-1}) \ .
\end{equation}
The first-order in $\hat{\Sigma}$ contribution is found by placing a small 
coefficient $a$ before the correlation potential, $\hat{\Sigma}\rightarrow 
a\hat{\Sigma}$. When $a\ll 1$, the linear in $\hat{\Sigma}$ contribution 
to $E_{PNC}$ dominates. Its extrapolation to $a=1$ gives the first-order 
in $\hat{\Sigma}$ contribution. So the non-linear in $\hat{\Sigma}$ 
contribution to $E_{PNC}$ is \cite{DFS1995}
\begin{equation}
E_{PNC}^{\rm non-lin}=\Big[ E_{PNC}(\hat{V}^{N-1}+\hat{\Sigma})-
E_{PNC}(\hat{V}^{N-1})\Big] 
-\frac{1}{a}\Big[ 
E_{PNC}(\hat{V}^{N-1}+a\hat{\Sigma})-
E_{PNC}(\hat{V}^{N-1})\Big] \ . 
\end{equation}    
%Although the wrong energy-dependence is used in the calculation of the 
%non-linear in $\hat{\Sigma}$ contribution,
%this contribution begins in second-order in $\hat{\Sigma}$, and 
%the energy-dependence itself is weak, so this effect is insignificant. 
%The nonlinear in $\hat{\Sigma}$ contribution to the Cs $6S$-$7S$
%PNC E1 amplitude is presented in Table \ref{table:cspnc}. 

To complete the calculation of corrections second-order in the 
residual Coulomb interaction the weak correlation 
potential, structural radiation, and normalization contributions 
to the PNC amplitude must be included. 
%These contributions are suppressed by the small parameter 
%$E_{\rm ext}/E_{\rm core}\sim 1/10$, where  $E_{\rm ext}$ and 
%$E_{\rm int}$ are excitation energies of the external and core electrons,
%respectively.

The weak correlation potential is calculated 
by direct summation over intermediate states.
See Section \ref{sssec:srnorm} for the approximate form 
for structural radiation in length form. 
%(the derivation is presented in Appendix \ref{appendixB})
%??????? however, in the recent calculations sum-over-states 
%       method was used ????????
and for the form for the 
normalization of the many-body states. 
Due to parity violation there is an opposite-parity correction to the 
orbitals $\alpha$ and $\beta$, $\tilde{\alpha}=\alpha +\delta \alpha$ 
and $\tilde{\beta}=\beta + \delta \beta$, and to the correlation 
potential $\hat{\Sigma}$, $\tilde{\Sigma}=\hat{\Sigma}+\delta \hat{\Sigma}$.

Structural radiation is then given by
\begin{equation}
\label{eq:strapplp}
\tilde{\bf M}_{\rm str}=-\frac{1}{2}\langle \tilde{\beta}|{\bf D}
\frac{\partial \tilde{\Sigma}}{\partial \epsilon} +
\frac{\partial \tilde{\Sigma}}{\partial \epsilon}{\bf D}|\tilde{\alpha}
\rangle .
\end{equation}
There are two contributions to structural radiation for the 
PNC E1 amplitude: one in which 
the electromagnetic vertex is parity conserving, the weak interaction 
included in the external lines:
\begin{equation}
\tilde{\bf M}_{\rm Fig. \ref{fig:pncint1}(b)}=
-\frac{1}{2}\langle \tilde{\beta}|{\bf D}
\frac{\partial \hat{\Sigma}}{\partial \epsilon} +
\frac{\partial \hat{\Sigma}}{\partial \epsilon}{\bf D}|\tilde{\alpha}
\rangle .	
\end{equation}
(see diagram (b) Fig. \ref{fig:pncint1});
and the other in which the weak interaction 
is included in the electromagnetic vertex (we call this structural radiation 
and not weak correlation potential):
\begin{equation}
{\bf M}_{\rm Fig. \ref{fig:pncint1}(c)}=-\frac{1}{2}\langle \beta|{\bf D}
\frac{\partial \tilde{\Sigma}}{\partial \epsilon} +
\frac{\partial \tilde{\Sigma}}{\partial \epsilon}{\bf D}|\alpha\rangle .
\end{equation}
(see diagram (c) of Fig. \ref{fig:pncint1}).
Note that in each case the amplitude first-order in the weak interaction 
is considered.

%Structural radiation in velocity form can be obtained from length form 
%using the transformation (\ref{eq:trans}), giving
%\begin{equation}
%\tilde{M}^{(v)}_{\rm str}=\tilde{M}^{(l)}_{\rm str}+(e/\omega)
%\langle\tilde{\beta}|\tilde{\Sigma}(\epsilon _{\beta}){\bf r} -
%{\bf r}\tilde{\Sigma}(\epsilon _{\alpha})|\tilde{\alpha}\rangle .
%\end{equation}
%Again, we can break this up into contributions to the diagrams of 
%Fig. \ref{fig:pncint1}(b,c):
%\begin{eqnarray}
%&&\tilde{M}^{(v)}_{\rm Fig. \ref{fig:pncint1}(b)}=
%\tilde{M}^{(l)}_{\rm Fig. \ref{fig:pncint1}(b)}+(e/\omega)
%\langle\tilde{\beta}|\hat{\Sigma}(\epsilon _{\beta}){\bf r} -
%{\bf r}\hat{\Sigma}(\epsilon _{\alpha})|\tilde{\alpha}\rangle \\ 
%&&\tilde{M}^{(v)}_{\rm Fig. \ref{fig:pncint1}(c)}=
%\tilde{M}^{(l)}_{\rm Fig. \ref{fig:pncint1}(c)}+(e/\omega)
%\langle\beta|\delta\hat{\Sigma}(\epsilon _{\beta}){\bf r} -
%{\bf r}\delta\hat{\Sigma}(\epsilon _{\alpha})|\alpha\rangle 
%\end{eqnarray} 

The normalization contribution is 
\begin{equation}
\tilde{\bf M}_{\rm norm}=\frac{1}{2}\langle\tilde{\beta}|{\bf D}
%+ \delta V
|\tilde{\alpha}\rangle 
\Big( \langle \beta |\frac
{\partial \hat{\Sigma}}{\partial \epsilon}|\beta\rangle +
 \langle \alpha |\frac
{\partial \hat{\Sigma}}{\partial \epsilon}|\alpha\rangle \Big) . 
\end{equation}

%Let us consider the relative size of the contributions arising from the 
%linear in $\hat{\Sigma}$ correlation corrections to the E1 PNC amplitude.
%Now, the term $\Delta _{\rm corr}(\hat{\Sigma})$ corresponds to 
%diagrams with the photon field attached to an external line 
%(Figs. \ref{fig:pncdom} and \ref{fig:pncint1}(a)).
%Due to the small energy denominator associated with the small excitation 
%energy of an external electron (Eq. \ref{eq:pncdom1a}), these 
%diagrams are enhanced by a factor $\Delta E_{\rm int}/\Delta E_{\rm ext}$ 
%compared to structural radiation diagrams (Figs. \ref{fig:pncint1}(b,c) 
%in which the energy in the denominator corresponds to the large excitation 
%energy of a core electron.
%Here $\Delta E_{\rm int}$ is the typical excitation energy from the core
%and $\Delta E_{\rm ext}$ is the typical excitation energy of the external 
%electron.      
%In the same way for the weak interaction, the contributions 
%$\Delta _{\rm corr}(\hat{\Sigma})$ are enhanced by the same factor 
%$\Delta E_{\rm int}/\Delta E_{\rm ext}$ compared to 
%$\Delta _{\rm corr}(\delta\hat{\Sigma})$ 
%(see Figs. \ref{fig:pncdom} and \ref{fig:pncint1}(a)).

The results of the calculation \cite{DFG02} for the  $6S-7S$ PNC amplitude 
are presented in Table~\ref{tab:pnci}.
Taking into account all corrections discussed in this section, the 
following value is obtained for the $6S-7S$ PNC amplitude in cesium
\begin{equation}
\label{eq:subtotal}
E_{PNC}=0.9078\times 10^{-11}iea_{B}(-Q_{W}/N) \ .
\end{equation}
This corresponds to ``Subtotal'' of Table \ref{tab:pnci}. 
This is in agreement with the 1989 result \cite{DFS1989pnc}.
Notice the stability of the PNC amplitude. 
The time-dependent Hartree-Fock value gives a 
contribution to the total amplitude of about $98\%$. 
The point is that there is a strong cancellation of the 
correlation corrections. 

The mixed-states approach has also been performed in 
\cite{BJS1990} and \cite{kozlovcs} to determine the PNC amplitude 
in cesium. 
However, in these works the screening of the 
electron-electron interaction was included in a simplified way. 
In \cite{BJS1990} empirical screening factors were placed before 
the second-order correlation corrections $\hat{\Sigma}^{(2)}$ to fit the 
experimental values of energies. 
Kozlov {\it et al.} \cite{kozlovcs} introduced screening factors 
based on average screening factors calculated for the Coulomb 
integrals between valence electron states. 
The results obtained by these groups 
(without the Breit interaction, i.e., corresponding to the 
Subtotal of Table~\ref{tab:pnci}) 
are $0.904$ \cite{BJS1990} and $0.905$ \cite{kozlovcs}.\footnote{
The numbers differ from those presented in Table \ref{tab:pnccalcs} due 
to the Breit interaction. In \cite{BJS1990} a value for Breit of $-0.2\%$ of the PNC 
amplitude was included (this value was underestimated), while in \cite{kozlovcs} the magnetic 
(Gaunt) part of the Breit interaction was included and calculated to be $-0.4\%$. 
See Section \ref{sssec:Breit}.} 
%To be sure that we understand the difference between these values and 
As a check, a pure second-order (i.e., using $\hat{\Sigma}^{(2)}$) calculation with 
energy-fitting was also performed in \cite{DFG02} 
(in the same way as \cite{BJS1990}), and the result $0.904$ was reproduced.

Contributions of the Breit interaction, the neutron distribution, 
and radiative corrections to $E_{PNC}$ are considered in the following sections.

\subsubsection{Inclusion of the Breit interaction}
\label{sssec:Breit}

The Breit interaction is a two-particle operator
\begin{equation}
\label{eq:breit}
\hat{H}_{\rm Breit}=-\frac{e^{2}}{2}\sum_{i<j}\frac{\mbox{\boldmath$\alpha$} _{i}
\cdot \mbox{\boldmath$\alpha$}_{j}+(\mbox{\boldmath$\alpha$}_{i}\cdot 
{\bf n}_{ij})
(\mbox{\boldmath$\alpha$}_{j}\cdot {\bf n}_{ij})}{|{\bf r}_{i}-{\bf r}_{j}|}
\ , 
\end{equation}
$\mbox{\boldmath$\alpha$}$ are Dirac matrices, 
${\bf n}=({\bf r}_{i}-{\bf r}_{j})/|{\bf r}_{i}-{\bf r}_{j}|$.
It gives magnetic (Gaunt) and retardation corrections to the 
Coulomb interaction. 
A few years ago it was thought that the correction to $E_{PNC}$ 
arising due to inclusion of the Breit interaction in the 
Hamiltonian (\ref{exacth}) is small (safely smaller than 1\%). 
In the work \cite{DFS1989pnc} the Breit interaction was neglected, 
and in \cite{BJS1990} it was only partially calculated. 
(Remember that these works claimed an accuracy of 1\%.)
The huge improvement in the experimental precision of the 
cesium PNC measurement in 1997 \cite{wieman} and the claim of 
Bennett and Wieman in 1999 \cite{BW} that the theoretical accuracy 
is 0.4\% prompted theorists to revisit their calculations. 
Naturally this also involves a consideration of previously neglected 
contributions which, while at the 1\% level could be neglected, 
are significant at the 0.4\% level. 
Derevianko \cite{derevianko00} calculated the contribution of the 
Breit interaction to $E_{PNC}$ and found that it is larger than had been expected. 
Its contribution to $E_{PNC}$ is -0.6\%. This result has been confirmed by 
subsequent calculations \cite{harabati,kozlovcs,DFG02}.

\subsubsection{Neutron distribution}
\label{sssec:neutron}

The weak Hamiltonian Eq. (\ref{pncham}) was used to obtain the result 
Eq. (\ref{eq:subtotal}) with $\rho(r)$ taken to be the charge 
density, parametrized according to Eq. (\ref{density}). 
However, as we mentioned in a footnote at the beginning of Section \ref{ssec:cscalcs}, 
the weak interaction is sensitive to the distribution of neutrons in the nucleus. 
Here we look at the effect of correcting for the neutron distribution.

For the neutron density the two-parameter Fermi model (\ref{density}) is used.
The result of Ref. \cite{r_{np}} was used in \cite{DFG02} for the difference 
$\Delta r_{np}=0.13(4)~{\rm fm}$ 
in the root-mean-square radii of the neutrons 
$\langle r_{n}^{2}\rangle ^{1/2}$ and protons 
$\langle r_{p}^{2}\rangle ^{1/2}$. 
Three cases which correspond to the same value of 
$\langle r_{n}^{2}\rangle$ were considered: (i) $c_{n}=c_{p}$, $a_{n}>a_{p}$; 
(ii) $c_{n}>c_{p}$, $a_{n}>a_{p}$; and (iii) $c_{n}>c_{p}$, $a_{n}=a_{p}$ 
(using the relation $\langle r_{n}^{2}\rangle \approx \frac{3}{5}c_{n}^{2} 
+\frac{7}{5}\pi ^{2}a_{n}^{2}$). 
It is found that $E_{PNC}$ shifts from $-0.18\%$ to $-0.21\%$ 
when moving from the extreme $c_{n}=c_{p}$ to the extreme $a_{n}=a_{p}$.
Therefore, $E_{PNC}$ changes by about $-0.2\%$ ($-0.0018$) 
due to consideration of the neutron distribution.  
This is in agreement with Derevianko's estimate,
$-0.19(8)\%$ \cite{derevianko02}.

\subsubsection{Strong-field QED radiative corrections}
\label{sssec:radcorr}

It was noted in Ref. \cite{sushkovbreitrad} that corrections 
to the PNC amplitude due to vacuum polarization by the strong Coulomb field of the 
nucleus could be comparable in size to the Breit correction. 
This has been confirmed by calculations, the strong-field 
radiative corrections associated with 
the Uehling potential (vacuum polarization) increase 
$E_{PNC}$ by $0.4\%$ \cite{JBS01,MS02,DFG02}. 
%It was thought that the Uehling potential gives the largest contribution to $E_{PNC}$ 
%from strong-field radiative corrections \cite{MS02}. 

In Ref. \cite{DFG02arxiv} it was pointed out that the self-energy correction 
can give a larger contribution to $^{133}$Cs PNC 
with opposite sign ($\sim -0.65\%$). The self-energy and vertex corrections were first 
calculated in \cite{KF2002} and found to be $-0.73(20)\%$ for $^{133}$Cs.
The relation between the PNC correction and radiative corrections to finite
nuclear size energy shifts was used in this work. 
This result was confirmed in direct analytical calculations using $Z \alpha$ expansion
 \cite{kuchiev2002,MST2002,MST2003a} and by all-orders in $Z \alpha$  
numerical calculations of the PNC matrix element of the 
$2S_{1/2}-2P_{1/2}$ transition in hydrogenic ions performed in \cite{sapirstein2003}.

Note that corrections occur at very small distances ($r \lesssim 1/m$) where the nuclear
Coulomb field is not screened and the electron energy is negligible. Therefore,
the relative radiative corrections to weak matrix elements in neutral atoms like Cs
are approximately the same as for the $2S_{1/2}-2P_{1/2}$ transition in hydrogenic ions.

There is good agreement between the different calculations for all values of $Z$; 
see \cite{KFarxiv} and the review \cite{KF2003}. 
For the strong-field self-energy and vertex contribution to PNC in $^{133}$Cs we will 
quote the value $-0.8\%$, which is the average value of \cite{KF2002,MST2003a} 
corresponding also to the value obtained in \cite{sapirstein2003}. 
%This correction has been calculated as $-0.73(20)\%$ \cite{KF2002} and 
%$-0.85\%$ \cite{MST2002}. We will quote the averaged value $-0.8\%$.

Above we discussed the radiative corrections to the weak matrix elements. However,
the sum-over-states expression for the PNC amplitude contains also energy denominators
and E1 electromagnetic amplitudes. It was shown in \cite{DFG02arxiv,DFG02} that for 
Cs the corrections to the energies ($-0.3 \%$) and E1 matrix elements ($+0.3 \%$) cancel.

The contributions of strong-field radiative corrections to $E_{PNC}$ of cesium are 
listed in Table \ref{tab:pnci}.      

\subsubsection{Tests of accuracy}
\label{sssec:acc}

%
%
%new measurement of static polarizability for Cs 6s
%Amini and Gould, hep-ph 
%

There are two main methods used to estimate the accuracy of the PNC 
amplitude $E_{PNC}$: 
(i) root-mean-square (rms) deviation of the calculated energy intervals, 
E1 amplitudes, and hyperfine structure constants 
from the accurate experimental values; 
(ii) influence of fitting of energies and hyperfine structure 
constants on the PNC amplitude. 

The PNC amplitude can be expressed as a sum over intermediate states 
(see beginning of Section \ref{ssec:cscalcs}). 
Notice that there are three dominating contributions to the $6S-7S$ PNC amplitude 
in Cs; see Eq. (\ref{sumcs}).
Each term in the sum is a product of E1 transition amplitudes, 
weak matrix elements, and energy denominators. 
Therefore, this amplitude is sensitive to the electron wave functions 
at all distances. (The weak matrix elements, energies, and E1 amplitudes 
are sensitive to the wave functions at small, intermediate, and large 
distances from the nucleus, respectively.)
While mixed-states calculations of PNC amplitudes do not involve a 
direct summation over intermediate states, it is instructive to 
analyze the accuracy of the weak matrix elements, energy intervals, 
and E1 transition amplitudes which contribute to Eq. (\ref{sumcs}) calculated 
using the same method as that used to calculate $E_{PNC}$. 
The accuracy of these quantities is determined 
by comparing the calculated values with experiment.
Note that we cannot directly compare weak matrix elements with experiment. 
However, like the weak matrix elements, hyperfine structure is determined 
by the electron wave functions in the vicinity of the nucleus, 
and this is known very accurately.

In Section \ref{section3} we presented calculations of the energies, E1 transition 
amplitudes, and hyperfine structure constants relevant to Cs $6S-7S$ $E_{PNC}$. 
The states that have been considered in these calculations are those relavant 
to $E_{PNC}$ [in the sum (\ref{sumcs})].

The calculated removal energies are presented in Table \ref{tab:energies}. 
The Hartree-Fock values deviate from experiment by $10\%$. 
Including the second-order correlation corrections ${\hat \Sigma} ^{(2)}$ 
reduces the error to $\sim 1\%$. 
When screening and the hole-particle interaction are included into 
${\hat \Sigma} ^{(2)}$ in all orders, the energies improve, $0.2-0.3\%$. 
The rms deviation between the calculated and experimental energy intervals 
$\epsilon_{6s}-\epsilon_{6p_{1/2}}$, $\epsilon_{7s}-\epsilon_{6p_{1/2}}$, 
and $\epsilon_{6s}-\epsilon_{7p_{1/2}}$ is $0.3\%$. 

We mentioned at the end of Section \ref{ssec:sum} that the experimental 
values for energies can be fitted exactly by placing a coefficient before 
the correlation potential $\hat{\Sigma}$. The stability of the amplitude 
$E_{PNC}$ (as well as the E1 amplitudes and hfs constants) 
with fitting gives us an indication of the size of omitted contributions.
Note that the accuracy for the energies is already very high and the remaining 
discrepancy with experiment is of the same order of magnitude as the Breit and 
radiative corrections. Therefore, generally speaking, we should not expect that
fitting of the energy will always improve the results for amplitudes and
hyperfine structure. In fact, as we will see below, some values do improve 
while others do not. The overall accuracy, however, remains at the same level.

Below we present results for $E_{PNC}$ obtained in three different approximations: 
with unfitted $\hat{\Sigma}$, and with $\hat{\Sigma}^{(2)}$ and $\hat{\Sigma}$ fitted with
coefficients to reproduce experimental removal energies.\footnote{
Note that fitting $\hat{\Sigma}^{(2)}$ is an empirical method to estimate 
screening corrections (which were accurately calculated in $\hat{\Sigma}$). 
Agreement between results with fitted $\hat{\Sigma}^{(2)}$ and ab initio $\hat{\Sigma}$ 
shows that the fitting procedure is a reasonable way to estimate omitted diagrams.} 
First, we analyse the E1 transition amplitudes and hfs constants 
calculated in these approximations.

The relevant E1 transition amplitudes (radial integrals) are presented in 
Table \ref{tab:e1}. 
These are calculated with the energy-fitted ``bare'' correlation 
potential $\hat{\Sigma}^{(2)}$ and the (unfitted and fitted) 
``dressed'' potential $\hat{\Sigma}$. 
Structural radiation and normalization contributions are also included. 
The rms deviations of the calculated E1 amplitudes from 
experiment are the following: without energy fitting, 
the rms deviation is $0.1\%$; fitting the energy gives a rms deviation of 
$0.2\%$ for $\hat{\Sigma}^{(2)}$ and $0.3\%$ for the complete $\hat{\Sigma}$. 
Note, these correspond to the deviations between the calculations and the central 
points of the measurements. The errors associated with the measurements are 
in fact comparable to this difference. So it is unclear if the theory is 
limited to this precision or is in fact much better. Regardless, the uncertainty 
in the theoretical accuracy remains the same.
  
The hyperfine structure constants calculated in different approximations 
are presented in Table \ref{tab:hfs}.
Corrections due to the Breit interaction, structural radiation,  and 
normalization are included. 
The rms deviation of the calculated hfs values from experiment 
using the unfitted ${\hat \Sigma}$ is $0.5\%$.  
With fitting, the rms deviation in the pure second-order approximation 
is $0.3\%$; with higher orders it is $0.4\%$. 
The point is to estimate the accuracy of the $s-p_{1/2}$ weak 
matrix elements. It seems reasonable then to use the square-root 
formula, $\sqrt{{\rm hfs}(s){\rm hfs}(p_{1/2})}$. 
Notice that by using this approach the deviation is smaller.
Without energy fitting, the rms deviation is $0.5\%$. 
With fitting, the rms deviation in the second-order calculation 
($\hat{\Sigma} ^{(2)}$) is $0.2\%$ and in the full calculation
($\hat{\Sigma}$) it is $0.3\%$.    

From the above consideration it is seen that the rms deviation for the relevant 
parameters is $0.5\%$ or better. 
Note that from this analysis the error for a sum-over-states 
calculation of $E_{PNC}$ would be larger than this, as
the errors for the energies, hfs constants, and E1 amplitudes 
contribute to each of the three terms in Eq. (\ref{sumcs}). 
However, in the mixed-states approach, the errors do not add in this 
way.

We now consider calculations of the PNC amplitude performed in \cite{DFG02} 
in different approximations (with unfitted $\hat{\Sigma}$, and with energy-fitted 
$\hat{\Sigma}^{(2)}$ and $\hat{\Sigma}$). 
The spread of the results can be used to estimate the error.
The results are listed in Table \ref{tab:pncii}.
It can be seen that the PNC amplitude is very stable. 
The PNC amplitude is much more stable than hyperfine structure. 
This can be explained by the much smaller correlation corrections 
to $E_{PNC}$ ($\sim 2\%$ for $E_{PNC}$ and $\sim 30\%$ for hfs;
compare Table \ref{tab:pnci} with Table \ref{tab:hfs}).
One can say that this small value of the correlation correction is 
a result of cancellation of different terms in (\ref{eq:pnc-cor})
but each term is not small (see Table \ref{tab:pnci}). However,
this cancellation has a regular behaviour. 
The stability of $E_{PNC}$ may be compared to the stability of the 
usual electromagnetic amplitudes where the error is very small 
(even without fitting).\footnote{
Note that different methods also give different signs of the 
errors for hfs. This is one more argument that the true value of 
$E_{PNC}$ is somewhere in the interval between the results of 
different calculations in Table \ref{tab:pncii}.}

In \cite{DFG02} the fitting of hyperfine structure was also considered, 
using different coefficients before each $\hat{\Sigma}$.
The first-order in $\hat{\Sigma}$ correlation correction (\ref{eq:pnc-cor}) 
changes by about $10\%$. It was found that the PNC amplitude changes 
by about $0.4\%$. 

It is also instructive to look at the spread of $E_{PNC}$ obtained 
in different schemes. 
%(This has already been discussed in some detail in 
%Section \ref{sec:results}.) 
The result of the work \cite{DFG02} (the number we present here) 
is in excellent agreement with the earlier
result \cite{DFS1989pnc} while the calculation scheme is significantly different.
The only other calculation of the $E_{PNC}$ in Cs which is as 
complete as \cite{DFS1989pnc,DFG02} is that of Blundell {\it et al.} 
\cite{BJS1990}. 
Their result in the all-orders sum-over-states approach is 0.909 
(without Breit) and is very close to the value of 0.908 
(corresponding to ``Subtotal'' of Table \ref{tab:pnci}). 

A note on the sum-over-states procedure.
The authors of reference \cite{BJS1990} include single, double, and 
selected triple excitations into their wave functions.
Note, however, that even if wave functions of $6S$, $7S$, and intermediate 
$NP$ states are calculated exactly 
(i.e., with all configuration mixing included) 
there are still some missed contributions in this approach.
Consider, e.g., the intermediate state $6P\equiv 5p^{6}6p$. 
It contains an admixture of states $5p^{5}ns6d$:
$6P=5p^{6}6p + \alpha 5p^{5}ns6d+...$.
This mixed state is included into the  sum (\ref{sum}).
However, the sum (\ref{sum}) must include all many-body states
of opposite parity. 
This means that the state $\widetilde{5p^{5}ns6d}=
5p^{5}ns6d- \alpha 5p^{6}6p+...$ should also be included into 
the sum. Such contributions to $E_{PNC}$ have never
been estimated directly within the sum-over-states approach. 
However, they are included into the mixed-states calculations 
\cite{DFS1989pnc,BJS1990,kozlovcs,DFG02}.

It is important to note that the omitted higher-order many-body corrections 
are different in the sum-over-states \cite{BJS1990} and 
mixed-states \cite{DFS1989pnc,DFG02} calculations.
This may be considered as an argument that the omitted many-body corrections 
in both calculations are small. 
Of course, here it is assumed that the omitted many-body corrections to both 
values (which, in principle, are completely different) do not 
``conspire'' to give exactly the same magnitude.

A comparison of calculations of $E_{PNC}$ in second-order with 
fitting of the energies is also useful in determining the accuracy 
of the calculations of $E_{PNC}$. 
(Remember that this value is in agreement with results of similar 
calculations performed in \cite{BJS1990,kozlovcs}; 
see Section \ref{sssec:mixedpnc}.)
One can see that replacing
the all-order $\hat \Sigma$ by its very rough second-order (with fitting)
approximation changes $E_{PNC}$ by less than 0.4\% only. On the other hand,
if the higher orders are included accurately, the difference between the 
two very different approaches is 0.1\% only. 

The maximum deviation obtained in the above analysis is $0.5\%$. 
This is the error claimed in the $E_{PNC}$ calculation \cite{DFG02}. 

\subsection{The vector transition polarizability}

The determination of the nuclear weak charge from the Stark-PNC interference 
measurements also requires knowledge of the vector transition polarizability 
$\beta$.
% no direct measurements 
This can be found in a number of ways: 

%
%??? Hoffnagle, Telegdi, Weis, Phys. Lett. 86A, 457 (1981).	
%	- determination of beta first discussed in this ref (according 
%	to Gilbert, Wieman)

(i) from a direct calculation of $\beta$. 
$\beta$ can be expressed as a sum over intermediate states 
and experimental E1 transition amplitudes and energies can 
be used \cite{bouchiats1} (see also \cite{BJS1990,DFS1997}). 
However, this calculation is 
unstable due to strong cancellations of different terms in the 
sum (see Ref. \cite{DFS1997}). These cancellations are explained 
by the fact that $\beta$ is proportional to the spin-orbit interaction, 
therefore for zero spin-orbit interaction the sum for $\beta$ must be zero; 

(ii) from the measurement of the ratio of the off-diagonal hyperfine 
amplitude  to the vector transition polarizability, $M_{hfs}/\beta$ 
\cite{bouchiatsm1}. $\beta$ is then extracted from the ratio 
using a theoretical determination of $M_{hfs}$; 

(iii) from the measurement of the ratio of the scalar to vector 
polarizabilities, $\alpha/\beta$. $\alpha$ can be calculated 
accurately using experimental values for E1 transition amplitudes 
and energies in the sum-over-states approach 
(the calculation of $\alpha$ is much more stable than that of $\beta$ 
\cite{DFS1997}).

There are currently two very precise determinations of $\beta$.  
One was obtained from 
the analysis \cite{DF2000} (calculation of $M_{hfs}$) of the 
measurement \cite{BW} of the ratio 
$M_{hfs}/\beta$, $\beta =26.957(51)a_{B}^{3}$, and 
another is from an analysis \cite{DFG02} (semi-empirical 
calculation of $\alpha$; see \cite{DFS1997} for details, 
where a similar calculation was performed)
of the measurement \cite{alpha/beta97} of the ratio $\alpha/\beta$ 
using the most accurate experimental data for E1 transition amplitudes 
including the recent measurements of Ref. \cite{vasilyev2002}, 
$\beta =27.15(11)a_{B}^{3}$.   
An average of these values gives
\begin{equation}
\label{eq:beta_av}
\beta =26.99(5)a_{B}^{3} \ .
\end{equation}

\subsection{The final value for the Cs nuclear weak charge $Q_{W}$ and implications}
\label{ssec:finalvalue}

Combining the measurement \cite{wieman}
\begin{equation}
-\frac{{\rm Im}(E_{PNC})}{\beta}=1.5935(56)\frac{\rm mV}{\rm cm}
\end{equation}
with the calculated value (see Table \ref{tab:pnci})
\begin{equation}
E_{PNC}=0.897(1\pm 0.5\%)\times 10^{-11}iea_{B}(-Q_{W}/N)
\end{equation}
(from the calculation \cite{DFG02} with the averaged value $-0.8\%$ of 
works \cite{KF2002,MST2002} for the self-energy and vertex radiative corrections) 
and the averaged value for $\beta$ [Eq. (\ref{eq:beta_av})], gives   
\begin{equation}
\label{eq:q_w(cs)}
Q_{W}=-72.74(29)_{\rm exp}(36)_{\rm theor}
\end{equation}
for the value of the nuclear weak charge for $^{133}$Cs. 
The difference between this value and that predicted by the standard model, 
$Q_{W}^{SM}(^{133}_{55}{\rm Cs})=-73.19\pm 0.13$ \cite{rosner2002},\footnote{
We use this value rather than the Particle Data Group value, Eq. (\ref{eq:Q_{W}SM}), 
since we use the new physics analysis of Ref. \cite{rosner2002}.} 
%[Eq. (\ref{eq:Q_{W}SM})] 
is 
\begin{equation}
\label{eq:q_w(cs)diff}
\Delta Q_{W}^{\rm new}\equiv Q_{W}-Q_{W}^{SM}=0.45(48) \ ,
\end{equation}
adding the errors in quadrature.

%It has been shown by Sandars \cite{sandars1990} that re-expressing 
%parity violation for heavy atoms in terms of the precisely known 
%$Z$-boson mass $M_{Z}$ as $G_{F}Q_{W}M_{Z}^{2}$ removes the dependence 
%on standard model parameters such as the Weinberg angle $\sin ^{2}\theta _{W}$ and 
%the top quark mass.
%This means that a deviation from the standard model value is due to new physics.

Let us briefly consider the constraints on physics beyond the standard 
model set by the nuclear weak charge of cesium [Eq. (\ref{eq:q_w(cs)diff})]. 
New physics corrections to the standard model are divided into two groups, 
those that originate through vacuum polarization corrections to gauge boson propagators 
and those that originate from all other mechanisms (such as new physics 
arising through vertex and self-energy diagrams and new tree-level physics). 
The former are termed ``oblique'' and the latter ``direct''. 

A formalism for oblique corrections has been devised by Peskin and Takeuchi 
\cite{peskintakeuchi} in terms of weak isospin conserving and weak isospin breaking parameters 
$S$ and $T$, respectively. Atomic parity violation (on a single isotope) is unique 
among other electroweak probes of new physics in its almost exclusive dependence on 
the parameter $S$. Indeed, the dependence of the nuclear weak charge on the parameter 
$T$ cancels almost exactly. For $^{133}$Cs \cite{rosner2002}
\begin{equation}
\label{eq:oblique}
\Delta Q_{W}^{\rm oblique}=-0.800S-0.007T \ .
\end{equation}
The standard model value corresponds to $S=T=0$ (no new physics) at values 
$m_{t}=174.3~{\rm GeV}$ for the top quark mass and $M_{H}=100~{\rm GeV}$ for 
the Higgs boson mass.\footnote{
The result is actually not very sensitive to $M_{H}$ which is currently 
$M_{H}>114.4~{\rm GeV}$ at the 95\% confidence level \cite{higgs}.} 
The constraint on $S$ from PNC in $^{133}_{55}{\rm Cs}$ [comparing Eqs. (\ref{eq:q_w(cs)diff}) 
and (\ref{eq:oblique})] is
\begin{equation}
S=-0.56(60) \ ,
\end{equation}   
adding the errors in quadrature.

In terms of direct new physics, a positive $\Delta Q_{W}^{\rm new}$ [Eq. (\ref{eq:q_w(cs)diff})] 
could be indicative of an extra $Z$ boson in the weak interaction. 
A lower bound for the $Z_{\chi}$ boson mass predicted in $SO(10)$ theories can be 
obtained from the deviation of the measured weak charge from theory, 
according to \cite{MR1990}
\begin{equation}
\Delta Q^{\rm new}_{W \ {\rm tree}}\approx 0.4(2N+Z)(M_{W}/M_{Z_{\chi}})^{2} \ .
\end{equation}
To one standard deviation, the lower bound on the mass $M_{Z_{\chi}}$ from 
parity violation in $^{133}_{55}{\rm Cs}$ is 
\begin{equation}
M_{Z_{\chi}}> 750~{\rm GeV} \ .
\end{equation}
A lower bound of about $600~{\rm GeV}$ has been obtained from a direct search 
at the Tevatron \cite{tevatron}.

For a discussion of atomic physics sensitivities to new physics, see, e.g., 
\cite{sandars1990,R-M1999,rosner2002} and references therein. 
For a recent analysis of electroweak tests of the standard model, including atomic parity violation, 
see, e.g., \cite{rosner2002,langacker2003,langacker2003b}; 
an earlier review on this topic \cite{LLM1992} is also very informative.  

%write explicit expression for weak charge in terms of new physics parameters? 
%explicitly show cancellation of parameter T

\subsection{Ongoing/future studies of PNC in atoms with a single valence electron}
\label{ss:further1pnc}

A PNC measurement in cesium is continuing at Paris \cite{guena2003}, 
where $\sim 1\%$ precision is expected to be reached. 
The possibility of performing PNC measurements on the single trapped ions Ba$^{+}$ and Ra$^{+}$ 
is being considered at Seattle \cite{fortson93,KSNF2003}. 
Preliminary atomic PNC calculations \cite{DFG2001} indicate that the accuracy for calculations 
of Ba$^{+}$  
could compete with that of cesium.  
An experiment to measure PNC in francium was discussed at Stony Brook \cite{sprouse2002,aubin}.
%expected accuracy for francium?
Francium is a heavier analog of cesium, and correspondingly the 
PNC effect is an order of magnitude (18 times \cite{DFS1995}) larger than that for cesium. 
However, there are no stable isotopes in francium. 
The group uses laser trapping to suspend the atoms.

A new facility (TRI$\mu$P) is under development at the Kernfysisch Versneller Instituut, 
Groningen. One aim is to measure parity violation in radioactive atoms and ions, 
attractive candidates being Ra$^{+}$ and Fr. See Ref. \cite{trimup}.

Note that the atomic theory is limited by the error associated with calculations of 
correlation corrections. It's expected that the relative corrections due to correlations are 
the same in analogous atoms, e.g., cesium and rubidium. Thus, taking ratios of PNC measurements in 
analogous atoms possibly provides a way to circumvent the troublesome correlations 
in the determination of the ratio of the nuclear weak charges. 
[Another way to exclude atomic theory is through isotope ratios, 
see Section \ref{ss:isotoperatios}.] 
Light atoms by themselves may provide some advantage from the point of view of atomic theory. 
Breit, neutron skin, and QED radiative corrections for rubidium are much smaller than for cesium. 
Therefore, atomic PNC calculations for rubidium could, in principle, reach a precision better 
than for cesium, however the amplitude is much smaller \cite{johnson8586,DFSS1987b}, 
about $6$ times \cite{DFSS1987b}.

\section{Atoms with several electrons in unfilled shells}
%\label{ssec:mbpt+ci}
\label{section3iic}

There are some advantages in measuring PNC effects in heavier atoms (where the PNC signal is 
expected to be larger) with relatively complicated electronic structure. 
However, the interpretation of the measurements in such a case is strongly impeded by 
the poor knowledge of the atomic wave functions. Even for Tl, which has 
a relatively simple structure, the atomic theory has an error of 
2.5-3\% \cite{DFSS1987b,kozlovtl}. 

The method we discussed in Sections \ref{section3} and \ref{section3iib} is applicable 
for heavy alkali-metal atoms. However, this method may not be accurate for atoms with 
more than one electron in unfilled shells. This is because the correlations between 
the electrons in the unfilled shells may be important. 

The most effective many-body method for atoms with several 
electrons above closed shells was developed in the works 
\cite{DFK1996,DKPF1998} (see also \cite{DFSS88}). 
It is a combination of many-body perturbation theory 
and configuration interaction methods (it is called ``MBPT+CI''). 
An effective Hamiltonian is constructed for the valence electrons using 
many-body perturbation theory for the interaction of the valence electrons 
with the core.   
In this way the correlations between the external electrons and the core are 
taken into account using MBPT while the correlations between 
the external electrons are calculated using the CI method. 

A detailed description of the MBPT+CI method is beyond the scope of this 
review. We refer the interested reader to the works \cite{DFK1996,DKPF1998}.

\subsection{Parity nonconservation in thallium}

High-precision measurements of parity violation in the 
$6P_{1/2}-6P_{3/2}$ transition in thallium 
have been performed by the Oxford and Seattle groups, 
\begin{equation}
{\cal R}\equiv {\rm Im}\{ E_{PNC}/M1 \}= \left \{
\begin{array}{lll} 
\label{eq:tlmeasurements}
(-15.68\pm 0.45)\times 10^{-8} & {\rm \qquad Oxford} & \cite{edwards95} \\
(-14.68\pm 0.17)\times 10^{-8} & {\rm \qquad Seattle} & \cite{vetter95} \ .
\end{array}
\right .
\end{equation}
We will use the latter, most precise measurement to extract the value for 
$Q_{W}$. A recent measurement of the ratio $E2/M1$ indicates that the 
Oxford result should be rescaled to ${\cal R}=-15.34\pm 0.45$ \cite{majumder1999}, 
reducing the discrepancy between the two measurements (\ref{eq:tlmeasurements}). 

Let us move on now to the calculations. 
The electronic configuration for the ground state of thallium is $6s^{2}6p_{1/2}$. 
This system can be treated as a one-electron or a three-electron system 
above closed shells. However, the $6p$ electron is energetically close to the 
$6s$ electrons, and correlations between $6s$ and $6p$ are significant. 
This means that the method described in Sections \ref{section3},\ref{section3iib} 
will not be as effective for thallium as for cesium. 
This method was employed in the 
work \cite{DFSS1987b} to the transition $6p_{1/2}-6p_{3/2}$, 
yielding the following result:
\begin{equation}
\label{eq:e_pnctl1987}
E_{PNC}=(-2.70 \pm 0.08)\times 10^{-10}iea_{B}(-Q_{W}/N) \ .
\end{equation}
This result includes core polarization of E1 and weak fields calculated in 
the TDHF method and all second order correlation corrections 
(higher-order diagrams -- electron-electron screening and the hole-particle interaction -- 
are not included). Finally, empirical fitting of the 
energies is used to take into account some missed higher-order correlations.   

Consideration of the Breit interaction, Eq. (\ref{eq:breit}), gives a correction 
\cite{dzubabreit}
\begin{equation}
\label{eq:e_pncbreittl}
\Delta E_{PNC}^{\rm Breit}/E_{PNC}=-0.0098 \ .
\end{equation} 
The size of radiative corrections to Tl PNC is the following. 
The Uehling correction is $0.94\%$ \cite{DFuehlingtl} and  
we take the self-energy and vertex corrections to be $-1.51\%$ 
(average of $-1.61\%$ from \cite{KF2002} and $-1.41\%$ from \cite{MST2002}),
\begin{equation}
\Delta E_{PNC}^{\rm rad.}/E_{PNC}=-0.0057  \ .
\end{equation}
The correction for the neutron distribution is small 
(with a relatively large error) \cite{kozlovtl},
%
%cite Derevianko or Bunny ..... instead???
%
\begin{equation}
\Delta E_{PNC}^{\rm neutron}/E_{PNC}=-0.003  \ .
\end{equation} 
With these corrections, Eq. (\ref{eq:e_pnctl1987}) becomes
\begin{equation}
\label{eq:e_pnctlfin}
E_{PNC}=(-2.65\pm 0.08)\times 10^{-10}iea_{B}(-Q_{W}/N) \ .
\end{equation}
(the errors associated with the corrections to Eq. (\ref{eq:e_pnctl1987}) 
are well below 1\%.)

More recently the method MBPT+CI was applied in Ref. \cite{kozlovtl}, with the result
\begin{equation}
\label{eq:e_pnctl2001}
E_{PNC}= (-2.72 \pm 0.07)\times 10^{-10}iea_{B}(-Q_{W}/N) \ .
\end{equation}
This method is well-suited for a system like thallium, where there are three electrons 
above a compact core, with the correlations between the three external electrons treated 
non-perturbatively. 
In this work the Gaunt interaction (the dominant, magnetic part of the Breit interaction, 
Eq. (\ref{eq:breit})) was included self-consistently at every stage of the calculation. 
Inclusion of radiative corrections shifts the value (\ref{eq:e_pnctl2001}) slightly,   
$E_{PNC}=(-2.70\pm 0.07)\times 10^{-10}iea_{B}(-Q_{W}/N)$. 
The difference between this value and Eq. (\ref{eq:e_pnctlfin}) is within the assigned errors.

We need a value for the $6P_{1/2}-6P_{3/2}$ magnetic dipole transition amplitude, $M1$. 
Calculations of this quantity are very stable. In Ref. \cite{kozlovtl} the values 
$M1=1.692\times 10^{-3}{\rm ~a.u.}$ and $M1=1.694\times 10^{-3}{\rm ~a.u.}$ were calculated, 
the former using the MBPT+CI method and the latter in the MBPT method (treating thallium as 
a single-electron atom). 

The Seattle measurement (\ref{eq:tlmeasurements}), the PNC calculation (\ref{eq:e_pnctlfin}), 
and the M1 amplitude calculated in \cite{kozlovtl} lead to the following value for the 
nuclear weak charge of thallium, 
\begin{equation}
Q_{W}(^{205}{\rm Tl})= -116.3(1.3)_{\rm exp}(3.5)_{\rm theor} \ ,
%kozlov et al. value is -114.1(1.3)_{exp}(2.9)_{theor}
\end{equation} 
in agreement with the standard model value \cite{PDG}
\begin{equation}
Q_{W}^{SM}(^{205}{\rm Tl})=-116.67(5) 
\end{equation}

\subsection{A method to exclude the error from atomic theory: isotope ratios and 
the neutron distribution}
\label{ss:isotoperatios}

In Ref. \cite{DFK86} it was pointed out that atomic theory 
(and hence the large associated error) 
can be excluded by taking ratios of atomic PNC measurements along an isotope chain. 
Indeed, from a simple non-relativistic consideration, the measured parity violating 
amplitude can be expressed as 
\begin{equation}
A_{PNC}=\xi Q_{W} \ ,
\end{equation}
where the atomic theory is included into $\xi$.
It is seen that in this approximation taking ratios of measurements of different 
isotopes with neutron numbers $N$ and $N'$  
\begin{equation}
\frac{A_{PNC}(N')}{A_{PNC}(N)}= 
\frac{Q_{W}(N')}{Q_{W}(N)}
\end{equation}
removes the dependence on $\xi$, 
assuming that the atomic structure does not change significantly from isotope to isotope. 
However, a consideration of relativistic effects (the variation of the 
electron wave functions inside the nucleus) changes this simple formulation. 

It was shown in Ref. \cite{FPW90} that while the atomic structure cancels 
in the isotope ratios, there is an enhanced sensitivity to the neutron distribution 
$\rho _{n}(r)$, and the uncertainty associated with $\rho _{n}(r)$ may limit 
the extraction of new physics. 
The dependence of the PNC amplitude on nuclear structure effects can be 
incorporated through the correction $Q^{\rm nuc}$ \cite{FPW90}, 
\begin{equation}
A_{PNC}=\xi (Q_{W}+Q_{W}^{\rm nuc}) \ ,
\end{equation}
where 
\begin{equation}
Q_{W}^{\rm nuc}=-N(q_{n}-1)+Z(1-4\sin^{2}_{\theta _{W}})(q_{p}-1)
\end{equation}
and 
\begin{equation}
q_{n}=\int \rho _{n}(r)f(r)d^{3}r \ , \qquad q_{p}=\int \rho _{p}(r)f(r)d^{3}r \ . 
\end{equation}
The variation of the electron wave functions inside the nucleus is given by 
$f(r)$, which is normalized to $f(0)=1$.
The isotope ratio 
\begin{equation}
R\equiv \frac{A_{PNC}(N')}{A_{PNC}(N)}\approx \frac{Q_{W}(N')}{Q_{W}(N)})[1+\Delta q_{n}] \ ,
\end{equation}
where $\Delta q_{n}\equiv q_{n}-q_{n}'$.
It is seen that $R$ is sensitive, in particular, to the difference 
in the neutron distributions, and it is the error associated with this 
difference, $\delta \big(\Delta q_{n}\big)$, that could limit the interpretation 
of the ratio $R$ of the PNC measurements in heavy atoms in terms of new physics.

The isotope ratios are sensitive to a combination of new physics that is different 
from that probed by measurements of parity violation in a single isotope.
While measurements of atomic parity violation in a single isotope are 
sensitive to the weak isospin conserving parameter $S$ 
%(the sensitivity to $T$ cancels for the values of $N$ and Z$ for atoms of interest 
%in PNC experiments) 
and to new tree level physics, 
the isotope ratios are sensitive to both S and T through $\sin^{2}\theta _{W}$ 
as well as to new tree level physics.  
The sensitivity of the isotope ratios to oblique new physics has been 
investigated in \cite{MR1990,rosner1996,PFW92}, and the sensitivity to direct 
new physics in \cite{R-M1999,PFW92}. 

%
%discuss sensitivity to new couplings to protons
%
%show why it is most beneficial to search for isotopes separated by large \Delta N
%

Nuclear structure corrections to the weak charge distribution have been calculated for 
different isotopes of lead \cite{PFW92,VLR2000}, 
cesium \cite{CV93,VLR2000,PD2000}, barium \cite{VLR2000}, 
and ytterbium \cite{VLR2000}.   
The calculations are model-dependent, and it is seen from the spread in the 
results that at this time our knowledge of the neutron distribution would 
preclude a determination of new physics competitive with other probes. 
The problem is that there is a very limited amount of empirical information 
on neutron distributions. In Ref. \cite{DP2002neutron} new data on neutron 
distributions from experiments with antiprotonic atoms was used to re-analyze 
the impact of the nuclear structure uncertainties on the extraction of new physics. 
It was found that the errors associated with nuclear structure are slightly reduced 
compared with those associated with nuclear calculations. It was suggested that 
isotope ratios of PNC measurements with atoms having $Z\lesssim 50$ may be more 
effective at searching for new physics, 
since the nuclear structure uncertainty in the extraction of (direct) new physics 
increases roughly as $Z^{8/3}$ and the required experimental uncertainty is less strict 
for lighter atoms \cite{DP2002neutron}. Since the PNC signal $\propto Z^{3}$, this would 
only be feasible if there is an accidental enhancement of the PNC effect 
(for example, very close levels of opposite parity). 

The high sensitivity of isotope ratios of PNC measurements to variations in the 
neutron distribution could be used to determine the nuclear structure and test 
nuclear models \cite{FPW90}. 
Parity violating electron scattering measurements appear to be very promising 
for extracting information on the neutron distribution (see, e.g, \cite{HPSM2001}). 
It appears that such measurements could be used to significantly reduce the 
nuclear structure uncertainties to the level where the new physics sensitivity 
of PNC isotope ratios is significant.
 
%For cesium isotopes the uncertainty in the neutron distribution does not appear to 
%be a limiting factor \cite{CV93}.
%However, our current understanding of nuclear structure 
%precludes a determination of $\sin ^{2}\theta$ to better than 1\% for lead \cite{FPW90,PFW92}. 
%Due to the high sensitivity to the neutron distribution, it was suggested in \cite{FPW90,PFW92} 
%that accurate measurements of parity violation could be used to determine the neutron distribution 
%in heavy atoms.  
% barium, P. Vogel, in Nuclear Shapes and Nuclear Structure, .....(see R-M1999 paper)

%\subsection{New physics sensitivity of isotope ratios}

%For illustrative purposes, let us consider the nuclear weak charge at tree level and 
%express it in terms of the proton and neutron weak charges $Q_{W}^{p}$ and $Q_{W}^{n}$, 
%\begin{equation}
%Q_{W}(N)=NQ_{W}^{n}+ZQ_{W}^{p} \ .
%\end{equation}
%The ratio of the weak charges measured on different isotopes, $Q_{W}(N)$ and 
%$Q_{W}(N')$, where $N'=N+\Delta N$, is then 
%\begin{equation}
%\frac{Q_{W}(N')-Q_{W}(N)}{Q_{W}(N)}=\frac{Q_{W}(N')}{Q_{W}(N)}-1=
%\frac{\Delta N}{N+Z(Q_{W}^{p}/Q_{W}^{n})} \ .
%\end{equation}
%Clearly isotope ratios have a very different dependence on the proton and 
%neutron weak charges and hence on new physics. 
%

%use expression for weak charge which shows dependence on new physics parameters 
%show that parameter T does not cancel

%It was proposed in Ref. \cite{DFK86} to study rare earth atoms, which have 
%anomalously close levels of opposite parity and a number of stable isotopes. 
% current experiments - Yb, Ba II, ?Dy

\subsection{Ongoing/future studies of PNC in complex atoms}
\label{ss:further2+pnc}

A thallium PNC measurement is continuing at Seattle \cite{cronin98} in which 
sub-1\% accuracy is expected to be reached.
Experimental studies of rare-earth elements dysprosium and ytterbium 
are in progress at Berkeley \cite{english}.
Studies with dysprosium are continuing despite the small upper limit 
found for the PNC amplitude \cite{nguyen97}. 
Dysprosium appeared to be particularly attractive for PNC studies due to 
the presence of a state ($4f^{9}5d^{2}6s~J=10$) that is nearly degenerate 
with the state $4f^{10}5d6s~J=10$ with opposite parity and the same angular momentum.  
However, the weak interaction does not mix the dominant electronic configurations of the 
nearly degenerate states, a non-zero weak matrix element shows up only through configuration 
mixing and core polarization which makes it very small.
An improvement in the statistical 
sensitivity of a few orders of magnitude is anticipated in the current search 
\cite{nguyen97,nguyen1999}. If the effect is of the same order of 
magnitude as the current upper limit, then dysprosium will provide an important test 
of the standard model and allow the determination of the nuclear anapole moment 
\cite{nguyen97,nguyen1999}.  
A measurement of PNC in ytterbium is in progress also at Berkeley \cite{demille95,stalnaker02}. 
In ytterbium there are seven stable isotopes in the range $A=168-176$ with two of these isotopes 
having non-zero nuclear spin (possibility to observe the nuclear anapole moment). 
The atomic theory in this case should be more reliable than in the other 
rare-earth atoms, since here there is a closed $4f$-shell.

PNC studies with samarium appear to have no advantages over other systems. 
Preliminary work showed that PNC optical rotation on the forbidden M1 transition 
$4f^{6}6s^{2}~^{7}F_{0}-4f^{6}5d6s~^{7}G_{1}$, 
which could have benefited from having close ($\sim 10~{\rm cm}^{-1}$) 
opposite parity states, is not feasible \cite{davies1989}.
Moreover, it has recently been found, through a term reassignment 
of several of the odd-parity states in Sm \cite{rochester1999}, that an  
estimate of the weak matrix element made in the original proposal for the 
experiment \cite{gongora} should be reduced by about $40$ times.
The null PNC results of the Oxford group reported in Refs. \cite{wolfenden93,lucas98} 
for M1 transitions between the ground state configuration show that the possible 
enhancement due to the relatively small energy denominators ($\sim 200~{\rm cm}^{-1}$) 
is not realized due to very small $E1$ and weak matrix elements.

There are some interesting proposals for new experiments where large enhancements are 
expected. 
E.g., atomic calculations \cite{flambaum1999,DFG2000} indicate that parity violation in 
the electronic transitions $7s^{2}~{^{1}S}_{0} - 7s6d~{^{3}D}_{1}$ of the radium 
isotopes $^{223}$Ra and $^{225}$Ra are enhanced due to a very small interval 
between the states $7s7p~{^{3}P}_{1}$ and $7s6d~{^{3}D}_{1}$. 
The parity violating amplitude is calculated to be 100 times larger than that of Cs.
(Note that the electronic structure of radium is relatively simple and so it is expected that 
the PNC amplitude will not suffer from large suppressions that appeared in the PNC 
amplitudes of the rare earth atoms.)
  
\section{The nuclear anapole moment and measurements of P-odd nuclear forces
in atomic experiments}
\label{section6}

\subsection{The anapole moment}

The notion of the anapole moment was
introduced by Zel'dovich \cite{zeldovich1957}
just after the discovery of parity violation.
He noted that a particle may have a parity violating
electromagnetic form factor, in addition to the usual
electric and magnetic form factors.
The first realistic example, the anapole moment of the
nucleus, was
considered in Ref. \cite{FK1980} and calculated
in Ref. \cite{FKS1984}.
In these works it was also demonstrated that atomic and molecular
experiments could detect anapole moments.
Subsequently, a number of experiments
were performed in Paris, Boulder,
Oxford, and Seattle
\cite{bouchiat84,gilbert85,noecker88,edwards95,vetter95} and some limits on the 
magnitude of the anapole moment were established. 
However, it was only recently
that a nuclear anapole moment was unambiguously detected
-- in 1997 a group in Boulder measured a nuclear anapole
moment in $^{133}$Cs (using atomic experiments) to an
accuracy of 14\% \cite{wieman} (see also the recent review 
\cite{HW2001}).
This is the
first observation of an electromagnetic moment that violates
fundamental discrete symmetries.

Besides the usual electric and magnetic monopole, dipole
and quadrupole (and so on) moments,
there are also other electromagnetic multipole moments, which
are not usually dealt with in multipole moment expansions
as they give rise to contact, rather
than long-range, potentials.
The anapole moment is
such a moment. It obeys time reversal
invariance but violates parity conservation
and charge conjugation invariance
(i.e., $T$-even and $P$- and $C$-odd).
The anapole moment arises out of an expansion of the
vector potential as a series in $R^{-1}$, where $R$ is the
distance from the center of the charge distribution
(see, e.g., \cite{SFK84,khriplovichpnc}).
The part of the vector potential that is due to the
anapole moment is\footnote{
In the gauge $\mbox{\boldmath$\nabla$}\cdot {\bf A}=0$ there is a long-range 
term in the anapole vector-potential which may be removed by a gauge 
transformation.} 
\begin{equation}
\label{eq:anaA}
{\bf A}^a ({\bf r}) = {\bf a} \delta ^{3}({\bf r}) \ ,
\end{equation}
%??? this form??? or density?
% definition not unique - see e.g. definition in spherical tensor notations 
%in Ref. HHM
where
\begin{equation}
\label{eq:anaj}
{\bf a} = -\pi \int r^2 {\bf j} ({\bf r}) d^{3}r \ ,
\end{equation}
${\bf j}$ is the electromagnetic current density. 
Eq. (\ref{eq:anaj}) can be taken as the definition of the anapole moment.
Notice the contact form of the potential
-- this is true for any $T$-even, $P$-odd moment; see, e.g.,
\cite{FK1980,khriplovichpnc} for a proof.

\begin{figure}
\psfig{figure=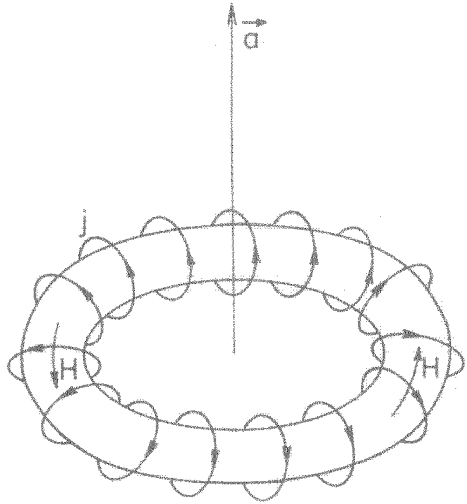,height=3in}
\caption{Diagram showing the anapole moment, ${\bf a}$,
the toroidal current that produces it, ${\bf j}$, and the
magnetic field that the current creates, ${\bf H}$.}
\label{f1}
\end{figure}

What kind of current distribution corresponds to an anapole
moment? Such a current distribution is shown in Fig. \ref{f1}.
This will give a nonzero anapole moment, because the
places at which the current is pointing downwards
are further from the center than the places where
the current is pointing upwards. Since the current in
Eq. (\ref{eq:anaj}) is weighted by a factor of $r^2$ this
current distribution will produce an anapole moment
pointing perpendicular to the plane of the doughnut.
A magnetic field is produced inside the current distribution,
as shown in the figure.

The expression for the anapole moment (\ref{eq:anaj}) contains
the current vector ${\bf j}$, which changes its sign under
reflection of co-ordinates.
The anapole moment is directed along
the nuclear spin ${\bf I}$:
$\langle \hat{\bf a} \rangle = -\pi \langle r^2 {\bf j} \rangle
= a {\bf I} / I$. However, the spin ${\bf I}$ does not change its
sign under co-ordinate reflection. The
different behavior of the right and left hand sides of the
relation $\langle r^2 {\bf j} \rangle \propto {\bf I}$ under
reflection of co-ordinates means that the existence of the
anapole moment violates parity, i.e., symmetry under the
reflection of co-ordinates (but it does not violate
time reversal invariance).

\subsection{Origin  of the nuclear anapole moment}

We now turn to the question of how an anapole moment
can actually be produced.
A $T$-even, $P$-odd moment like
the anapole can only arise if there is some kind of
$P$-odd force present; for this the weak interaction
is needed. 
The two-body $P$-odd nucleon-nucleon interaction, in the 
contact limit, has the form (see, e.g., \cite{khriplovichpnc})
\begin{eqnarray}
\label{eq:2bodyPNCnn}
\hat{W}_{ab}&=&
\frac{G}{\sqrt{2}}\frac{1}{2m}
\Big( \Big\{ ( g_{ab}\mbox{\boldmath$\sigma$}_{a} 
-g_{ba}\mbox{\boldmath$\sigma$}_{b} )
\cdot ({\bf p}_{a}-{\bf p}_{b}), 
\delta ({\bf r}_{a} - {\bf r}_{b})\Big\} \nonumber \\
&&+g^{\prime}_{ab}[\mbox{\boldmath$\sigma$}_{a} 
\times \mbox{\boldmath$\sigma$}_{b}] \cdot 
\mbox{\boldmath$\nabla$}\delta ({\bf r}_{a}-{\bf r}_{b}) 
\Big) \ ,
\end{eqnarray}
where $\{\ ,\ \}$ is an anticommutator, 
$G$ is the Fermi constant of the weak interaction, 
$m$ is the nucleon mass, 
and $\mbox{\boldmath$\sigma$}$, ${\bf p}$, and ${\bf r}$ 
are the spins, momenta, and coordinates of the nucleons $a$ and $b$. 
The dimensionless constants $g_{ab}$, $g_{ba}$, $g^{\prime}_{ab}$ give the strength 
of the weak interaction between nucleons. 
The effective one-body $P$-odd weak interaction between 
an unpaired nucleon and the nuclear core can then be obtained, 
\begin{equation}
\label{eq:poddvalence}
\hat{W} = \frac{G}{2 \sqrt{2} m} g [\bbox{\sigma}
\cdot {\bf p} \rho(r) + \rho(r) \bbox{\sigma}
\cdot {\bf p}] \ , 
\end{equation}
where $\rho(r)$ is the number density of core nucleons and 
\begin{equation}
\label{eq:gZN}
g=\frac{Z}{A}g_{ap}+\frac{N}{A}g_{an} \ .
\end{equation}
Here $a=p,n$ denotes the unpaired nucleon.  
For the $^{133}$Cs atom
the unpaired nucleon is a proton and so $g \equiv g_p$.

\begin{figure}
\psfig{figure=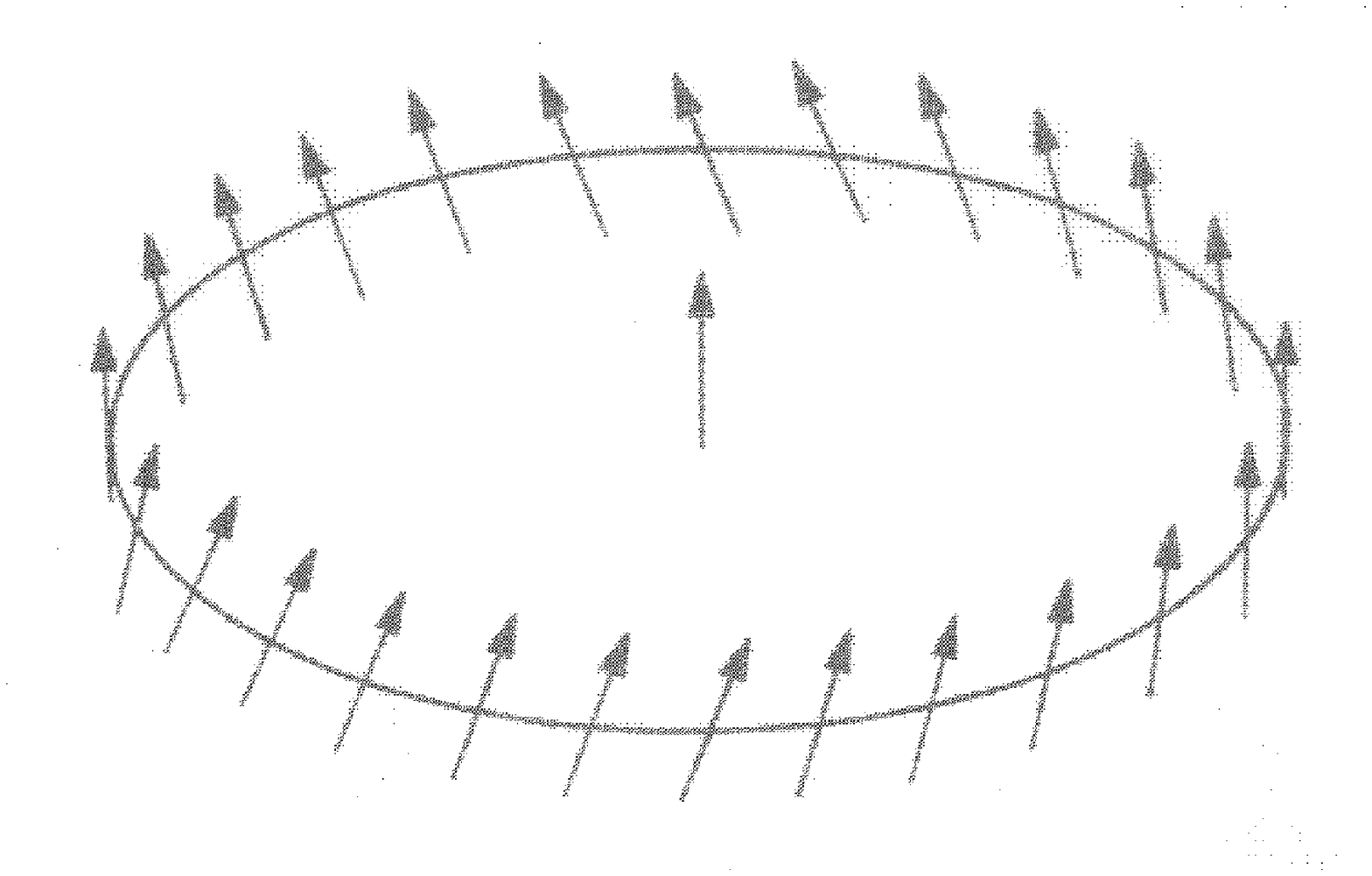,height=2.5in}
\caption{Diagram showing the spin helix that occurs due
to the parity violating nucleon-nucleus interaction.
The degree of spin rotation is proportional to the distance from
the origin and the strength of the weak interaction.}
\label{f2}
\end{figure}

The interaction (\ref{eq:poddvalence}) perturbs the 
wave function of the unpaired nucleon, resulting 
in the mixing of opposite parity states: 
$\psi = \psi_0 + \delta \psi$,
where $\psi_0 \equiv | 0 \rangle$ is the
unperturbed wave function and $\delta \psi = \sum_n | n \rangle
\langle n | \hat{W} | 0 \rangle (E_0 - E_n)^{-1}$.
An approximate analytical solution for
the perturbed Schr\"{o}dinger equation $(\hat{H}_0 + \hat{W}) \psi
= E \psi$ (which assumes that the nuclear density is constant)
gives (see, e.g., \cite{FKS1984,michel1964})
\begin{equation}
\label{eq:rot}
\psi = e^{i \theta \mbox{\boldmath $\sigma$}
\cdot {\bf r}} \psi_0,
\end{equation}
where $\theta = - g G \rho / {\sqrt{2}}$. What this means
is that the spin (${\bf s} = \frac{1}{2} \bbox{\sigma}$) 
of the unperturbed wave function will be rotated around the
vector ${\bf r}$ by an angle of $2 \theta r$. If, for
example, the unperturbed wave function was in a spin up state,
the spin at different points for the perturbed wave function
will be as shown in Fig. \ref{f2}.
Thus we have a spin
helix, with a definite chirality, i.e., right- or
left-handedness (see, e.g.,\ \cite{khriplovichpnc}).
This means that the parity symmetry has
been broken. Now consider the current
and magnetic field produced by such a spin
helix. The electromagnetic current of the
unpaired nucleon in the non-relativistic limit has the form 
\begin{equation}
\label{eq:emcurrent}
{\bf j} = -\frac{i e}{2 m} q [\psi^{\dagger} \bbox{\nabla} \psi
- (\bbox{\nabla} \psi^{\dagger}) \psi]
+ \frac{e \mu}{2 m} \bbox{\nabla} \times
(\psi^{\dagger} \bbox{\sigma} \psi),
\end{equation}
where $q = 0$ ($1$) for a neutron (proton) and 
$\mu$ is the nucleon magnetic moment in nuclear
magnetons. The first term comes from the orbital
motion of the nucleon (convection or orbital current), 
while the second term is a magnetic
moment current term, which produces the
dominating contribution.
The current distribution and the magnetic field produced
by the wave function $\psi$ of Eq. (\ref{eq:rot}) have
been calculated, e.g., in Ref. \cite{FH1993}.
%?? check
%Cross-sections of these are shown in Figs. \ref{f3} and \ref{f4}.
%Note their toroidal shapes; this means that they will
%produce an anapole moment.

Using the expression for the electromagnetic current (\ref{eq:emcurrent})
in Eq. (\ref{eq:anaj}), the operator of the anapole moment, $\hat{\bf a}$
(${\bf a} = \langle \psi | \hat{\bf a} | \psi \rangle$) can be
written as
\begin{equation}
\label{eq:anaop}
\hat{\bf a} = (\pi e / m) [ \mu ({\bf r} \times
\bbox{\sigma}) - (q/2) ({\bf p} r^2 + r^2 {\bf p}) ],
\end{equation}
where
${\bf r}$ and ${\bf p}$ are the position and momentum operators of
the nucleon.
The dominant contribution
to the nuclear anapole comes from the first, spin term and so
we can express the anapole moment operator in terms of the
magnetic dipole moment operator
$\hat{\bf M} = \bbox{\sigma} {(e \mu)}/{(2 m)}$ as
$\hat{\bf a} \approx 2 \pi ({\bf r} \times \hat{\bf M})$.

The anapole moment is usually described by a dimensionless
parameter, $\kappa_a$, defined by the following equation 
\begin{equation}
\label{eq:ana}
{\bf a} = \frac{1}{e} \frac{G}{\sqrt{2}}
\frac{K {\bf I}}{I (I+1)} \kappa_a,
\end{equation}
where $K = (I+\frac{1}{2}) (-1)^{I+1/2-l}$ and $l$ is the
orbital angular momentum of the external nucleon. 

In Ref. \cite{FKS1984} an approximate analytical result
for $\kappa_a$
(in terms of the parameter $g_p$)
was obtained by using the wave function
(\ref{eq:rot}) to calculate the mean value of the anapole
moment operator (\ref{eq:anaop}). The result is
\begin{eqnarray}
\label{eq:kappa_ag_p}
\kappa_a &=& \frac{9}{10} \frac{\alpha \mu}{m r_0} A^{2/3} g_p \\
\label{eq:kappa_ag_p2}
&=& 0.08 g_p {\rm \quad for \ }^{133}{\rm Cs} \ , 
\end{eqnarray}
where $\alpha = 1/137$, $\mu$ is the magnetic moment of the external nucleon in
nuclear magnetons, and $r_0 = 1.2 \mbox{ fm}$. 
It is the dependence on $A^{2/3}$ that makes the anapole 
moment the dominant nuclear spin-dependent effect in 
heavy atoms.

Single-particle calculations of $\kappa _{a}$ for Cs in the 
Woods-Saxon and harmonic oscillator potentials have been performed 
in Refs. \cite{FKS1984,BPzpc1991,BPplb1991,DKTnp1994} 
(note that in Refs. \cite{BPzpc1991,BPplb1991} the effects of 
configuration mixing were also included semi-empirically) 
and many-body calculations in 
Refs. \cite{HHM1989,DT1997,DT2000,HLR-M2001}.
%cite ab1999 for Tl
%cite bowman for statistical methods for calculations in compound nuclei
The results of these calculations are presented in Table \ref{tab:kappa_a(Cs)}; 
a compilation of these results in terms of DDH ``best values'' 
(see Section \ref{ssec:pvnuclearforces}) can also be found in 
Ref. \cite{DK2002}.
Single-particle calculations for the anapole moment are remarkably stable 
with the choice of potential \cite{DKTnp1994}. 
We will quote the value 
\begin{equation}
\label{eq:kappa_ag_pCs}
\kappa ^{SP}_{a}= 0.06g_{p}
\end{equation}
for $^{133}$Cs, obtained from numerical calculations 
in the Woods-Saxon potential with spin-orbit interaction \cite{FKS1984,DKTnp1994}.
We will leave the discussion of many-body effects to 
Section \ref{ssec:pvnuclearforces}.

\subsection{Parity violating effects in atoms dependent on the 
nuclear spin}

The nuclear anapole moment interacts with an atom's electrons due
to its magnetic field. The interaction is
[using Eqs. (\ref{eq:anaA}) and (\ref{eq:ana})]
\begin{equation}
\label{eq:anaint}
%?? should be consistent with definition in earlier section
\hat{h}_a = e \bbox{\alpha} \cdot {\bf A} = e \bbox{\alpha} \cdot {\bf a}
\delta ^{3}({\bf r})
= \frac{G}{\sqrt{2}}  \frac{K {\bf I} \cdot
\bbox{\alpha}}{I (I+1)} \kappa_a \delta ^{3}({\bf r}),
\end{equation}
where ${\bf A}$ is the anapole vector-potential and
$\bbox{\alpha}$ is the relativistic velocity operator (Dirac
matrices).
The magnetic field corresponding to the anapole is 
localized inside the nucleus. Therefore the interaction with 
atomic electrons occurs only if the electron wave functions 
penetrate the nucleus. 

The anapole moment produces parity violating nuclear spin-dependent 
effects in atoms. 
In heavy atoms it is the dominant mechanism producing such effects. 
However, its effects are indistinguishable 
from the smaller parity violating neutral current effects, and 
these need to be accounted for in the extraction of $\kappa _{a}$ 
from atomic measurements (we mentioned these effects briefly in 
Section \ref{sssec:NSDPNC}).   
One contribution to parity violating nuclear spin-dependent effects 
arises from $Z^{0}$ exchange in an electron-nucleus interaction, 
with an axial-vector $Z^{0}$-nucleus coupling and a vector 
$Z^{0}$-electron coupling
% - this contribution is suppressed in the standard model
[Eq. (\ref{eq:NSDVeAn})]; see Ref. \cite{NSFK1977}. 
Another contribution arises from the perturbation of the nuclear 
spin-independent contribution, corresponding to the nuclear weak charge, 
by the hyperfine interaction 
%- this contribution is of the order of $\alpha G_{F}A^{2/3}$
(Eq. (\ref{eq:NSDQW})); see \cite{FK85,BPplb1991,JSS2003}. 
%For cesium, these contributions are roughly of the same size.
The total effective nuclear spin-dependent interaction can be 
expressed in the form\footnote{Note that different definitions 
for $\kappa$, $\kappa _{a}$, $\kappa _{2}$, $\kappa _{Q}$ are 
used in different works. For example, in \cite{HLR-M2001,HW2001,JSS2003} 
the effective Hamiltonian $\hat{h}^{I}_{eff}=\frac{G}{\sqrt{2}}\kappa 
\mbox{\boldmath$\alpha$}\cdot {\bf I}\rho (r)$, 
$\kappa = \kappa _{a}+\kappa _{2}+\kappa _{Q}$. 
Also different notations for the constants are used in these works.}
\begin{equation}
\hat{h}_{eff}^{I}=\frac{G}{\sqrt{2}}\kappa\frac{K}{I(I+1)}
\mbox{\boldmath$\alpha$}\cdot {\bf I}\rho (r) \ ,
\end{equation}
where 
\begin{equation}
\label{eq:kappa}
\kappa=\kappa_{a}-\frac{K -1/2}{K}\kappa_{2} +\frac{I+1}{K}\kappa_{Q} \ ,
\end{equation}
and $K=(I+1/2)(-1)^{I+1/2-l}$.

\subsection{Measurement of nuclear spin-dependent effects in cesium 
and extraction of the nuclear anapole moment}   
%measurements of the nuclear anapole moment and parity violating 
%nuclear forces in atomic experiments. Comparison with theories of parity
% violating nuclear forces}

The anapole moment can be detected by observing the amplitudes of transitions
between atomic levels that violate parity. In the case of
the Boulder experiment \cite{wieman}, it was an E1 transition 
between the $6S$ and $7S$ states of the cesium atom. 
The nuclear spin-dependent $P$-odd effects can be separated from 
the dominant effect (the weak interaction between the electron 
and the weak charge of the nucleus; see Sections \ref{section3iia},\ref{section3iib}) 
by observing the dependence of the parity violating effects in two 
different hyperfine transitions. 
The nuclear spin has different relative orientations in the
different hyperfine states. For $^{133}$Cs, $F=3$ and $F=4$, where 
$F$ is the total angular momentum of the atom (${\bf F} =
{\bf J} + {\bf I}$, where ${\bf J}$ is the
electron's angular momentum).
By observing the transitions $6S_{F=4} \rightarrow 7S_{F=3}$ and
$6S_{F=3} \rightarrow 7S_{F=4}$, the anapole moment can be detected.
In the Boulder experiment \cite{wieman} these two
amplitudes were measured, and they were found to be
significantly different, indicating the presence of an anapole moment.

\subsubsection{Atomic calculations and extraction of $\kappa$}

Atomic many-body calculations are required to extract the 
effective constant $\kappa$ from experiment. 
Atomic calculations of nuclear spin-dependent effects  
in cesium have been performed in \cite{NSFK1977,frant88,kraft88,BPzpc1991,BJS1990,JSS2003}.  
In \cite{FM1997} the value for $\kappa$ was extracted from \cite{wieman} by taking 
the ratio of the nuclear spin dependent PNC amplitude to the main spin independent 
PNC amplitude, using the spin dependent calculation \cite{kraft88} and the 
spin independent calculation \cite{DFSS1987b}. The calculations \cite{kraft88,DFSS1987b} 
were performed in the relativistic Hartree-Fock approximation, with the effects of 
core polarization included using the time-dependent Hartree-Fock method 
and correlations included through the use of Brueckner orbitals 
(see Section \ref{section3}). 
These calculations were performed using the same method and computer codes, so the 
theoretical errors should cancel in the ratio. The value obtained in \cite{FM1997} is 
\begin{equation}
\kappa (^{133}{\rm Cs})= 0.442 (63) \ .
\end{equation}
In the recent work \cite{JSS2003} the value $\kappa (^{133}{\rm Cs})=0.462(63)$ 
was obtained. The calculations were performed in zeroth-order in the relativistic 
Hartree-Fock approximation, and core polarization was included using the 
random phase approximation. This result was extracted directly from the spin-dependent 
component measured in \cite{wieman}, with $\beta$ from \cite{BW}. The $4\%$ 
difference between the two values for $\kappa (^{133}{\rm Cs})$ is explained 
by correlation corrections: 
correlation corrections (Brueckner orbitals) are included in \cite{kraft88}, 
while they are not considered in \cite{JSS2003}. At the TDHF level, the results 
coincide.   

\subsubsection{Extraction of $\kappa _{a}$}

The value of $\kappa$ contains three contributions 
(see Eq. (\ref{eq:kappa})). 
The contributions to $\kappa$ from the nuclear spin-dependent neutral current 
$\kappa _{2}$ and from the combined neutral weak charge and hyperfine 
interaction $\kappa _{Q}$ must be subtracted to 
give the anapole constant $\kappa _{a}$. 
In the single-particle approximation, $\kappa_{2}=-C_{2p}\approx -0.05$ 
[see Section \ref{section2}, Eq. (\ref{eq:NSDVeAn})]. 
Large-basis nuclear shell-model calculations for $^{133}$Cs give 
$\kappa _{2}= -0.063$ \cite{HLR-M2001}. 
Atomic calculations for the combined weak charge and hyperfine interaction 
have been performed in \cite{FK85,BPplb1991,JSS2003}. 
The recent calculation \cite{JSS2003} is the most complete: it was performed 
in third-order perturbation theory, with core polarization taken 
into account in the random phase approximation. In the other calculations 
\cite{FK85,BPplb1991} the single-particle form for the operator 
[Eq. (\ref{eq:NSDQW})] was used. The result of Johnson {\it et al.} \cite{JSS2003} 
for $^{133}$Cs is $\kappa_{Q} = 0.017$. 
%$K=4$, $I=7/2$ for $^{133}$Cs 
Results of calculations for $\kappa _{2}$ and $\kappa _{Q}$ are compiled in 
Table \ref{tab:kappa_2,Q}. 
With $\kappa _{2}$ from \cite{HLR-M2001} and $\kappa _{Q}$ from \cite{JSS2003},
the anapole moment constant is 
\begin{equation}
\label{eq:ka.368}
\kappa _{a}=0.368(63) \ .
\end{equation}

The interaction (\ref{eq:anaint}) is for a point-like nucleus. 
However, a real nucleus has a finite size. 
In the works \cite{frant88,kraft88,BPzpc1991,BJS1990,JSS2003} the atomic 
calculations were performed by replacing $\delta ^{3}(r)$ with the 
nuclear density $\rho (r)$. 
A more accurate treatment of finite nuclear size effects was given  
in \cite{FH1993,FM1997}.
For Cs the correction is small. The corrected value for $\kappa _{a}$ 
[Eq. (\ref{eq:ka.368})] is
\begin{equation}
\label{eq:kappa_a}
\kappa_a = 0.362 (62) \ .
\end{equation}

\subsection{The nuclear anapole moment and parity violating nuclear forces}
\label{ssec:pvnuclearforces}

The weak potential is usually parametrized in terms of a one-boson 
exchange model. 
The proton-nucleus and neutron-nucleus constants, $g_p$ and $g_{n}$, can be 
expressed in terms of the following combination of meson-nucleon parity 
nonconserving interaction constants \cite{FKS1984,FPhysScr}
(using the notation of Ref. \cite{DDH1980}):
\begin{eqnarray}
\label{eq:g_p}
g_p &=& 2.0 \times 10^5 W_{\rho} \left[176 \frac{W_{\pi}}{W_{\rho}} f_{\pi}
-19.5 h_{\rho}^0 - 4.7 h_{\rho}^1 + 1.3 h_{\rho}^2
-11.3 (h_{\omega}^0 + h_{\omega}^1) \right] \\
\label{eq:g_n}
g_n &=& 2.0 \times 10^5 W_{\rho} \left[-118 \frac{W_{\pi}}{W_{\rho}} f_{\pi}
-18.9 h_{\rho}^0 + 8.4 h_{\rho}^1 - 1.3 h_{\rho}^2
-12.8 (h_{\omega}^0 - h_{\omega}^1) \right] \ ,
\end{eqnarray}
$f_{\pi} \equiv h_{\pi}^{1}$ and the $h$'s are
weak meson-nucleon coupling constants --- the subscript denotes
the type of meson involved and the superscript indicates
whether it is an isoscalar, isovector or isotensor interaction
($0$, $1$, or $2$).
These are the effective constants obtained in the contact limit of the nucleon-nucleon 
PNC interaction [Eq. (\ref{eq:2bodyPNCnn})], 
with short-range nucleon-nucleon repulsion and long-range effects 
taken into account through the parameters $W_{\rho}$ and $W_{\pi}$. 
Following \cite{FKS1984,FM1997}, we take $W_{\rho}=0.4$ and $W_{\pi}=0.16$ 
using calculations of PNC for neutron and proton scattering on $^{4}{\rm He}$ 
\cite{DFST1983}.  

The standard reference values for weak meson-nucleon couplings are those of 
Desplanques, Donoghue, and Holstein \cite{DDH1980} 
[we list the DDH ``best values'' for the weak couplings and the effective 
coupling constants (Eqs. (\ref{eq:g_p},\ref{eq:g_n},\ref{eq:gZN}); Refs. \cite{FKS1984,ST1993,DKTnp1994}) 
in terms of these values in Table \ref{tab:DDH} for easy reference]. 
% cite also more recent calculations?
However, there are large uncertainties in the possible values 
(known as the ``DDH reasonable ranges''). 
It is therefore important to determine the weak coupling constants 
experimentally. 
Information on these constants can be obtained from the measurement of 
the Cs anapole moment. 
%Using DDH ``best values'', the 
%following value for $g_{p}$ is obtained,
%\begin{equation}
%g_{p}=4.5 \ .
%\end{equation}  

\subsubsection{The cesium result and comparison with other experiments}

The measurement of the anapole moment $\kappa_a$ (\ref{eq:kappa_a}) and 
the single-particle calculation (\ref{eq:kappa_ag_pCs}) give the following 
value for the weak interaction constant between an unpaired proton and the 
nuclear core \cite{FM1997},  
\begin{equation}
\label{eq:gpSP}
g^{SP}_p = 6 \pm 1 \mbox{(exp.)} \ .
\end{equation}
(Note that only the experimental error is included here; 
there is also a theoretical error from the nuclear
calculation of $\kappa_a$ (\ref{eq:kappa_ag_pCs}).)
%The constants $h_{\rho}$ and $h_{\omega}$ are known fairly well \cite{brown}
%what about recent pp scattering experiment?
There is a lot of uncertainty about the value for $f_{\pi}$. 
Following \cite{FM1997}, we use DDH best values for $h_{\rho}$ and $h_{\omega}$ 
to obtain a value for $f_{\pi}$ from the measurement of the cesium anapole moment. 
%There is uncertainty about the value for $f_{\pi}$. 
%Information on this constant can be obtained from the Cs anapole measurement. 
%Using Eqs. (\ref{eq:gpSP}), (\ref{eq:g_p}), 
%and the ``best values'' for the $h_{\rho}$'s and $h_{\omega}$'s from \cite{DDH1980}, 
%a value for $f_{\pi}$ can be obtained from the Cs anapole measurement, 
Using Eqs. (\ref{eq:gpSP}), (\ref{eq:g_p}) it is found that 
\begin{equation}
\label{eq:fpi}
f^{SP}_{\pi} = (g^{SP}_p - 2) \times 1.8 \times 10^{-7}
= [7 \pm 2 \mbox{ (exp.)}] \times 10^{-7} \ .
%error so large?
\end{equation}
(The contribution of $\rho$ and $\omega$ to $g_{p}$ is 2.)
This result (\ref{eq:fpi}) agrees with QCD calculations 
\cite{khatsimovskii1985,kaplan1993} which give 
$f_{\pi} = 5\mbox{--} 6 \times 10^{-7}$ 
and is in agreement with the DDH best value 
$f_{\pi} = 4.6 \times 10^{-7}$ 
(note that the DDH ``reasonable range'' for $f_{\pi}$ is 
$0.0-11.4 \times 10^{-7}$).

However, there is a serious discrepancy between the weak meson-nucleon 
couplings extracted from the Cs anapole moment and those extracted from 
other experiments. 
The following experiments are thought to give reliable 
information on the weak meson-nucleon couplings 
(that is, their interpretation is not hindered too much by 
nuclear structure uncertainties) \cite{AHHH,HLR-M2001}:
the longitudinal analyzing power for $\vec{p}p$ scattering 
at 13.6~MeV \cite{pp1}, 45~MeV \cite{pp2}, and 221~MeV \cite{pp3} 
%\cite{pp1,pp2,pp3} 
and $\vec{p}\alpha$ scattering at 46~MeV \cite{palpha}, the 
circular polarization of $\gamma$-rays emitted from the 
1081~keV state in $^{18}$F \cite{18F}, 
and the asymmetry of $\gamma$-rays emitted in the 
decay of the state 110~keV in polarized $^{19}$F \cite{19F}. 
These experiments depend on different combinations of 
the coupling constants. They are consistent and favor a 
small value of $f_{\pi}$, roughly lying between $-1\times 10^{-7}$ 
and $1\times 10^{-7}$ (see \cite{HLR-M2001,HW2001}), in contradiction 
with the Cs anapole result.

The value for $f_{\pi}$ extracted above for Cs was obtained 
in the single particle model. Let us briefly discuss how this value 
changes when many-body effects are included. 
Core polarization effects have been calculated in 
Refs. \cite{DT1997,DT2000} using the random phase approximation. 
In Ref. \cite{DT2000} it was found that for $^{133}$Cs core polarization 
and pairing effects decrease the single-particle value for $\kappa _{a}$ 
%[Eq. (\ref{eq:kappa_ag_pCs})] 
by almost a factor of two.
% $\sim 2$.  
These calculations therefore increase the discrepancy of the cesium result 
with other experiments and with the DDH best value. 
Large-basis shell model calculations have been performed in 
Refs. \cite{HHM1989,HLR-M2001}. These calculations also 
reduce the size of $\kappa _{a}$, although not to the extent of the 
calculation \cite{DT2000}. 

In Table \ref{tab:kappa_a(Cs)} results of single-particle and many-body calculations 
of $\kappa _{a}$ for $^{133}$Cs are presented. 
It appears that consideration of many-body effects tends 
to exacerbate the disagreement between the meson-nucleon couplings extracted 
from the Cs anapole moment and those obtained from other experiments. 

Note that there may be other significant corrections to the anapole moment calculations 
that may resolve the discrepancy between theory and experiment. 
For instance, consideration of the strong renormalization of the weak potential 
\cite{BPzpc1991,fv1994}. 
Another correction to the calculations could stem from the use of an improved set 
of strong meson-nucleon coupling constants. 
(In the one-boson exchange model the weak potential is formed from a product of 
weak and strong meson-nucleon couplings. 
The strong couplings $g_{\pi}=13.45$, $g_{\rho}=2.79$, $g_{\omega}=8.37$ and the 
isoscalar $\chi _{S}=-0.12$ and isovector $\chi _{V}=3.7$ anomalous magnetic moments of the nucleon 
have been employed in the calculations; see, e.g., \cite{DDH1980,DFST1983,HLR-M2001,miller}. 
These values were used to arrive at Eqs. (\ref{eq:g_p},\ref{eq:g_n})).
In the recent work \cite{miller} it was shown that using strong coupling constants 
in the one-boson exchange model constrained by nucleon-nucleon phase shifts 
leads to an enhancement of parity violating effects by about a factor of two. 

The cesium anapole moment is not the only case in which parity violating effects 
appear to be enhanced. Parity violation in the Mossbauer transitions in 
$^{57}$Fe and $^{199}$Sn nuclei give a value for the circular polarization of 
$\gamma$-rays at least four orders of magnitude larger than calculations 
(see Ref. \cite{ST1993}). 
Statistical methods were applied in Refs. \cite{FV1993,TJHB2000} 
to study parity violating effects in polarized neutron scattering from 
compound nuclei. It was found in Ref. \cite{TJHB2000} that the experiments 
give parity violating effects $1.7-3$ times larger than those estimated 
using the DDH values for weak meson-nucleon couplings.    

The Cs anapole measurement is also inconsistent with the limit 
on the anapole moment of Tl \cite{vetter95},
\begin{equation}
\label{eq:tlseattle}
\kappa ({\rm Tl})= -0.22\pm 0.30 \ .
\end{equation}
In fact, the Tl anapole moment itself produces problems. 
Its central point is of opposite sign to that predicted by 
theory.\footnote{
Note that while the Oxford result \cite{edwards95},
$\kappa ({\rm Tl})=0.23\pm 1.20$ [using the single-particle atomic 
calculation \cite{khrippla95} used to obtain (\ref{eq:tlseattle})],
has a central point that is 
positive, its error is four times larger than the Seattle value \cite{vetter95}. 
The Oxford result has been misquoted in the literature.}
Results of calculations of $\kappa _{a}$ for Tl are presented in Table \ref{tab:kappa_a(Tl)} 
(these results in terms of DDH ``best values'' can also be found in \cite{DK2002}). 
Using DDH best values for the weak couplings, 
the most complete many-body calculations of the anapole moment of Tl give 
$\kappa _{a}({\rm Tl})=0.24$ within RPA \cite{DT2000} and 
$\kappa _{a}({\rm Tl})=0.24$ \cite{AB1999} and $\kappa _{a}({\rm Tl})=0.10$ \cite{HLR-M2001} 
in shell model calculations.  

Note that the anapole moment of Tl extracted above (\ref{eq:tlseattle}) 
was obtained from the nuclear spin-dependent component of the PNC optical rotation 
using a single-particle atomic calculation \cite{khrippla95}. 
In a recent many-body calculation \cite{kozlovkappa_a}, 
in which the MBPT+CI method was implemented (see Section \ref{section3iic} for a brief 
description and references), the central point for the Tl anapole moment 
is interpreted as $\kappa _{a}=-0.26$; that is, the disagreement with theory is more obvious.  

\subsection{Ongoing/future studies of nuclear anapole moments}

It is clear that further experimental investigation is required to 
resolve these inconsistencies. An improved measurement of the Tl 
anapole moment is important, as is an anapole measurement involving 
a nucleus with an unpaired neutron (Cs and Tl have unpaired protons). 
This latter experiment would give information on $g_{n}$ which depends 
on a different combination of weak meson-nucleon couplings 
that are roughly perpendicular to the current anapole measurements 
(compare Eqs. (\ref{eq:g_p},\ref{eq:g_n})) 
and so would provide a very important cross-check. 

We have already discussed in Sections \ref{ss:further1pnc},\ref{ss:further2+pnc} the 
PNC experiments underway. Nuclear spin-dependent effects in those atoms with 
non-zero nuclear spin will be measured. 

Enhancement of anapole moment effects in atoms can occur due to close levels of 
opposite parity. For example, calculations \cite{flambaum1999,DFG2000} show that 
the PNC amplitudes in the transitions 
$7s^{2}~{^{1}S}_{0} - 7s6d~{^{3}D}_{1}$ and $7s^{2}~{^{1}S}_{0} - 7s6d~{^{3}D}_{2}$ 
in $^{223}$Ra and $^{225}$Ra can be more than 1000 times larger than 
the nuclear spin-dependent transition amplitudes in Cs. Moreover, in the transition 
$7s^{2}~{^{1}S}_{0} - 7s6d~{^{3}D}_{2}$ there is no background nuclear spin-independent amplitude, 
since the large change in angular momentum $\Delta J=2$ forbids it. 
Note that measurements with $^{223,225}$Ra, sensitive to the neutron constant $g_{n}$, 
would provide a sought-after cross-check on the parity violating meson-nucleon couplings. 

Work towards measurements of anapole moments in diatomic molecules is underway 
\cite{demille}. Diatomic molecules are attractive for anapole moment searches because the nuclear 
spin-dependent effects, but not the (usually dominating) nuclear spin-independent ones, 
are enhanced compared to atoms due to the presence of close rotational levels 
of opposite parity \cite{labzovsky1978,SF1978mol}; see also the review \cite{KL1995}.

%Anapole moment measurements are being considered in francium at 
%Stony Brook \cite{aubin}, in Ba$^{+}$ at Seattle \cite{}. 

%Measurements with Dy, Fr, and Ba$^{+}$ have been proposed. 
%Experiments in progress include ........

%This seems to be in contradiction with the limit on
%$f_{\pi}$ derived from a $^{18}$F PNC measurement:
%$|f_{\pi}| < 1.3 \times 10^{-7}$
%(see, e.g., the review \cite{Desplanques}),
%but note that $^{18}$F is a nonspherical nucleus with
%a complex shape and this makes the interpretation of
%the experiment complicated.

%It was shown in Ref. \cite{fv1994} that RPA corrections to the weak 
%potential increase $g_p$ by 30\%, so the derived value of $f_{\pi}$ may 
%actually be smaller. However, other calculations
%\cite{HHM1989} 
%suggest that additional
%many body corrections may compensate for this increase.

%cite also Wilburn and Bowman work on ``consistency of PV pion-nucleon couplings...''

\section{Electric dipole moments: manifestation of time reversal
violation in atoms}
\label{section7}

Violation of $CP$ symmetry (combined symmetry of charge conjugation, $C$, 
and parity) was discovered in 1964 in the decays of the neutral 
K mesons \cite{ccft}. It is incorporated into the standard electroweak model 
as a single complex phase in the quark mixing matrix 
(Kobayashi-Maskawa mechanism \cite{kobmask}).
For a long time K mesons remained the only system in which $CP$ violation 
had been observed, until just recently, in 2001, when the collaborations 
BaBar and Belle detected it in the neutral B mesons \cite{B0}.
The $CP$ violation seen there is consistent with the standard model 
predictions. 

%Standard model $CP$ violation, however, seems somewhat contrived. The standard 
%model does not explain how $CP$ violation appears. 
%
%A more 
However, a striking problem arises from cosmology. 
Sakharov proposed that $CP$ violation, present at the time of the 
Big Bang, is a necessary ingredient in the asymmetry of matter and antimatter 
\cite{sakharov}. 
It is well-known that standard model $CP$ violation is insufficient 
to generate the level of matter-antimatter asymmetry in the Universe. 
Understanding the origin of $CP$ violation and searching for possible new sources 
is therefore a very interesting and fundamental problem.

%
%Sakharov suggested that CP violation may be a necessary ingredient in the
%asymmetry of matter and antimatter
%

If $CPT$ is a good symmetry, as it is in gauge theories, then $CP$ violation is 
accompanied by $T$ (time-reversal) violation. So far there has been no (undisputed) 
direct observation of $T$ violation and its detection is of interest in its own 
right. Also, detection of T-violation may shed light on the origin of $CP$ violation. 
The measurement of a permanent electric dipole moment (EDM) of, e.g., a neutron, atom, 
or molecule would be direct evidence of $T$-violation. 
This can be seen very simply: an EDM, ${\bf d}$, of a non-degenerate quantum system 
is directed along the total angular momentum ${\bf F}$ (the only vector specifying the 
system), and such a correlation can only occur if both $T$ and $P$ are violated 
(${\bf d}$ and ${\bf F}$ behave differently on reversal of ${\bf r}$ and $t$). 

$T$ violating electric dipole moments have an extraordinary sensitivity to new models 
of $CP$ violation. The reason is that the standard model $CP$ violation, appearing through 
just a single phase, is highly suppressed in such flavour-conserving phenomena. 
In particular, the standard model prediction for the neutron EDM is 
$d_{n}({\rm SM})\sim 10^{-34}~e~{\rm cm}$ \cite{PK1991}, with the proton EDM the same order of 
magnitude, and the electron EDM is even smaller 
$d_{e}({\rm SM})\lesssim 10^{-38}~e~{\rm cm}$ \cite{CK1997}. 
The EDMs appear only in high-order loops (3-loop diagrams for the neutron and 
3- to 4-loop
%check 
diagrams for the electron). The new sources of $CP$ violation appearing in new theories 
generate EDMs that are many orders of magnitude larger than the standard model values and 
the size of these EDMs are in reach of current experiments. Already the parameter space 
of popular theories such as supersymmetry, multi-Higgs models, and left-right symmetric 
models are strongly restricted by current measurements. 

A $CP$-violating phase also appears in quantum chromodynamics (QCD) and it can be constrained 
(or detected) from EDM measurements. 
It defines the strength of the $P$- and $T$-violating term in the QCD Lagrangian containing 
$G_{\mu \nu}\tilde{G}_{\mu \nu}$ (this is analogous to $F_{\mu \nu}\tilde{F}_{\mu \nu}$ 
in electrodynamics which reduces to ${\bf E}\cdot {\bf B}$, clearly a $P,T$-violating 
correlation; see, e.g, Ref. \cite{khriplovichpnc}).
The ``unnatural'' smallness of this phase ($\theta _{QCD}\lesssim 10^{-10}$, 
compare this to the Kobayashi-Maskawa angle $\delta _{KM}\sim 1$) is the famous 
``strong $CP$ problem''. 
One solution to this problem involves the introduction of a new Goldstone boson 
known as the ``axion'' \cite{axion}, and this particle has become a popular candidate 
for dark matter. 
However, after many experimental searches, there is no evidence of its existence. 
(For a review of experimental searches for axions, see, e.g.,  \cite{expaxions}.)

%Note that we are talking about EDMs that violate $P$ and $T$. 
%That is, they show up as a linear Stark shift in non-degenerate systems 
%when a small (compared to level spacings) electric field is applied.
%This is different to ordinary EDMs in molecules (and hydrogen) 
%where the levels are essentially degenerate and so the Stark 
%shift is linear. If the applied electric field is small enough (so that 
%it shifts the levels by an amount smaller than the level spacing) then 
%the shift is in fact quadratic.
%
%Because an experimental search for an EDM involves 
%the application of an electric field, a neutral 
%particle, atom, or molecule must be used, otherwise it 
%will be accelerated out of the experimental apparatus. 
%%(beam experiments may be an exception)
%%????

\subsection{Atomic EDMs}
\label{ss:atomicedms}

In this section we will consider the contributions to an atomic EDM 
arising from various $P,T$-odd mechanisms. 
Contributions to an atomic EDM arise from (i) the sum of intrinsic 
EDMs of the atomic constituents (averaged over the atomic state) and 
(ii) the mixing of opposite parity wave functions due to a 
$P,T$-odd interaction $\hat{H}_{PT}$.

The atomic EDM induced in an atomic state $K$ due to admixture with the 
opposite-parity wave functions $M$ has the form
\begin{equation}
\label{eq:atomicedm}
{\bf d} _{\rm atom}=
2\sum_{M}\frac{\langle K|\hat{\bf D}|M\rangle \langle M|\hat{H}_{PT}|K\rangle}
{E_{K}-E_{M}}=d_{\rm atom}({\bf F}/F)\ ,
\end{equation}
where $\hat{\bf D}=-e\sum_{i}{\bf r}_{i}$ is the electric dipole operator, 
$\hat{H}_{PT}$ is the $P,T$-odd operator that mixes $K$ with the set 
of wave functions $M$, and $F$ is the total angular momentum of the atom 
corresponding to the state $K$. 

We will use the following notations for the electron wave functions:
\begin{equation}
\label{psidef}
\psi ({\bf R})= 
\left( 
\begin{array}{c}
f(R)\Omega _{jlm} \\
-i(\mbox{\boldmath$\sigma$}\cdot {\bf n})g(R)\Omega _{jlm}
\end{array}
\right) \ ,
\end{equation}
where $\Omega _{jlm}$ is a spherical spinor, ${\bf n}={\bf R}/R$, 
and $f(R)$ and $g(R)$ are radial functions 
(see, e.g., \cite{khriplovichpnc}).

\subsubsection{Electronic enhancement mechanisms}

What are the mechanisms that lead to enhancement of 
parity and time invariance violating effects in atoms? 
Like the $P$-odd, $T$-even effects we discussed in the first 
part of this review, the effects of $P,T$-violation increase 
in atoms with:
\newline
\indent
(i) a high nuclear charge $Z$. $P,T$-odd effects in atoms increase rapidly with $Z$, 
\cite{sandars1965,flambaum76}.
Therefore, it is best to search for atomic EDMs in heavy atoms.
\newline
\indent 
(ii) close levels of opposite parity.
From Eq. (\ref{eq:atomicedm}) we can see that if $H_{PT}$ mixes 
opposite parity levels with energies $E_{1}\approx E_{2}$, 
then the atomic EDM induced will be enhanced, 
$d_{\rm atom}\propto 1/(E_{1}-E_{2})$. 
This enhancement occurs, for example, in rare-earth atoms, where there 
are anomalously close levels of opposite parity.

\subsection{Enhancement of $T$-odd effects in polar diatomic molecules}

Sandars was the first to point out the enhancement of $T$-odd effects 
in polar diatomic molecules compared to atoms in his proposal to measure the 
proton EDM in TlF \cite{sandars1967}. 
The idea is to polarize the molecules along an external electric field, 
thereby aligning the enormous intramolecular field, leading to an effective 
enhancement of the external field by several orders of magnitude. 
This enhancement mechanism is related to that arising from the 
presence of close rotational levels of opposite parity \cite{SF1978mol}.
The two cases correspond to the respective cases of strong and weak interactions with the 
electric field compared to the rotational level spacing. See \cite{KL1995} for a review.

\subsection{Limits on neutron, atomic, and molecular EDMs}

A typical EDM experiment is performed in parallel electric and magnetic fields.  
The corresponding Hamiltonian is
\begin{equation}
\hat{h}=-\mbox{\boldmath$\mu$}\cdot {\bf B}-{\bf d}\cdot {\bf E} \ .
\end{equation}
A linear Stark shift is measured by observing the change in 
frequency when the electric field is reversed 
(this is a measure of the $P,T$-odd correlation ${\bf E}\cdot {\bf B}$). 
%Because an EDM experiment involves the application of an electric field, the 
%natural choice is a neutral system. 
%
%In measurements of $T$-violating EDMs an electric field is applied and a 
%signal arising from a linear Stark shift is sought.\footnote{
%A linear Stark shift in a non-degenerate system, in the case when the electric 
%field is small compared to the level spacing, would be a signal of $T$-violation. 
%The ``ordinary'' ($T$-even) linear Stark shifts seen in hydrogen and molecules arise because 
%the level-spacing is very small, smaller than the shift due to the external field. 
%However, if the field is smaller than the level-spacing the shift is in fact quadratic.} 
%What is measured is a dependence on the $P,T$-odd correlation 
%${\bf E}\cdot {\bf B}$. 
%
%The application of an electric field makes neutral systems the natural choice for 
%EDM measurements.

To date, permanent EDMs in neutrons, atoms, and molecules have escaped detection. 
Null measurements of EDMs have been obtained for 
the paramagnetic atoms 
$^{85}{\rm Rb}$ \cite{ensberg,huang-hellinger},
%cite also Huang-Hellinger PhD thesis result for Rb (1987)????? 
${^{133}{\rm Cs}}$ \cite{sandars1964,weisskopf1968,murthy1989},
%are there more Cs measurements????
${^{205}{\rm Tl}}$ 
\cite{carrico1970,gould1970,abdullah1990,commins1994,tledm02}, 
and $^{129}{\rm Xe}$ in the metastable state $5p^{5}6s~^{3}P_{2}$ \cite{player1970}
and, very recently, the molecule YbF in the ground state \cite{hudson2002}.  
An experiment aimed to measure the EDM of the ion ${\rm Fe}^{3+}$ 
in the solid state was carried out a long time ago \cite{vasilev1978}. 
%though without much success. 
Interest to carry out measurements in solids, in particular in gadolinium gallium garnet 
and gadolinium iron garnet, has been sparked recently 
\cite{lam02,hunter01} (see Section \ref{section7iv}). 
%Here it is expected that the sensitivity to $T$-violating effects will be improved by 
%several orders of magnitude! The solid-state nature of the problem complicates the 
%calculations required for interpretation of the measurements. 
%A series of calculations have been performed in \cite{sushkovsolids}.
In diamagnetic systems, EDM experiments have been performed for 
atomic ${^{129}{\rm Xe}}$ \cite{vold1984,rosenberry2001}
and ${^{199}{\rm Hg}}$ \cite{lamoreaux1987,jacobs9395,hgedm} 
in the ground states and for the polar molecule TlF 
\cite{harrison1969,hinds1980b,wilkening1984,schropp1987,tlfedm}.

Measurements of EDMs in paramagnetic systems (that is, with non-zero total electron angular momentum) 
are most sensitive to leptonic sources of $P,T$-violation, 
in particular the electron EDM, while measurements of EDMs in diamagnetic systems 
(zero total electron angular momentum) 
are most sensitive to $P,T$-odd mechanisms in the hadronic sector. 
Both are sensitive to $P,T$-odd semi-leptonic processes (the electron-nucleon interaction), 
the former to those involving the electron spin, and the latter to those involving the 
nuclear spin.

The most precise measurement of an atomic EDM in a paramagnetic system has been obtained for 
$^{205}{\rm Tl}$ \cite{tledm02},
\begin{equation}
\label{eq:tledm}
d(^{205}{\rm Tl})=-(4.0\pm 4.3)\times 10^{-25}~e~{\rm cm} \ .
\end{equation}
With the use of atomic calculations, described in Section \ref{sss:electron}, 
this measurement can be expressed in terms of a limit on the EDM of the electron, 
for which it currently sets the tightest constraint. 

The most precise measurement of an atomic EDM has been carried out with 
$^{199}{\rm Hg}$ \cite{hgedm}, 
\begin{equation}
\label{eq:hgedm}
d(^{199}{\rm Hg})=-(1.06\pm 0.49\pm 0.40)\times 10^{-28}~e~{\rm cm} \ .
\end{equation}
This measurement sets the best limits on a number of $P,T$-violating mechanisms in the 
hadronic sector. The nucleus of $^{199}$Hg has an unpaired neutron. 
A limit on the neutron EDM extracted from (\ref{eq:hgedm}) is competitive with 
the best direct neutron EDM measurements\footnote{In line with the comments of 
Ref. \cite{LG2000}, we present the most recent 1999 result of the ILL group, rather 
than the final value cited in Ref. \cite{ILL} which is the average of their 1999 
and 1990 results.} performed at the Institut Laue-Langevin (ILL) 
and the Petersburg Nuclear Physics Institute (PNPI) 
\begin{eqnarray}
\label{eq:neutronlimits}
d_{n}=
\left\{
\begin{array}{lll}
(1.9\pm 5.4)\times 10^{-26}~e~{\rm cm} & \ \qquad {\rm ILL} & \ \quad \cite{ILL}\\
(2.6\pm 4.0\pm 1.6)\times 10^{-26}~e~{\rm cm} & \ \qquad {\rm PNPI} & \ \quad \cite{PNPI}
\end{array}
\right.
.
\end{eqnarray}
As we will see in Section \ref{section7iv}, at the most fundamental scale 
the limits on $P,T$-violating parameters from the measurement (\ref{eq:hgedm}) are 
more strict than those from the neutron measurements (\ref{eq:neutronlimits}) 
due to the presence of a nucleon-nucleon $P,T$-violating interaction that induces 
$P,T$-violating nuclear moments more efficiently than does an intrinsic EDM of a nucleon. 

It is interesting that although the $^{199}$Hg nucleus has an unpaired neutron, 
the measurement (\ref{eq:hgedm}) sets a constraint on the proton EDM (it contributes due to 
configuration mixing) that is tighter than that from the measurement of the 
EDM of the TlF molecule (Tl has an unpaired proton) used 
in the past to set the proton EDM upper limit. 
In TlF the $P,T$-odd interaction is given by 
$\hat{H}=-hd\mbox{\boldmath$\sigma$}\cdot\hat{\mbox{\boldmath$\lambda$}}$, 
where $\mbox{\boldmath$\sigma$}$ is the spin operator of the Tl nucleus, 
$\hat{\mbox{\boldmath$\lambda$}}$ is the unit vector along the internuclear axis, and 
$h$ is Planck's constant. 
The most precise limit on the $P,T$-odd coupling constant $d$ in TlF is \cite{tlfedm}
\begin{equation}
d=-(0.13\pm 0.22)~{\rm mHz} \ .
\end{equation}
The coupling constant $d$ can be expressed, e.g., in terms of permanent EDMs of 
the proton and electron and $P,T$-violating interactions; such calculations for 
TlF have been performed in 
\cite{sandars1967,HLS1976,hinds1980a,covsand,SFK84,FK85,parpia1997,quiney1998,petrov02}; 
%check calcs - any others?
see, e.g, Ref. \cite{tlfedm} for a general overview. 
The limit on the proton EDM following from the calculations of Ref. \cite{petrov02} 
is presented in Table \ref{tab:limitsI}.

A more detailed comparison of limits on $P,T$-violating parameters will be postponed 
until Section \ref{section7iv}. 
In the following sections we discuss the different mechanisms that induce atomic EDMs.

\subsection{Mechanisms that induce atomic EDMs}
\label{ssec:atomicEDMmech}

An external electric field acting on a neutral atom consisting 
of non-relativistic point-like charged particles with EDMs, 
interacting via electrostatic forces, is screened exactly at each particle 
\cite{PR1950,schiff}; see also \cite{khriplovichpnc}.\footnote{
Actually, for moving particles like electrons the {\it average} value of the screened 
field is zero. 
An explanation is the following: a neutral atom is not accelerated by the homogeneous 
external field. If a charged particle inside the atom is not accelerated the electric field 
acting on this particle is zero. Therefore, the EDM of the particle has nothing to interact 
with, ${\bf d}\cdot \big<{\bf E}\big>=0$.} 
This occurs due to polarization of the atomic electrons by the 
external field. 
%This corresponds to the fact that the constituent particles of the 
%neutral system do not leave the system on the application of an 
%external electric field.
An atomic EDM cannot be induced in such a case.
However, as shown by Schiff \cite{schiff}, if magnetic or finite-size effects are 
taken into account, there is incomplete shielding, 
and so atomic EDMs can in principle be measured.  

An atomic EDM can be induced from the following $P,T$-odd 
mechanisms:
\newline
\indent
(i) an intrinsic EDM of an electron 
-- the electron EDM can interact with the atomic 
field producing an atomic EDM that is many times larger than 
the single electron EDM \cite{sandars1965}; 
\newline
\indent
(ii) a $P,T$-odd electron-nucleon interaction \cite{bouchiat1975,HLS1976}; 
\newline
\indent
(iii) an intrinsic EDM of an external nucleon \cite{sandars1967,khriplovich76};
\newline
\indent
(iv) a $P,T$-odd nucleon-nucleon interaction -- 
this interaction can induce $P,T$-odd nuclear moments 
that can greatly exceed the moments of single nucleons \cite{haxhen,SFK84}.

The latter two mechanisms (iii),(iv) can be grouped together at the 
nuclear scale since they both produce $P,T$-odd nuclear multipole moments. 

One may consider other exotic $T$-violating mechanisms such as dyon\footnote{
A dyon is a particle with both 
electric and magnetic charges.} 
vacuum polarization \cite{dyon}. 
 
There are in fact other mechanisms that can lead to 
an atomic EDM that are not $P,T$-violating at the fundamental level. 
For example, $T$-odd, $P$-even interactions can have $P,T$-odd 
effects due to $P$-odd radiative corrections 
(see, e.g., \cite{cpoddbook}).
%Also, a $P$-odd, $T$-even anapole moment of the nucleus
%can lead to an imitation of an EDM in atomic experiments 
%due to macroscopic $T$-violating effects. This occurs due 
%to the interaction between the leakage current (created from 
%the external electric field) and the anapole moment \cite{cpoddbook}.

\begin{figure}[b]
\centerline{\psfig{file=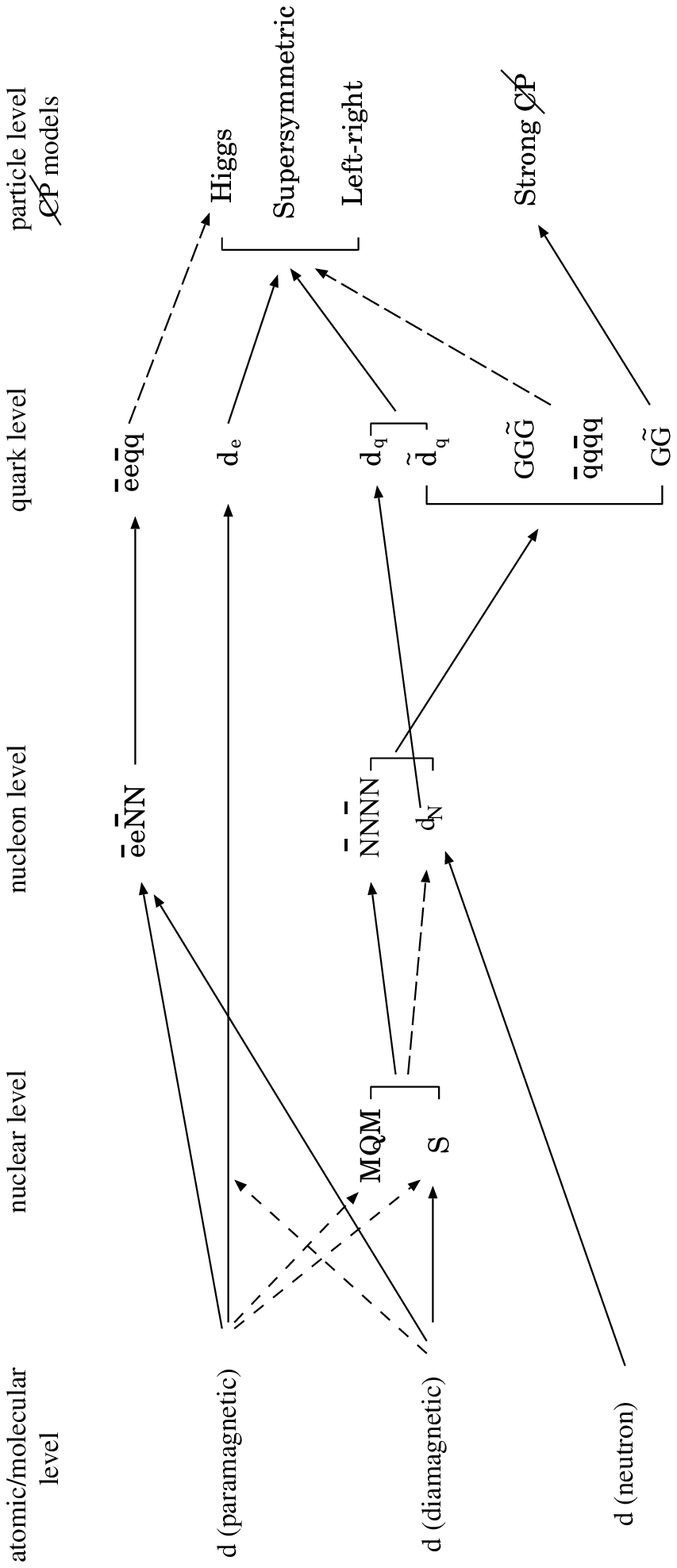,clip=}}
\caption{Flow diagram of CP-violation mechanisms at different levels 
that induce neutron, atomic, and molecular EDMs.}
\label{fig:flowdiagram}
\end{figure}

In Fig. \ref{fig:flowdiagram} we present a flowdiagram (slightly modified 
from the review Ref. \cite{barrreview}) showing the CP-violating mechanisms 
at different energy scales that induce neutron, atomic, and molecular EDMs. 
Read from left to right, it is seen clearly which measurements constrain which 
$CP$-violating parameters at smaller distances (and which popular $CP$-violating models). 
(Read in the other direction, it is seen which small-distance mechanisms induce 
large-distance $CP$-violating effects.) 
Calculations are required to relate the $CP$-violating parameters at different scales.
Solid lines indicate the parameters that are most strongly constrained (induced) by 
the parameters to the left (right), while dashed lines indicate a weaker constraint 
(inducing mechanism).   

In this review we focus primarily on atomic EDMs induced 
by $P,T$-odd nuclear moments that originate from  $P,T$-odd nuclear forces, 
since there have been several recent developments in this area. 
We will first discuss atomic EDMs induced by the 
$P,T$-violating electron-nucleon interaction in Section \ref{sss:e-n} 
and by the electron EDM in Section \ref{sss:electron}. 
For the sake of completion, we briefly introduce nuclear $P,T$-violating 
moments in Section \ref{sss:nuclear} before a comprehensive overview of 
these moments is given in Section \ref{section7ii}.

\subsubsection{The $P,T$-violating electron-nucleon interaction}
\label{sss:e-n}

The $P,T$-violating electron-nucleus interaction has the following form 
(see, e.g. \cite{cpoddbook}): 
\begin{equation}
\label{eq:eN}
\hat{h}=i\frac{G}{\sqrt{2}}\sum _{N}\Big[ 
C^{SP}_{N}\bar{N}N\bar{e}\gamma _{5}e +
C^{PS}_{N}\bar{N}\gamma _{5}N\bar{e}e +
C^{T}_{N}\bar{N}\gamma_{5}\sigma _{\mu \nu}
N\bar{e}\sigma _{\mu \nu}e 
\Big] \ .
\end{equation}
%
%in particular, check tensor contribution
%
%\begin{eqnarray}
%\label{eq:eNsp}
%&&C^{SP}i\frac{G}{\sqrt{2}}\bar{N}N\bar{e}\gamma _{5}e \quad ,\\
%\label{eq:eNps}
%&&C^{PS}i\frac{G}{\sqrt{2}}\bar{N}\gamma _{5}N\bar{e}e \quad ,\\
%\label{eq:eNt}
%&&C^{T}i\frac{G}{\sqrt{2}}\bar{N}\gamma_{5}\sigma _{\mu \nu}
%N\bar{e}\sigma _{\mu \nu}e \quad .
%\end{eqnarray}
The real, dimensionless constants $C^{SP}_{N}$, $C^{PS}_{N}$, and $C^{T}_{N}$ 
give the strength of the scalar-pseudoscalar, pseudoscalar-scalar, 
and tensor $P,T$-odd electron-nucleon interactions for the nucleon $N$. 
Upper limits on the constants $C^{SP}$, $C^{PS}$, and $C^{T}$ 
can be obtained from measurements of atomic EDMs.

In the limit of an infinitely heavy nucleon the following form for the   
electron-nucleus interaction is obtained \cite{cpoddbook}
\begin{equation}
\label{eq:eNsp,t}
\hat{h}^{SP,T}=\hat{h}^{SP}+\hat{h}^{T}=i\frac{G}{\sqrt{2}}\delta ({\bf r})
\Big[
(ZC^{SP}_{p}+NC^{SP}_{n})\gamma _{0}\gamma _{5}
+2(C^{T}_{p}\sum_{p}\mbox{\boldmath$\sigma$}_{p}+ 
C^{T}_{n}\sum_{n}\mbox{\boldmath$\sigma$}_{n}) \cdot \mbox{\boldmath$\gamma$}
\Big] \ .
\end{equation}
In this approximation the term containing $C^{PS}$ vanishes.

Notice the similarity between this expression (\ref{eq:eNsp,t}) and 
Eqs. (\ref{eq:pncnp}), (\ref{eq:NSDnc}).  
However, here the matrix element is real 
(the factor $i$ is placed here to make the operator Hermitian; 
another factor of $i$ arises due to the mixing of the upper and lower 
components of the electron wave function (\ref{psidef}) due to 
$\gamma _{5}$ and $\gamma _{i}$). 
Therefore the interaction (\ref{eq:eNsp,t}) mixes atomic states of 
opposite parity and induces static electric dipole moments in atoms. 

Like their $P$-odd, $T$-even analogues (\ref{eq:pncnp}) and (\ref{eq:NSDnc}), 
the nuclear spin-independent term in (\ref{eq:eNsp,t}) receives coherent 
contributions from the nucleons inside the nucleus, whereas the 
smaller second term, dependent on the nuclear spin, arises from the unpaired nucleons.    

Each of the interactions $\hat{H}^{SP}$ and $\hat{H}^{T}$ can induce EDMs 
in paramagnetic atoms. 
However, only $\hat{H}^{T}$ can open up the closed 
electron shells of diamagnetic atoms and thus induce an EDM; 
this term is dependent on the nuclear spin, 
while the scalar-pseudoscalar term is not. 
Therefore, the interaction $\hat{H}^{SP}$ cannot {\it by itself} 
contribute to EDMs in diamagnetic atoms.

However, by allowing for the hyperfine interaction, 
measurements of EDMs of diamagnetic atoms {\it can} place limits on 
$C^{SP}$ which are, in fact, as competitive as those obtained from 
experiments with paramagnetic atoms.    
The atomic EDM induced by the $P,T$-odd scalar-pseudoscalar 
electron-nucleus interaction $\hat{H}^{SP}$ along with the hyperfine interaction 
$\hat{H}_{hf}$
%=|e|{\bf M}\cdot
%\frac{{\bf r}\times \mbox{\boldmath$\alpha$}}{r^{3}}$, where 
%${\bf M}=(|e|\mu/2m_{p}){\bf I}/I$ is the nuclear magnetic moment, 
arises in the third order of perturbation theory \cite{FK85},
\begin{equation}
\label{eq:3hf}
{\bf d}_{\rm atom}=\sum_{mn}\frac{\langle 0|\hat{\bf D}|m\rangle 
\langle m|\hat{H}_{hf}|n\rangle \langle n|\hat{H}^{SP}|0\rangle }
{(E_{0}-E_{m})(E_{0}-E_{n})} + {\rm permutations} \ .
\end{equation} 
The matrix element of an effective operator constructed from 
$\hat{H}^{SP}$ and $\hat{H}_{hf}$, 
\begin{equation}
\hat{H}_{\rm eff}^{SP}=\sum_{n}\frac{\Big( \hat{H}^{SP}|n\rangle \langle n|\hat{H}_{hf}\Big) + 
\Big(\hat{H}_{hf}|n\rangle \langle n|\hat{H}^{SP}\Big)}{E_{0}-E_{n}} \ ,
\end{equation}
has the form $\langle p_{1/2}|\hat{H}_{\rm eff}^{SP}|s_{1/2}\rangle \propto 
(ZC^{SP}_{p}+NC^{SP}_{n}){\bf j}\cdot {\bf I}$. 
This matrix element can be related to that of 
$\langle p_{1/2}|\hat{H}^{T}|s_{1/2}\rangle \propto 
{\bf j}\cdot \big< C^{T}_{p}\sum_{p}\mbox{\boldmath$\sigma$}_{p} + 
C^{T}_{n}\sum_{n}\mbox{\boldmath$\sigma$}_{n} \big>$,  
%for which direct calculations have been performed, 
the brackets $\big< \ \big>$ denote averaging over the nuclear state. 
The correspondence between $d_{\rm atom}(C^{SP})$ and $d_{\rm atom}(C^{T})$ 
\cite{FK85,kozlov88,cpoddbook}, 
\begin{equation}
\label{eq:SP<->T}
\big( \frac{Z}{A}C^{SP}_{p}+\frac{N}{A}C^{SP}_{n}\big) \frac{\bf I}{I}
\leftrightarrow
1.9\times 10^{3}(1+0.3Z^{2}\alpha ^{2})^{-1}A^{-2/3}\mu ^{-1}
\langle C^{T}_{p}\sum_{p}\mbox{\boldmath$\sigma$}_{p}+
C^{T}_{n}\sum_{n}\mbox{\boldmath$\sigma$}_{n} \rangle \ ,
\end{equation}
where $\mu$ is the nuclear magnetic moment in nuclear magnetons, 
can be used to obtain the sensitivity of $d_{\rm atom}$ to $C^{SP}$ from 
calculations of $d_{\rm atom}(C^{T})$ for diamagnetic atoms. 

Calculations of atomic EDMs induced by $\hat{H}^{SP}$ and $\hat{H}^{T}$ are 
presented in Table \ref{tab:e-n}.

The current best limits for $C^{SP}$ have been obtained from the 
$^{205}{\rm Tl}$ and $^{199}{\rm Hg}$ measurements, 
Eqs. (\ref{eq:tledm}) and (\ref{eq:hgedm}), 
\begin{eqnarray}
\label{eq:limitCsp}
(0.40C^{SP}_{p}+0.60C^{SP}_{n})=
\left\{ 
\begin{array}{ll}
(6\pm 6)\times 10^{-8} & \ \qquad {\rm Tl} \\ 
(1.8\pm 0.8\pm 0.7)\times 10^{-7} & \ \qquad {\rm Hg}
\end{array}
\right.
,
\end{eqnarray}
and for $C^{T}$ from Hg, 
\begin{equation}
\label{eq:limitCt}
C^{T}_{n}=-(5.3\pm 2.5\pm 2.0)\times 10^{-9} \ .
\end{equation}
Here we have used the simple shell model of the nucleus, 
$\big<\mbox{\boldmath$\sigma$}_{n}\big>=-(1/3){\bf I}/I.$

Let us now consider the contribution to an atomic EDM arising 
from the pseudoscalar-scalar component of the electron-nucleus interaction (\ref{eq:eN}). 
In the lowest non-vanishing approximation in $m_{p}^{-1}$, 
the Hamiltonian of the electron-nucleus interaction reduces to the form \cite{cpoddbook} 
\begin{equation}
\hat{h}^{PS}=-\frac{G}{\sqrt{2}}\frac{1}{2m_{p}}
(C^{PS}_{p}\sum _{p}\mbox{\boldmath$\sigma$}_{p}+
C^{PS}_{n}\sum _{n}\mbox{\boldmath$\sigma$}_{n})
\mbox{\boldmath$\nabla$}\delta ({\bf r})\gamma _{0} \ .
\end{equation}  
Again, the matrix element of this interaction 
$\langle p_{1/2}|\hat{H}^{PS}|s_{1/2}\rangle \propto 
{\bf j}\cdot \big< C^{PS}_{p}\sum_{p}\mbox{\boldmath$\sigma$}_{p} + 
C^{PS}_{n}\sum_{n}\mbox{\boldmath$\sigma$}_{n} \big>$ 
can be related to that of $\hat{H}^{T}$. 
The correspondence is \cite{FK85,khriplovichpnc,cpoddbook}
\begin{equation}
\label{eq:PS<->T}
C^{PS}\leftrightarrow 4.6\times 10^{3}\frac{A^{1/3}}{Z}C^{T} \ .
\end{equation}

See Table \ref{tab:e-n} for calculations of the sensitivities 
of atomic EDMs to $C^{PS}$. It is seen that this interaction 
induces EDMs much less efficiently than $\hat{H}^{SP}$ and $\hat{H}^{T}$.  
 
It is the $^{199}{\rm Hg}$ EDM measurement (\ref{eq:hgedm}) 
that currently places the tightest constraint on $C^{PS}$, 
\begin{equation}
\label{eq:limitCps}
C^{PS}_{n}=-(1.8\pm 0.8\pm 0.7)\times 10^{-6} \ .
\end{equation} 

\subsubsection{The electron EDM}
\label{sss:electron}

%
%cite Salpeter 1958 and Sachs and Schwebel 1959 ???? 
%

The best limits on the electron electric dipole moment are derived from 
measurements of atomic EDMs. 
Salpeter first noted the possibility of an enhancement of 
the electron EDM in atoms through consideration of the metastable $2s$ state in 
hydrogen \cite{salpeter1958}.

As noted above, when magnetic effects are considered, the screening 
of the EDMs in an atom (Schiff theorem) is lifted \cite{schiff}.
Sandars \cite{sandars1965} pointed out that due to relativistic magnetic effects, 
the atomic EDM induced in heavy atoms can be strongly enhanced 
compared to the electron EDM inducing it.
The value of the atomic EDM compared to the electron EDM is expressed 
through an enhancement factor 
\begin{equation}
\label{eq:K}
K=d_{\rm atom}/d_{e} \ . 
\end{equation} 
The enhancement factor $K$ increases with nuclear 
charge $Z$ faster than $Z^{3}$.

We will look briefly at how an electron EDM induces an EDM of an atom as a whole; 
for a more detailed consideration we refer the reader to the works 
\cite{sandars1968,flambaum76,khriplovichpnc,cpoddbook,comminsreview}. 
As we mentioned at the beginning of Section \ref{ss:atomicedms}, there 
are two types of contributions to an atomic EDM arising from constituent EDMs. 
For the case of electron EDMs these are the following:  
(i) the sum of the intrinsic EDMs of the electrons 
$\langle 0 |d_{e}\sum _{i=1}^{N}\gamma_{0}^{i}\Sigma _{z}^{i}|0\rangle$; and 
(ii) the admixture of opposite parity atomic states due to the pseudoscalar interaction 
$\hat{H}_{e}=-d_{e}\sum _{i=1}^{N}\gamma_{0} ^{i} {\bf \Sigma}^{i}\cdot {\bf E}^{i}_{\rm int}$; 
see Eq. (\ref{eq:atomicedm}). 
Here $\gamma_{0}$ and $\mbox{\boldmath$\Sigma$}=\gamma_{0}\gamma_{5}\mbox{\boldmath$\gamma$}$ 
are Dirac matrices defined in Eq. (\ref{eq:diracmatrices}), 
$|0\rangle$ is the unperturbed state of the atom (it is an eigenstate of the 
$P,T$-even Hamiltonian with no external electric field), 
and ${\bf E}_{\rm int}$ is the internal atomic electric field.
%These generate a Stark shift linear in the external electric field ${\bf E}_{e}$, 
%$\Delta E=-{\bf d}\cdot {\bf E}_{e}$. 

Let us consider for a moment the Stark shift generated by the presence of the 
electron EDMs. It has the form 
$\Delta E=\langle \tilde{0}|-d_{e}
\sum_{i=1}^{N}\gamma_{0}^{i}\mbox{\boldmath$\Sigma$}^{i}\cdot {\bf E}^{i}|\tilde{0}\rangle$, 
where $|\tilde{0}\rangle$ is an eigenstate of the $P,T$-even Hamiltonian $\hat{H}$ which includes 
the external electric field and ${\bf E}={\bf E}_{\rm int}+{\bf E}_{\rm ext}$ is the 
total electric field. 
(When the external field is treated perturbatively and only terms first-order in this field 
are considered, this reduces to the linear Stark shift generated by the two 
contributions to the atomic EDM mentioned above.)
It is convenient to break up the pseudoscalar interaction 
$\gamma_{0}\mbox{\boldmath$\Sigma$}\cdot{\bf E}=\mbox{\boldmath$\Sigma$}\cdot{\bf E}+
(\gamma_{0} -1)\mbox{\boldmath$\Sigma$}\cdot{\bf E}$, since the first term on the right-hand-side 
does not contribute to a linear Stark shift, 
${\bf \Sigma}\cdot {\bf E}=(1/e)[{\bf \Sigma}\cdot \mbox{\boldmath$\nabla$},\hat{H}]$. 

The enhancement factor for the atom can then be written as 
\begin{equation}
\label{eq:K12}
K(J_{z}/J)=\langle 0|\sum_{i=1}^{N}(\gamma_{0} ^{i}-1)\Sigma_{z}^{i}|0\rangle +
2e\sum_{M}\frac{\langle 0|\sum_{i=1}^{N}(\gamma_{0} ^{i}-1)
{\bf \Sigma}^{i}\cdot{\bf E}^{i}_{\rm int}|M\rangle 
\langle M|\sum_{i=1}^{N}z^{i}|0\rangle}{E_{0}-E_{M}} \ ,
\end{equation}
where $J$ is the electron angular momentum of the state $|0\rangle$.
The operator
\begin{eqnarray}
(\gamma_{0} -1)\mbox{\boldmath$\Sigma$}=
\left(
\begin{array}{cc}
0 & 0 \\
0 & -2\mbox{\boldmath$\sigma$}
\end{array}
\right)
\end{eqnarray}
mixes the lower components of the wave functions (\ref{psidef}). 
So it is seen that this is a purely relativistic effect. 
In heavy atoms the first term in Eq. (\ref{eq:K12}) is small compared to the second 
and so can be omitted in the calculations.

The first analytical formulation of the enhancement factor was performed in 
Ref. \cite{flambaum76}. The following expression was obtained for alkaline 
atoms, in terms of quantities that can be determined experimentally, 
\begin{equation}
K=\sum_{m}\frac{4(Z\alpha)^{3}r_{0m}\hbar c}{(J+1)a_{B}^{2}\gamma(4\gamma ^{2}-1)
(\nu _{0}\nu _{m})^{3/2}(E_{m}-E_{0})} \ . 
\end{equation}
The sum over $m$ is taken over the excited states of the external 
electron, $\nu _{0}$, $\nu _{m}$ are effective principal quantum numbers 
for the ground and excited states, 
$\gamma =\sqrt{(J+1/2)^{2}-(Z\alpha)^{2}}$, 
and $r_{0m}$ is the electric dipole radial integral. 
%$r_{0m}=\int _{0}^{\infty}(f_{0}f_{m}+g_{0}g_{m})r^{3}dr$, 
%$f$ and $g$ are the upper and lower radial components of the 
%Dirac wave function.  
Taking into account mixing with only the nearest level, and 
assuming the values $r_{0m}=5a_{B}$, $\nu_{0}=\nu_{m}=2$, 
and $E_{m}-E_{0}=(1/10){\rm Ry}$, 
we obtain:
\begin{equation}
\label{eq:Ksimple}
|K|\sim 10\frac{Z^{3}\alpha ^{2}}{J(J+1/2)(J+1)^{2}}R \ ,
\end{equation}
where $R$ is a relativistic enhancement factor that increases with $Z$ and 
is $1.2$ for Rb and $2.8$ for Fr in the ground states 
(R tends to unity when $J$ is large).

This simple formula (\ref{eq:Ksimple}) illustrates the dependence of the 
enhancement factor on $Z^{3}$ and on the angular momentum $J$ and can 
be used to obtain order of magnitude estimates. 
It is seen that $K$ is large for high $Z$ and low $J$. 
%When $J$ is large the denominator is small and the relativistic factor goes to unity. 

For atoms with more complex configurations these formulae are not applicable 
and numerical calculations of the enhancement factor are required.  

As we mentioned earlier, experiments with paramagnetic atoms and molecules 
are most sensitive to the electron EDM.  
The current best limit on the electron EDM 
comes from the Tl measurement (\ref{eq:tledm}).
The sensitivity of atomic thallium to the electron EDM is \cite{liu92}
\begin{equation}
d(^{205}{\rm Tl})=-585 d_{e} 
\end{equation}
and accordingly the measurement (\ref{eq:tledm}) of the electron EDM is
\begin{equation}
\label{eq:electronedmlimit} 
d_{e}=(6.9\pm 7.4)\times 10^{-28}~e{\rm ~cm} \ .
\end{equation}

There is some sensitivity of diamagnetic systems to the electron EDM 
\cite{fortson1983}, although this sensitivity is very weak. 
The dominant contribution appears in third-order perturbation theory due to 
consideration of the hyperfine interaction, 
Eq. (\ref{eq:3hf}) with $\hat{H}^{SP}$ replaced with 
the interaction $\hat{H}_{e}$ \cite{FK85}. 
(The second-order ``bare'' contribution, 
$2\sum_{n}\langle 0|{-d_{e}\sum_{i=1}^{N}\gamma_{0}^{i}\mbox{\boldmath$\Sigma$}^{i}}|n\rangle 
\langle n|\hat{H}_{hf}|0\rangle/(E_{0}-E_{n})$, is significantly smaller and can be 
neglected \cite{FK85}.)
Another contribution \cite{FK85} comes from the direct interaction of the 
nuclear magnetic field ${\bf B}$ (arising from the nuclear magnetic moment) 
with the electron EDM \cite{salpeter1958}, 
$\hat{H}_{PT}=-id_{e}\sum_{i=1}^{N}\mbox{\boldmath$\gamma$}^{i}\cdot {\bf B}^{i}$. 
This latter interaction contributes in second-order, Eq. (\ref{eq:atomicedm}). 
The following relation between $d_{e}$ and the $P,T$-odd tensor electron-nucleon 
interaction has been obtained \cite{FK85,khriplovichpnc,cpoddbook}, 
\begin{equation}
d_{e}{\bf I}/I \leftrightarrow \frac{3}{7}\frac{Gm_{p}e}{\sqrt{2}\pi \alpha \mu} 
\frac{R}{(R-1)}\big< C_{p}^{T}\sum_{p}\mbox{\boldmath$\sigma$}_{p} 
+C^{T}_{n}\sum_{n}\mbox{\boldmath$\sigma$}_{n} \big> \ ,
\end{equation}
$R$ is a relativistic enhancement factor.
Using the result of Ref. \cite{martpend85} for the calculation of $d_{\rm atom}(C^{T})$ 
for $^{199}{\rm Hg}$ and the above relation, the enhancement factor 
$K(^{199}{\rm Hg})=-0.014$ was found in Ref. \cite{FK85}. 
A TDHF self-consistent calculation performed in \cite{M-P1987} yielded the result 
\begin{equation}
d(^{199}{\rm Hg})=1.16\times 10^{-2}~d_{e}\ . 
\end{equation}
Due to huge polarization corrections, it is of opposite sign, and the same order of magnitude, 
as the lowest-order result, and its value may change with inclusion of correlation 
corrections \cite{M-P1987}.

While the enhancement factors are small for diamagnetic systems, 
the extraordinary precision that has been achieved in the Hg measurement 
(\ref{eq:hgedm}) makes the corresponding measurement of the 
electron EDM, $d_{e}=-(9.1\pm 4.2\pm 3.4)\times 10^{-27}~e{\rm ~cm}$, 
%change value - update using M-P result
comparable with those from the best paramagnetic EDM measurements.

In Table \ref{tab:enhancement} we list enhancement factors for both 
paramagnetic and diamagnetic atoms 
% and molecules 
of experimental interest.

Paramagnetic polar diatomic molecules are attractive for electron EDM studies, 
in particular those with electron states $^{2}\Sigma _{1/2}$ and $^{2}\Pi _{1/2}$ 
(see, e.g., \cite{SF1978mol,FK85mol} and the review \cite{KL1995}).
%cite others??!!
In \cite{SF1978mol} an analytical estimate for the energy shift in such molecules ($\Omega =1/2$) 
was made,
\begin{equation}
\label{eq:Kmol}
\langle\beta |-d_{e}(\gamma _{0} -1)
\mbox{\boldmath$\Sigma$}\cdot{\bf E}_{\rm tot}|\beta\rangle =
\kappa _{d}\mbox{\boldmath$\sigma$}\cdot \hat{\mbox{\boldmath$\lambda$}}\ ,\qquad 
|\kappa _{d}|\sim \frac{Z^{3}\alpha ^{2}e d_{e}}{\gamma (4\gamma ^{2}-1)a_{B}^{2}} \ ,
\end{equation} 
where $|\beta\rangle$ are molecular orbitals built up from atomic orbitals mixed by 
the strong internal electric field. 
Molecules with electron ground state $^{2}\Sigma _{1/2}$ 
include BaF, YbF, HgF, PbF. The same estimate (\ref{eq:Kmol}) is valid for the metastable 
$a(1)~^{3}\Sigma$ state in PbO with which an EDM experiment is in progress \cite{demille2000}; 
more refined calculations can be found in Refs. \cite{KD2002,isaev2003}.

The recent measurement of the electron EDM in YbF yielded the result \cite{hudson2002}
\begin{equation}
d_{e}=(-0.2\pm 3.2)\times 10^{-26}~e~{\rm cm}
\end{equation} 
(calculations for the effective electric fields have been performed in 
Refs. \cite{YbF}).
This is the first measurement of an EDM in a paramagnetic molecule, and 
while the limit on the electron EDM is not as impressive as that from Tl 
or even Hg, it is limited only by statistics.

\subsubsection{$P,T$-violating nuclear moments}
\label{sss:nuclear}

Atomic EDMs can be induced if the nucleus possesses $P,T$-odd 
nuclear moments. These moments arise at the nucleon scale 
due to a $P,T$-violating interaction between nucleons or 
due to intrinsic nucleon EDMs.
The induced nuclear moments can be electric or magnetic. For 
example, the following moments violate parity and time-reversal invariance: 
electric dipole, magnetic quadrupole, electric octupole. 
For the electric case, the interaction Hamiltonian that mixes opposite 
parity electron states, and induces an EDM of the atom, is of the form 
$\hat{h}_{PT}=-e\varphi$, where $\varphi$ is the electrostatic potential of the 
nucleus corresponding to a $P,T$-odd charge distribution. 
In fact, due to Schiff's theorem \cite{schiff}, there is an additional screening 
term which we will look at in Section \ref{section7ii}. 
In the magnetic case Schiff's theorem is not valid, and the interaction 
Hamiltonian of a relativistic electron with the vector potential ${\bf A}$ corresponding 
to a $P,T$-odd current distribution 
is simply 
$\hat{h}_{PT}=e\mbox{\boldmath$\alpha$}\cdot {\bf A}$. 
See formula (\ref{eq:atomicedm}).

The operator $\hat{h}_{PT}$ has electronic and nuclear components. 
While the overall operator $\hat{h}_{PT}$ is a scalar, 
the electronic and nuclear operators can be of any (equal) rank. 
Accordingly, the triangle rule for addition of angular momenta 
imposes restrictions on the angular momenta of the electron and 
nuclear states for non-zero matrix elements.  
For example, the electron interaction with the nuclear magnetic quadrupole 
moment cannot mix $s$ and $p_{1/2}$ electron states, since we must 
have $|j_{1}-j_{2}|\leq 2\leq j_{1}+j_{2}$. 
Similarly, a static magnetic quadrupole moment 
of the nucleus cannot arise in nuclei with total angular momentum $I<1$.

%In atoms with closed electron subshells, the electrons cannot produce a magnetic field 
%for magnetic moments to interact with. So measurements of EDMs in diamagnetic atoms cannot 
%detect MQMs of the nucleus. 

We leave a detailed consideration of $P,T$-odd nuclear moments for the 
next section.

%\subsection{Enhancement of $P$- and $T$-violation in diatomic molecules}

\section{$P,T$-violating nuclear moments and the atomic EDMs they induce}
\label{section7ii}
%\label{ssec:nuclmom}

This section is devoted to a consideration of the $P,T$-odd nuclear moments 
that can induce atomic EDMs. In Sections \ref{ssec:schiff},\ref{ssec:MQM} 
we look at the form of the $P,T$-violating electric and magnetic moments. 
In Section \ref{ssec:nucleonPTodd} we discuss how $P,T$-violating nuclear moments 
are induced by $P,T$-violating mechanisms at the nucleon scale. 
Enhancement mechanisms for nuclear moments are reviewed in Section \ref{ss:nuclearenhancement}.
Finally, in Section \ref{ss:nmatomic} we look at calculations of atomic EDMs induced 
by nuclear moments.

\subsection{Electric moments; the Schiff moment}
\label{ssec:schiff}

When considering the $P,T$-odd electric moments of the nucleus, 
we must take note of an important screening phenomenon - 
the Schiff theorem 
(we mentioned this at the beginning of Section \ref{ssec:atomicEDMmech}).
The electron screening is taken into account by using the 
following (screened) electrostatic potential of the nucleus (for a derivation, 
see, e.g., \cite{SAF}):
\begin{equation}
\label{eq:screlec}
\varphi({\bf R})=\int \frac{e\rho ({\bf r})}{|{\bf R}-{\bf r}|}d^{3}r+
\frac{1}{Z}({\bf d}\cdot \mbox{\boldmath$\nabla$})\int \frac{\rho ({\bf r})}
{|{\bf R}-{\bf r}|}d^{3}r~,  
\end{equation}  
where $\rho ({\bf r})$ is the nuclear charge density, 
$\int \rho ({\bf r})d^{3}r=Z$, and 
\begin{equation}
\label{eq:d1}
{\bf d}=\int e{\bf r}\rho({\bf r})d^{3}r=d{\bf I}/I
\end{equation}
is the $P,T$-odd nuclear EDM.\footnote{
The screening term appears as a result of a unitary transformation of the Hamiltonian 
which does not change the linear Stark shift. For exact atomic wave functions the result 
for the atomic EDM must be the same with and without the screening term. However, in real 
(approximate) calculations inclusion of the screening term is a must.}

We are interested in the contributions of the first and second terms to 
$\varphi$ that are first order in the $P,T$-odd interaction. 
The first term on the right-hand-side of Eq. (\ref{eq:screlec}) 
is $P,T$-odd if the charge density is distorted due to a $P,T$-odd interaction. 
The density in the second term can be considered spherical, since 
the nuclear EDM is $P,T$-violating \big(it arises due to a $P,T$-violating 
component of the density in Eq. (\ref{eq:d1})\big).

If we consider the nucleus to be point-like, then we can perform a 
multipole expansion of the potential (\ref{eq:screlec}) in terms of $r/R$.
According to Schiff's theorem, the nuclear electrostatic potential 
is screened by atomic electrons such that the dominant nuclear $P,T$-odd 
moment, the nuclear EDM, of a point-like nucleus cannot generate an 
atomic EDM. It is easy to see this from Eq. (\ref{eq:screlec}): 
\begin{equation}
-\int e\rho ({\bf r})\big( {\bf r}\cdot \mbox{\boldmath$\nabla$}\frac{1}{R}\big)d^{3}r 
+\frac{1}{Z}({\bf d}\cdot \mbox{\boldmath$\nabla$})\frac{1}{R}\int \rho ({\bf r})d^{3}r =0 \ .
\end{equation} 
The first non-zero $P,T$-odd term in Eq. (\ref{eq:screlec}) is then 
\begin{equation}
%\label{eq:phi3}
\varphi ^{(3)}=-\frac{1}{6}\int e\rho ({\bf r})r_{\alpha}r_{\beta}r_{\gamma}d^{3}r 
\nabla_{\alpha}\nabla_{\beta}\nabla_{\gamma}\frac{1}{R}+
\frac{1}{2Z}({\bf d}\cdot \mbox{\boldmath$\nabla$})\nabla_{\alpha}\nabla_{\beta}
\frac{1}{R}\int \rho({\bf r})r_{\alpha}r_{\beta}d^{3}r \ .
\end{equation}
The first term $r_{\alpha}r_{\beta}r_{\gamma}$ on the right-hand-side of the equation 
is a reducible rank-3 tensor, while the second $r_{\alpha}r_{\beta}$ is a 
reducible rank-2 tensor. 
Separating the trace,
\begin{eqnarray}
r_{\alpha}r_{\beta}r_{\gamma} &=& 
\Big[ r_{\alpha}r_{\beta}r_{\gamma}-\frac{1}{5}r^{2}(r_{\alpha}\delta _{\beta \gamma} 
+r_{\beta}\delta _{\alpha \gamma}+r_{\gamma}\delta _{\alpha \beta})\Big] 
+\frac{1}{5}r^{2}(r_{\alpha}\delta _{\beta \gamma} 
+r_{\beta}\delta _{\alpha \gamma}+r_{\gamma}\delta _{\alpha \beta}) \\
r_{\alpha}r_{\beta}&=&\Big[ r_{\alpha}r_{\beta}-\frac{1}{3}r^{2}\delta _{\alpha \beta}\Big] 
+ \frac{1}{3}r^{2}\delta _{\alpha \beta} \ ,
\end{eqnarray}
it is seen that $\varphi ^{(3)}$ is comprised of a rank-3 octupole potential 
$\varphi _{\rm octupole}$ and a rank-1 ``Schiff'' potential $\varphi _{\rm Schiff}$, 
\begin{equation}
\varphi ^{(3)}=\varphi _{\rm octupole}+ \varphi _{\rm Schiff}~,
\end{equation}
where
\begin{eqnarray}
\label{eq:phioct}
\varphi _{\rm octupole}&=&-\frac{1}{6}O_{\alpha \beta \gamma}
\nabla _{\alpha}\nabla _{\beta}\nabla _{\gamma}\frac{1}{R}
+\frac{1}{e}\frac{1}{2Z}Q_{\alpha \beta}\big( {\bf d}\cdot \mbox{\boldmath$\nabla$}\big) 
\nabla _{\alpha}\nabla _{\beta}\frac{1}{R} \\
\label{eq:phischiff}
\varphi _{\rm Schiff}&=&4\pi {\bf S}\cdot \mbox{\boldmath$\nabla$}\delta ({\bf R}) \ ,
\end{eqnarray}
and we have used $\nabla ^{2}\big(1/R\big)=-4\pi \delta ({\bf R})$. 
${\bf S}$ is the $P,T$-odd nuclear Schiff moment, 
$O_{\alpha \beta \gamma}$ is the $P,T$-odd nuclear electric octupole moment, 
and $Q_{\alpha \beta}=
\int e\rho ({\bf r})(r_{\alpha}r_{\beta}-\frac{1}{3}r^{2}\delta _{\alpha \beta})d^{3}r$ 
is the $P,T$-even nuclear electric quadrupole moment.
The second term in Eq. (\ref{eq:phioct}) is small since only protons in the 
external shell contribute to $Q_{\alpha \beta}$ and there is a factor $\frac{1}{Z}$.
The nuclear octupole and Schiff moments are given by
\begin{eqnarray}
\label{eq:oct}
O_{\alpha \beta \gamma}&=
&\int e\rho ({\bf r})\Big[ r_{\alpha}r_{\beta}r_{\gamma}-\frac{1}{5}r^{2}
\Big( r_{\alpha}\delta _{\beta \gamma}+r_{\beta}\delta _{\alpha \gamma}+
r_{\gamma}\delta _{\alpha \beta}\Big) \Big] 
d^{3}r~,\\
\label{eq:schiff}
{\bf S}&=&\frac{1}{10}\Big[ \int e\rho ({\bf r}){\bf r}r^{2}d^{3}r-
\frac{5}{3}{\bf d}\frac{1}{Z}\int \rho ({\bf r})r^{2}d^{3}r\Big] =S{\bf I}/I \ . 
\end{eqnarray}

Because the octupole moment $O_{ijk}$ carries $3$ units of angular 
momentum it can only arise in nuclei with spin $I\geq3/2$, 
whereas the Schiff moment can arise in nuclei with spin $I\geq1/2$ 
(due to the triangle rule for the addition of angular momenta).
Of the atomic EDM measurements performed so far, only Cs ($I=7/2$) has a 
nuclear spin large enough to have a static octupole moment; all 
other nuclei have spin $I=1/2$.

However, for these moments to induce an atomic EDM they must 
satisfy electronic angular momentum requirements.
Due to the higher rank of the octupole moment it mixes electronic 
states of higher angular momentum than the Schiff moment. 
This means that the atomic EDM induced by the octupole moment 
is smaller than that induced by the Schiff moment because 
the wave functions of electrons with higher angular momentum 
penetrate the vicinity of the nucleus less due to the greater centrifugal barrier.
The lowest value for the angular momentum of the electrons that 
can induce an atomic EDM due to mixing by the octupole moment is $j=3/2$. 
[The conditions imposed on the allowed electronic angular momentum 
for mixing by the octupole moment is $|j_{1}-j_{2}]\leq 3 \leq j_{1}+j_{2}$ 
and that allowed by the electric dipole mixing is  
$|j_{1}-j_{2}]\leq 1 \leq j_{1}+j_{2}$, so the conditions for inducing an atomic EDM 
are $|j_{1}-j_{2}]\leq 1$ and $j_{1}+j_{2}\geq 3$.]
This means that $s$ states cannot contribute to the EDM produced by the 
electric octupole moment, so that in fact the octupole moment of the Cs 
nucleus cannot induce an atomic EDM in Cs in the ground state, 
as this state corresponds to a configuration with a single electron 
in an $s$-state above closed shells.
Static nuclear octupole moments and the atomic EDMs they induce have been 
considered in detail in Ref. \cite{FMO}.   
The EDMs they induce in atoms are very small, so we will consider them no further.  

It is therefore obvious that the Schiff moment is the only $P,T$-odd moment 
that induces an EDM in atoms with closed electron subshells such as Xe and Hg. 
%and the molecule TlF 
(The nuclear magnetic quadrupole moment, 
which will be discussed in the next section, cannot induce an atomic 
EDM in systems with zero electron angular momentum, since there is 
no magnetic field of the electrons for the MQM to interact with. 
The same conclusion also follows from the triangle rule applied to 
Eq. (\ref{eq:atomicedm}).)
In fact, all the atoms for which EDM measurements have been performed 
\Big($^{133}$Cs ($I=7/2$, $J=1/2$), $^{205}$Tl ($I=1/2$, $J=1/2$), 
$^{129}$Xe $^{3}P_{2}$ ($I=1/2$, $J=2$), $^{129}$Xe $^{1}S_{0}$ ($I=1/2$, $J=0$), 
$^{199}$Hg ($I=1/2$, $J=0$)\Big) 
can have contributions from the nuclear Schiff moment, however it is only 
Cs that can have an EDM arising due to a static magnetic quadrupole moment. 

Let us consider the form of the atomic EDM (\ref{eq:atomicedm}) induced by 
the interaction of electrons with the Schiff moment (\ref{eq:phischiff}).
The contact interaction 
$\hat{H}_{PT}=-e\sum_{i}\varphi _{\rm Schiff}^{i}$ mixes $s$- and $p$-wave 
electron orbitals and produces EDMs in atoms.
The expression (\ref{eq:phischiff}) is consistently defined for 
non-relativistic electrons. 
Using integration by parts, it is seen that the matrix element 
$\langle s|-e\varphi _{\rm Schiff}|p\rangle$ is finite,
\begin{equation}
\langle s|-e\varphi _{\rm Schiff}|p\rangle =
4\pi e{\bf S}\cdot 
(\mbox{\boldmath$\nabla$}
\psi _{s}^{\dagger}
\psi _{p})_{R=0}={\rm constant}\ .
\end{equation}

However, atomic electrons near the nucleus are ultra-relativistic, 
the ratio of the kinetic or potential energy to $mc^{2}$ in heavy atoms 
is about $100$. 
For the solution of the Dirac equation, 
$(\nabla \psi _{s}^{\dagger}\psi _{p})_{R\rightarrow 0}\rightarrow \infty$ 
for a point-like nucleus. Usually this problem is solved by a cut-off of 
the electron wave functions at the nuclear surface. However,
even inside the nucleus 
$\nabla \psi_{s}^{\dagger}\psi_{p}$ varies significantly, 
$\approx Z^{2}\alpha^{2}$, where $\alpha$ is the fine-structure constant, 
$Z$ is the nuclear charge. In Hg ($Z=80$), $Z^{2}\alpha^{2}=0.34$.

A more accurate treatment requires the calculation of a new nuclear 
characteristic which has been termed the {\it local dipole moment} (LDM) 
\cite{fgschiff}.
This moment takes into account relativistic corrections to the nuclear 
Schiff moment which originate 
%due to a consideration of finite-nuclear 
%size. The corrections come 
from the electron wave functions. 
So in the non-relativistic limit, $Z\alpha \rightarrow 0$, the LDM $L=S$. 
For ${^{199}{\rm Hg}}$, 
$L\approx S(1-0.8 Z^{2}{\alpha}^{2})\approx 0.75S$.
When considering the interaction of atomic electrons with the LDM it 
is defined as placed at the center of the nucleus, that is the 
electrostatic potential is
\begin{equation}
\varphi ({\bf R})=4\pi {\bf L}\cdot {\nabla}\delta ({\bf R}) \ .
\end{equation}
See Ref. \cite{fgschiff} for the explicit form for ${\bf L}$.

It is more convenient to use a real electric field distribution produced 
by a $P,T$-odd perturbation. 
In \cite{fgschiff} it was shown (by considering several nuclear models) 
that the natural generalization 
of the Schiff moment potential for a finite-size nucleus is
\begin{equation}
\label{phigen}
\varphi ({\bf R})=-\frac{15 {\bf S}\cdot {\bf R}}{R_{N}^{5}}n(R-R_{N})\ ,
\end{equation}
where $R_{N}$ is the nuclear radius and $n(R-R_{N})$ is a smooth 
function which is $1$ for $R<R_{N}-\delta$ and 
$0$ for $R>R_{N}+\delta$; 
$n(R-R_{N})$ can be taken as proportional to the nuclear density.
This form for the electrostatic potential has no singularities and is 
suitable for relativistic atomic calculations.
%
%Note that this form for the electrostatic potential (\ref{phigen}) 
%was calculated in two models: (i) an external nucleon in state $s_{1/2}$; 
%(ii) an octupole deformed nucleus. 

\subsubsection{The $P,T$-odd electric field distribution in nuclei 
created by the nuclear Schiff moment}
\label{schiff_field}

\begin{figure}[b]
\centerline{\psfig{file=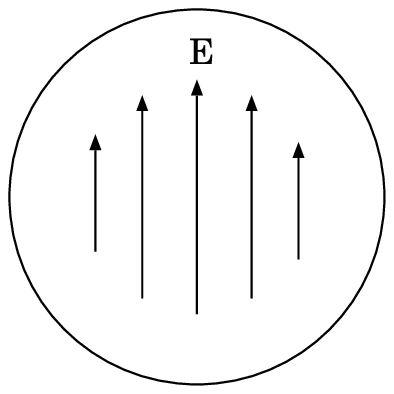,height=1.4in}}
\caption{Constant electric field ${\bf E}$ inside the nucleus produced by 
the $P,T$-odd interaction (Schiff moment field).
${\bf E}$ is directed along the nuclear spin ${\bf I}$.}
\label{fig:schiff}
\end{figure}

From the new form for the electrostatic potential Eq. (\ref{phigen}) 
it can easily be seen that the Schiff moment gives rise to a constant 
electric field inside the nucleus (see Fig. \ref{fig:schiff}), 
${\bf E} = - \mbox{\boldmath$\nabla$}\varphi$. 
The correlation between the electric field and the nuclear spin, 
${\bf E}\propto {\bf I}$, is naturally $P,T$-odd. 
This electric field polarizes atomic electrons, producing an EDM of the atom.

\subsection{Magnetic moments; the magnetic quadrupole moment}
\label{ssec:MQM}

In the gauge $\mbox{\boldmath$\nabla$}\cdot {\bf A}=0$ the vector potential 
produced by a steady current is 
\begin{equation}
\label{eq:A}
{\bf A}({\bf R})
=\int \frac{{\bf j}({\bf r})}{|{\bf R}-{\bf r}|}d^{3}r \ ,
\end{equation}
where ${\bf j}$ is the vector current density. 
The lowest-order term in the multipole expansion of Eq. (\ref{eq:A}) is the 
$P,T$-even magnetic dipole moment, and we have no interest in this.   
The lowest-order $P,T$-odd moment arises in second-order 
and is the rank-2 magnetic quadrupole moment (MQM); 
it appears alongside the $P$-odd, $T$-even anapole moment ${\bf a}$  
(see Section \ref{section6}) \cite{SFK84},  
\begin{eqnarray}
A^{(2)}_{\gamma}&=&\frac{1}{2}\int j_{\gamma}r_{\delta}r_{\alpha}d^{3}r\nabla _{\delta} 
\nabla _{\alpha}\frac{1}{R}\\ 
&=&\Big[ \frac{1}{4\pi}(\delta _{\gamma \alpha}a_{\delta}-
\delta _{\delta \alpha}a_{\gamma})-
\frac{1}{6}\epsilon _{\gamma \delta \beta}M_{\alpha \beta}\Big] 
\nabla _{\delta}\nabla _{\alpha}\frac{1}{R}\\ 
&=&A^{a}_{\gamma}+A^{MQM}_{\gamma} \ ,
\end{eqnarray}
where the anapole moment is given by Eq. (\ref{eq:anaA}) and the MQM is 
\begin{equation}
\label{eq:MQMdef}
M_{\alpha \beta}=
-\int(r_{\alpha}\epsilon _{\beta \xi \eta}+r_{\beta}\epsilon _{\alpha \xi \eta})
j_{\xi}r_{\eta}d^{3}r \ .
\end{equation}
The $P,T$-odd component of the current density ${\bf j}$ will give rise to a 
non-zero magnetic quadrupole moment. 
Its form is specific to the $P,T$-odd mechanism creating it. 
For instance, if we consider that it is produced by an external nucleon 
perturbed by $P,T$-odd nuclear forces, then we can use the 
current (\ref{eq:emcurrent}), 
\begin{equation}
\label{eq:MQM1}
M_{\alpha \beta}=\frac{e}{2m}\int \Big[ 3\mu 
\big( r_{\alpha}\sigma _{\beta}+r_{\beta}\sigma _{\alpha}-
\frac{2}{3}\delta _{\alpha \beta}\mbox{\boldmath$\sigma$}\cdot {\bf r}\big)+ 
2q(r_{\alpha}l_{\beta}+r_{\beta}l_{\alpha})\Big] \rho ({\bf r})d^{3}r \ .
\end{equation}
The magnetic quadrupole moment can now be calculated using 
a $P,T$-odd perturbed density $\rho ({\bf r})$.
  
A general expression for the MQM can be constructed in terms of the 
total angular momentum of the system ${\bf I}$,
\begin{equation}
\label{eq:MQM2}
M_{\alpha \beta}=\frac{3}{2}\frac{M}{I(2I-1)}\Big[
I_{\alpha}I_{\beta}+I_{\beta}I_{\alpha}-\frac{2}{3}I(I+1)\delta _{\alpha \beta}\Big] \ .
\end{equation}
The quantity $M$ is conventionally referred to as the MQM and is 
defined as the maximum projection of $M_{\alpha \beta}$ on the nuclear axis, 
$M=M_{zz}$.
It is easily seen from a comparison of Eqs. (\ref{eq:MQMdef}), (\ref{eq:MQM2}) 
that the magnetic quadrupole moment violates parity and time-reversal invariance.  

The interaction of electrons with a nuclear MQM induces an atomic 
EDM typically an order of magnitude larger than that induced by the nuclear 
Schiff moment \cite{khriplovich76,SFK84}. 
The ratio of the $s-p$ electronic matrix elements is \cite{SFK84} 
\begin{equation}
\label{eq:MSratio}
\frac{\langle s |\mbox{\boldmath$\alpha$}\cdot {\bf A}^{MQM}|p\rangle}
{\langle s |\varphi _{\rm Schiff}|p\rangle} 
\sim 10^{2}A^{-2/3}\frac{R_{M}}{R_{S}} \ ,
\end{equation}
$R_{M}$, $R_{S}$ are relativistic enhancement factors for 
the magnetic quadrupole and Schiff moments, respectively, 
$R\rightarrow 1$ as $Z\alpha \rightarrow 0$. 
It is seen that for atoms with light nuclei the 
contribution of the MQM dominates. 
The relativistic factor $R_{S}$ grows faster than $R_{M}$ with 
increase of $Z$. For example, at $Z=80$, 
$R_{S}=7~[p_{1/2}],5~[p_{3/2}]$ while $R_{M}=1.8$; 
explicit formulae can be found in Ref. \cite{SFK84}. 
(In the square brackets the angular momentum of the 
$p$ electron state is specified; 
for the MQM, $p\equiv p_{3/2}$.) 
At $Z=80,~A=200$, the ratio (\ref{eq:MSratio}) reaches $\sim 1$.
%However, the MQM can only work in a system with non-zero electron 
%angular momentum (when the magnetic field of the electrons is not 
%zero).  

\subsubsection{The spin hedgehog}
\label{sss:hedgehog}

%this is collective! In wrong section?!
%
%

In a spherically symmetric system the $P,T$-odd interaction induces 
a ``spin-hedgehog'' whereby the spin density is proportional to the 
radial vector, 
$\mbox{\boldmath$\sigma$}\propto {\bf r}$ \cite{ryndin,flambaum94}. 
The $P,T$-odd nucleon-nucleon interaction 
(leading to the perturbed wave functions (\ref{eq:pertwav}), see below) 
produces the following distributions of the spins for protons and neutrons 
in the nucleus,
\begin{equation}
\label{eq:spinhh}
\mbox{\boldmath$\sigma$}_{p}({\bf r})=
\xi _{p}\mbox{\boldmath$\nabla$}\rho _{p}({\bf r}) \ , 
\qquad \mbox{\boldmath$\sigma$}_{n}({\bf r})=
\xi _{n}\mbox{\boldmath$\nabla$}\rho _{n}({\bf r}) \ ,
\end{equation}
the unperturbed nuclear density $\rho =\sum |\psi|^{2}$.
This collective spin distribution, however, produces no current 
\Big(${\bf j}({\bf r})=\mu \mbox{\boldmath$\nabla$}\times \mbox{\boldmath$\sigma$}
({\bf r})\propto \mbox{\boldmath$\nabla$}\times\mbox{\boldmath$\nabla$} 
\rho({\bf r})=0$\Big) and hence no magnetic field \cite{flambaum94}.
One may think that because the spin-hedgehog has no magnetic field 
it produces no effects. This is not the case. 
The spin-dependent part of the strong interaction is 
sensitive to this spin structure.
It reduces the constants of the $P,T$-odd nucleon-nucleon interaction: 
for the case of an external proton interacting with the 
spin-hedgehog, $\eta _{p}\rightarrow \eta _{p}/1.5$, while for a neutron, 
$\eta _{n}\rightarrow \eta_{n}/1.8$ \cite{flambaum94,fv1994}. 
A distorted spin hedgehog in deformed nuclei produces a collective magnetic 
quadrupole field (see Section \ref{sss:collective}).
 
Note that the spin-hedgehog is not specific to nuclei. 
For example, the $P,T$-odd electron-nucleon interaction 
(Section \ref{sss:e-n}) mixes atomic states of opposite parity 
and induces a spin-hedgehog of the atom \cite{ryndin}; 
see, e.g., \cite{khriplovichpnc,cpoddbook} for details.

\subsection{What mechanisms induce $P,T$-odd nuclear moments at the 
nucleon scale?}
\label{ssec:nucleonPTodd}

$P,T$-odd nuclear moments can arise due to an intrinsic EDM of an  
external nucleon or due to $P,T$-odd nuclear forces. 
The $P,T$-odd nuclear forces induce larger 
nuclear moments than a single nucleon EDM (Section \ref{sss:comparison}). 
This is what makes atomic experiments so competitive compared 
to neutron experiments in probing $CP$-violation in the 
hadron sector. As we will see in Section \ref{section7iv}, atomic experiments are 
more sensitive than neutron experiments to many underlying $CP$-violating 
mechanisms.  

\subsubsection{The $P,T$-odd nucleon-nucleon interaction}

The $P,T$-odd nucleon-nucleon interaction is the dominating 
nuclear mechanism inducing atomic EDMs in diamagnetic atoms and 
molecules.

The $P,T$-odd nucleon-nucleon interaction, to first-order in the 
velocities $p/m$, can be presented as \cite{SFK84}
\begin{equation}
\label{wfull}
\hat{W}_{ab}=\frac{G}{\sqrt{2}}\frac{1}{2m}
\left( (\eta_{ab}\mbox{\boldmath$\sigma$}_{a}-
\eta_{ba}\mbox{\boldmath$\sigma$}_{b})\cdot \mbox{\boldmath$\nabla$}_{a}
\delta({\bf r}_{a}-{\bf r}_{b})+
\eta '_{ab}\left[ \mbox{\boldmath$\sigma$}_{a}
\times \mbox{\boldmath$\sigma$}_{b}\right]
\left\{ 
({\bf p}_{a}-{\bf p}_{b}),\delta({\bf r}_{a}-{\bf r}_{b})
\right\} 
\right),
\end{equation}
where $\{\ ,\ \}$ is an anticommutator, $G$ is the 
Fermi constant of the weak interaction, $m$ is the nucleon mass, 
and \mbox{\boldmath$\sigma$}, ${\bf r}$, and ${\bf p}$ are the spins, 
coordinates, and momenta of the nucleons $a$ and $b$.
The dimensionless constants $\eta _{ab}$ and $\eta '_{ab}$ 
characterize the strength of the $P,T$-odd nuclear interaction 
(experiments on EDMs are aimed to measure these constants).  

If we consider the $P,T$-odd interaction between a single unpaired 
nucleon and a heavy spherical core, then
we can average the two-particle interaction (\ref{wfull}) 
over the core nucleons to obtain the effective single-particle $P,T$-odd 
interaction between the nucleon and core \cite{SFK84},
\begin{equation}
\label{w}
\hat{W}=\frac{G}{\sqrt{2}}\frac{\eta _{a}}{2m}
\mbox{\boldmath$\sigma$}\cdot \mbox{\boldmath$\nabla$}
\rho _{A}({\bf r}) \ . 
\end{equation}
Here it has been assumed that the proton and neutron densities 
are proportional to the total nuclear density $\rho _{A}({\bf r})$; 
the dimensionless constant 
\begin{equation}
\eta _{a}=\frac{Z}{A}\eta _{ap} +
\frac{N}{A}\eta _{an} \ .
\end{equation}
Notice that there is only one surviving term from the $P,T$-odd 
nucleon-nucleon interaction (\ref{wfull}); this is because all 
other terms contain the spin of the internal nucleons for which 
$\langle \mbox{\boldmath$\sigma$}\rangle =0$.

The shape of the nuclear density $\rho _{A}$ and the strong potential 
$U$ are known to be similar; we therefore take
\begin{equation}
\label{eq:rhoU} 
\rho _{A}({\bf r})=\frac{\rho _{A}(0)}{U(0)}U({\bf r}) \ .
\end{equation} 
Then Eq. (\ref{w}) can be rewritten in the following form:
\begin{equation}
\label{eq:w,xi}
\hat{W}=\xi \mbox{\boldmath$\sigma$}\cdot \mbox{\boldmath$\nabla$}U \ , \qquad
\xi=\eta \frac{G}{2\sqrt{2}m}\frac{\rho _{A}(0)}{U(0)}=
-2\times 10^{-21}\eta~{\rm cm} \ .
\end{equation} 
Now it is easy to find the solution of the Schr{\"o}dinger equation 
including the interaction $\hat{W}$ \cite{SFK84}, 
\begin{eqnarray}
&(\hat{H}+\hat{W})\tilde {\psi}=E\tilde{\psi} & \ , \\
\label{eq:pertwav}
&\tilde{\psi}=(1+\xi \mbox{\boldmath$\sigma$}
\cdot \mbox{\boldmath$\nabla$})\psi & \ ,
\end{eqnarray}
where $\psi$ is the unperturbed solution ($\hat{H}\psi =E\psi$).
The density arising from the wave function (\ref{eq:pertwav}) 
is 
\begin{equation}
\label{eq:ptodddensity}
\rho =\tilde{\psi}^{\dagger}\tilde{\psi}=\psi ^{\dagger}\psi 
+\xi\mbox{\boldmath$\nabla$}\cdot (\psi ^{\dagger}\mbox{\boldmath$\sigma$}\psi)
\end{equation}
The second term is the $P,T$-odd part of the density 
which generates the nuclear $P,T$-odd moments.  

The electric dipole (\ref{eq:d1}), Schiff (\ref{eq:schiff}), 
and magnetic quadrupole (\ref{eq:MQM1},\ref{eq:MQM2}) 
nuclear moments induced by the $P,T$-odd nucleon-nucleon interaction, 
through the perturbed density (\ref{eq:ptodddensity}), are \cite{SFK84}
\begin{eqnarray}
\label{eq:d}
d &=& -e\xi \Big( q-\frac{Z}{A}\Big)t_{I} \ , \\
\label{eq:S}
S &=& -\frac{eq}{10}\xi\Big[ \Big( t_{I}+\frac{1}{I+1} \Big) r_{\rm ex}^{2}
-\frac{5}{3}t_{I}r_{q}^{2}\Big] \ , \\
\label{eq:M}
M &=& \frac{e}{m}\xi (\mu -q)(2I-1)t_{I} \ ,
\end{eqnarray}
where $q=0~(1)$ for an external neutron~(proton),
$r_{q}^{2}$ and $r_{\rm ex}^{2}$ are the mean-square radii 
of the nuclear charge and external nucleon, respectively, and 
\begin{equation}
\label{eq:tI}
t_{I}=
\left\{
\begin{array}{ll}
1 \qquad & I=l+1/2 \\
-\frac{I}{I+1} \qquad & I=l-1/2
\end{array}
\right. \ ,
\end{equation}
$l$ is the orbital angular momentum of the external nucleon.   
The recoil effect for the electric moments (the motion of the nuclear core 
around the center-of-mass; see Ref. \cite{SFK84}) has been taken into account 
($q\rightarrow q-Z/A$ in the expression for $d$).
In the single-particle model, the recoil effect for the Schiff moment 
disappears due to the cancellation of its contributions to the 
first and second (screening) terms in Eq. (\ref{eq:screlec}).

Note that for nuclei with an unpaired nucleon in the state $s_{1/2}$ 
(in the simple shell model), such as $^{203,205}$Tl, 
the Schiff moment is reduced to the difference of two approximately equal terms, 
\begin{equation}
\label{eq:S1/2}
S\big( s_{1/2}\big)\propto \Big( r_{\rm ex}^{2}-r_{q}^{2}\Big) \ .
\end{equation}
In obtaining numerical values for the Schiff moment in an analytical calculation 
it is usually assumed that $r_{\rm ex}^{2}=r_{q}^{2}=\big( 3/5\big)R_{N}^{2}$, 
where $R_{N}=r_{0}A^{1/3}$, 
$r_{0}=1.1~{\rm fm}$. Then the Schiff moment Eq. (\ref{eq:S1/2}) vanishes. 
This cancellation makes calculations for $^{203,205}$Tl unstable.  

The moments we have discussed so far are produced by a valence nucleon. 
In the work \cite{FKS86} it was shown that core nucleons make a contribution 
to $P,T$-odd moments that is comparable to that of a valence nucleon. 

Of particular interest is $^{199}$Hg, which gives the best limit on the 
nuclear Schiff moment. 
In the $^{199}$Hg nucleus the unpaired nucleon is a neutron. 
It doesn't contribute to the Schiff moment directly (see Eq. (\ref{eq:S})). 
The nuclear Schiff moment arises due to the 
polarization of the protons of the core by the $P,T$-odd field of the external neutron. 
(The charge distribution must 
be distorted to give a $P,T$-odd correction to the charge density.) 
The strength of the $P,T$-odd interaction is defined by the parameter $\eta _{np}$. 
A numerical calculation of the Schiff moment for $^{199}$Hg in the Woods-Saxon 
potential with spin-orbit interaction gives \cite{flambaum85,FKS86}
\begin{equation}
\label{eq:SHg}
S(^{199}{\rm Hg})=-1.4\times 10^{-8}\eta _{np}~e~{\rm fm}^{3} \ .
\end{equation}
An analytical treatment of the electron relativistic corrections to the Schiff moment 
(that is, calculation of the local dipole moment) of $^{199}$Hg shows that these 
corrections are small -- they reduce the Schiff moment $S$ only by about $25\%$. 
A many-body treatment of the $^{199}$Hg Schiff moment has been performed recently 
in the work \cite{DS2003}. A finite-range $P,T$-odd nucleon-nucleon interaction was 
used and core polarization was calculated in the random-phase approximation. 
The result of Ref. \cite{DS2003} is
\begin{equation}
\label{eq:SHgDS}
S(^{199}{\rm Hg})=-0.0004~g\bar{g}_{0}-0.055~g\bar{g}_{1}+0.009~g\bar{g}_{2}~e~{\rm fm}^{3} \ ,
\end{equation}
where $g\equiv g_{\pi NN}$ is the strong pion-nucleon coupling constant, 
$\bar{g}\equiv \bar{g}_{\pi NN}$ are the $P,T$-violating isoscalar ($i=0$), isovector ($i=1$), 
and isotensor ($i=2$) pion-nucleon couplings; 
see, e.g., Refs. \cite{haxhen,herczeg}
%check reference! anything better?
for the form of the finite-range $P,T$-violating interaction. 
Eqs. (\ref{eq:SHg},\ref{eq:SHgDS}) can be compared by using the 
relation \cite{DS2003} 
$\eta _{np}\sim (Gm_{\pi}^{2}/\sqrt{2})^{-1}g(\bar{g}_{0}+\bar{g}_{1}-2\bar{g}_{2}) 
\sim 7\times 10^{6}~g(\bar{g}_{0}+\bar{g}_{1}-2\bar{g}_{2})$, 
so Eq. (\ref{eq:SHg}) gives $S(^{199}{\rm Hg})\sim 
-0.09~g(\bar{g}_{0}+\bar{g}_{1}-2\bar{g}_{2})~e~{\rm fm}^{3}$.
It is seen that while the contribution of the isovector channel does not change much 
from the value in Eq. (\ref{eq:SHg}), the isoscalar 
channel is suppressed by two orders of magnitude and the isotensor channel by 
one order of magnitude. These corrections are due largely to the 
inclusion of core polarization \cite{DS2003}.

As we mentioned earlier, of the EDM experiments performed so far, only 
the $^{133}$Cs measurement can be interpreted in terms of a magnetic 
quadrupole moment of the nucleus. A calculation in the Woods-Saxon potential 
with the spin-orbit interaction gives \cite{SFK84}
\begin{equation}
M(^{133}{\rm Cs})=1.7\times 10^{-7}\eta_{p} ~\frac{e}{m}~{\rm fm}
\end{equation}
Single-particle calculations have been performed in \cite{DKT1994} for 
the external nucleon contribution and in \cite{DTFD1996} for the 
core contribution.
These calculations show that the MQM is very sensitive to 
the shape of the $P,T$-odd potential relative to the shape of the 
central field potential and to the spin-orbit potential. 
It was found that the core contributions arising from the 
interaction proportional to $\eta$ is comparable to that 
of the valence contribution, and that if $\eta \sim \eta'$ 
then the core contribution is several times larger than the 
valence contribution. 

Results of calculations of nuclear $P,T$-odd moments of current interest 
are presented in Table \ref{tab:nuclearmoments}.

\subsubsection{The external nucleon EDM}

Even though the nucleon-nucleon interaction may be more effective in 
inducing nuclear $P,T$-odd moments and hence atomic EDMs, measurements of 
nucleon EDMs are interesting in their own right. 
Also, nucleon EDMs measured from atomic experiments may be compared 
to those from direct measurements. Currently, the limit on the 
neutron EDM from the Hg measurement (\ref{eq:hgedm}) is competitive with those 
from direct neutron EDM searches (\ref{eq:neutronlimits}); see 
Tables \ref{tab:limitsI},\ref{tab:limitsII}.  
Here we will merely quote the results of the work \cite{khriplovich76} for the 
nuclear Schiff and magnetic quadrupole moments induced by a single unpaired nucleon 
(neutron or proton) with an intrinsic EDM $d_{n,p}$,  
\begin{eqnarray}
\label{eq:S(d)}
S&=&\frac{1}{10}d_{n,p}\Big[
r_{\rm ex}^{2}\Big(
\frac{1}{I+1}+t_{I}\Big)
-\frac{5}{3}r_{q}^{2}t_{I}
\Big] \ , \\
M&=&\frac{d_{n,p}}{m}(2I-1)t_{I} \ .
\end{eqnarray}
See, e.g., \cite{khriplovich76,cpoddbook} for details. 
Notice the similarity between these expressions and those 
for the corresponding moments induced by $P,T$-odd nuclear 
forces (\ref{eq:S}),(\ref{eq:M}). 
In the approximations used, there is a simple correspondence 
between the $P,T$-violating parameters. 
For the Schiff moment, $-eq\xi\leftrightarrow d_{n,p}$, 
while for the MQM, $-e\xi(q-\mu)\leftrightarrow d_{n,p}$.
The second contribution ($\propto \mu$) to the MQM from the 
$P,T$-odd nuclear forces appears from the spin term in (\ref{eq:emcurrent}). 
This has no analog in the case of $d_{n,p}$ since a stationary 
EDM cannot induce a MQM.  
It is the orbital motion of a nucleon with an intrinsic EDM 
that induces the nuclear MQM \cite{khriplovich76}. 

In the simple shell model the unpaired neutron in the $^{199}$Hg 
nucleus carries the nuclear spin $I$, and the Schiff moment 
is induced by the EDM of just this neutron. 
In this picture, the induced Schiff moment can be calculated 
using Eq. (\ref{eq:S(d)}) and the simplifying assumptions 
$r_{\rm ex}^{2}=r_{q}^{2}=(3/5)R_{N}^{2}$, $R_{N}\approx 1.1A^{1/3} ~{\rm fm}$, 
giving $S(^{199}{\rm Hg})\approx 2.2~d_{n}~{\rm fm}^{2}$. 
Proton EDMs also contribute due to configuration mixing. It is possible 
to estimate their contribution by comparing the experimental value of the 
magnetic moment of $^{199}$Hg with that of the simple shell model; see Ref. \cite{dzuba85}.  
In this way, it is found \cite{atomicedms} that $d_{n}$ can be replaced 
by ($d_{n}+0.1d_{p}$),
\begin{equation}
\label{eq:S(Hg,d)}
S(^{199}{\rm Hg})=(2.2~d_{n}+0.2~d_{p})~{\rm fm}^{2} \ .
\end{equation}

A numerical calculation of the nuclear Schiff moment of $^{199}$Hg induced 
by neutron and proton EDMs has recently been performed, with core polarization 
accounted for in the RPA approximation \cite{DS2003b}. The result is \cite{DS2003b}
\begin{equation}
\label{eq:S(Hg,d)RPA}
S(^{199}{\rm Hg})=(1.9~d_{n}+0.2~d_{p})~{\rm fm}^{2} \ .
\end{equation}
(See \cite{DS2003b} for the discussion of uncertainty.)

\subsubsection{Comparison of the size of nuclear moments induced by 
the nucleon-nucleon interaction and the nucleon EDM}
\label{sss:comparison}

Here we consider the enhancement of the nuclear EDM induced by $P,T$-odd 
nuclear forces compared to that induced by an external valence nucleon 
using the simple one-boson exchange model, 
following \cite{flambaum85,FKS86,flambaumreview87}.
The largest contribution to the constant $\eta$ is probably 
given by the lightest $\pi ^{0}$-meson, 
\begin{equation}
\frac{G}{\sqrt{2}}\eta \approx \frac{g_{\pi NN}\bar{g}^{0}_{\pi NN}}{m_{\pi}^{2}} \ ,
\end{equation}
$g_{\pi NN}$ and $\bar{g}_{\pi NN}$ are the constants of the strong and 
$T$-odd $\pi$ meson-nucleon interactions,
\begin{equation}
(ig_{\pi NN}\bar{n}\gamma_{5}n+\bar{g}^{0}_{\pi NN}\bar{p}p)\pi ^{0}+ 
\sqrt{2}(ig_{\pi NN}\bar{p}\gamma_{5}n+\bar{g}^{-}_{\pi NN}\bar{p}n)(\pi ^{-})^{\dagger} 
+ ... \ .
\end{equation}
A neutron EDM is induced through virtual creation of a $\pi ^{-}$ meson \cite{crewther79},
\begin{equation}
d_{n}=\frac{e}{m}\frac{g_{\pi NN}\bar{g}^{-}_{\pi NN}}{4\pi ^{2}}\ln \frac{M}{m_{\pi}} \ .
\end{equation}
Here $M\sim m_{\rho}\sim 700~{\rm MeV}$ is the scale at which 
the $\pi$-meson loop converges. 
The values of $\eta$ and $d_{n}$ are expressed in terms of different 
quantities, $\bar{g}^{0}_{\pi NN}$ and $\bar{g}^{-}_{\pi NN}$, respectively. 
However, for example, in the model of $T$-violation with the 
$\theta$-term, $|g_{\pi NN}\bar{g}^{0}_{\pi NN}|=
|g_{\pi NN}\bar{g}^{-}_{\pi NN}|=0.37|\bar{\theta}|$ 
\cite{crewther79}. 
Taking $|g_{\pi NN}\bar{g}^{0}_{\pi NN}|\sim |g_{\pi NN}\bar{g}^{-}_{\pi NN}|$, it is found that 
\cite{SFK84,flambaum85,FKS86,flambaumreview87} 
\begin{equation}
\frac{d}{d_{n}}\sim \frac{e\xi}{d_{n}}\sim 
2\pi \Big(m_{\pi}^{2}r_{0}^{3}|U(0)|\Big)^{-1}\sim 40 \ ,
\end{equation}
that is, the nuclear EDM exceeds the nucleon EDM by one to two orders 
of magnitude. 
Similarly, $P,T$-odd nuclear forces generate all $P,T$-odd nuclear moments, 
such as Schiff and MQM moments, $10-100$ times larger than those 
generated by the presence of a nucleon EDM \cite{flambaumreview87}.

%
%\begin{equation}
%d_{\rm n}=\frac{e}{m_{p}}\frac{g_{\pi NN}\bar{g}_{\pi NN}}{4\pi ^{2}}
%\ln \Big( \frac{m_{\rho}}{m_{\pi}} \Big) \ .
%\end{equation}
%
%\begin{equation}
%\eta =\frac{g_{\pi NN}\bar{g}_{\pi NN}\sqrt{2}}{Gm_{\pi ^{2}}} \ .
%\end{equation}

\subsection{Nuclear enhancement mechanisms}
\label{ss:nuclearenhancement}

So far we have considered $P,T$-odd nuclear moments in spherical nuclei.
However, in non-spherical nuclei there is the possibility of 
enhancement due to 
(i) the presence of a low-lying level with opposite 
parity and the same angular momentum with respect to the ground state; 
and (ii) collective effects.

%In deformed nuclei, the nuclear calculations are carried out in a ``frozen'' 
%frame (rotating with the nucleus). The nuclear moments can be transformed 
%into the laboratory frame using the formulae
%\begin{equation}
%\label{eq:lab}
%d_{\rm lab}=\frac{I}{I+1}d_{z} \ ,
%\quad 
%S_{\rm lab}=\frac{I}{I+1}S_{z} \ ,
%\quad
%M_{\rm lab}=\frac{I}{I+1}\frac{2I-1}{2I+3}M_{zz} \ .
%\end{equation} 
%Here the $z$-axis is directed along the (rotating) nuclear axis.

\subsubsection{Close-level enhancement}

It is known that nuclei with non-spherical symmetry have close 
levels of opposite parity. It was pointed out in 
Ref. \cite{feinberg} that due to the existence of a 
close level of opposite parity with the same angular momentum as the ground 
state, the nuclear EDM can be enhanced; 
calculations of enhanced EDMs and MQMs were performed in \cite{haxhen}, 
Schiff moments in \cite{SFK84}.
In the frozen frame, the contribution to the $z$-components of the 
electric dipole, Schiff, and magnetic quadrupole nuclear moments
in the ground state $\Omega$ due to the close opposite parity state 
$\bar{\Omega}$ is
\begin{equation}
T=2\frac{\langle \Omega |\hat{H}_{PT}|\bar{\Omega}\rangle 
\langle \bar{\Omega}|\hat{T}|\Omega\rangle}{E_{\Omega}-E_{\bar{\Omega}}} \ ,
\end{equation}
where $T=d_{z},~S_{z},~M_{zz}$.
The magnetic quadrupole moment in heavy 
stable nuclei can be enhanced by an order of magnitude due 
to the ``close level'' mechanism, while for the
electric dipole and Schiff moments this 
enhancement hardly exceeds $\sim 5-10$ \cite{SFK84}.
However, this ``close-level'' enhancement is not regular: 
it gives contributions to $P,T$-odd moments with different 
magnitudes and signs even in the ``nearest'' nuclei, 
the results are unstable.\footnote{
The reason is explained in \cite{SFK84}: 
taking Eq. (\ref{eq:w,xi}) for the $P,T$-odd interaction $\hat{H}_{PT}$, 
it is seen that $\hat{W}=\xi \mbox{\boldmath$\sigma$}\cdot 
\mbox{\boldmath$\nabla$}U\propto [\hat{H},\mbox{\boldmath$\sigma$}\cdot {\bf p}]$, 
where $\hat{H}$ is the single-particle Hamiltonian, 
has small matrix elements between close levels, 
$\langle \Omega |[H,\mbox{\boldmath$\sigma$}\cdot {\bf p}]|\bar{\Omega}\rangle
\propto E_{\Omega}-E_{\bar{\Omega}}$.}

\subsubsection{Collective enhancement}
\label{sss:collective}

While the ``close-level'' mechanism enhances the contribution of the 
external nucleon to the $P,T$-odd nuclear moments, there can also 
be a ``collective'' enhancement of the nuclear moments that occurs 
due to the contribution of many nucleons.
In the work \cite{flambaum94} it was shown that a collective magnetic 
quadrupole moment can be produced in deformed nuclei 
by $P,T$-odd nuclear forces. 
Unlike the close-level enhancement, this collective enhancement is regular: 
in deformed nuclei about $A^{2/3}$ nucleons belong to open shells 
that contribute to the MQM. 
The $P,T$-odd nuclear forces create a spin hedgehog [Eq. (\ref{eq:spinhh})] 
as in the case of spherical nuclei (Section \ref{sss:hedgehog}), 
however in the deformed case there is a non-zero magnetic field associated 
with it. 
The MQM of a deformed nucleus (in the rotating frame) 
can basically be calculated as a summation of the single-particle 
MQMs (\ref{eq:M}) of all nucleons in the open shells. 
Notice, from Eqs. (\ref{eq:M},\ref{eq:tI}),
that spin-orbit pairs $I=l+1/2$ and $I=l-1/2$ make contributions 
to the collective MQM of opposite sign. A sufficiently large spin-orbit 
splitting is therefore required to avoid cancellation, and this is satisfied 
in nuclei.
This collective mechanism gives an order of magnitude enhancement of 
the nuclear MQM in deformed nuclei compared to spherical nuclei. 

\subsubsection{Octupole deformation; collective Schiff moments}
\label{sss:octdef}

We will now move on to the collective $P,T$-odd nuclear moments 
produced by $P,T$-odd nuclear forces that arise in nuclei with 
static octupole deformation \cite{AFS,SAF,FMO}.  
There is an enhancement of these collective moments, compared to 
single-particle nuclear moments, due to the collective nature 
of the intrinsic moments and the small energy separation between 
members of parity doublets. This enhancement can be as large as 
1000 times. 

Static octupole deformation in the ground state has been demonstrated 
to exist in nuclei in the regions Ra-Th and Ba-Sm. It produces effects 
such as parity doublets, large dipole and octupole moments in the 
intrinsic frame of reference and enhanced electric dipole and octupole 
transitions; see the review Ref. \cite{octrev}.

While it has been shown that the Schiff and electric dipole and octupole 
moments are enhanced in nuclei with octupole deformation, we will focus 
our attention on the nuclear Schiff moment (the EDM is not of direct 
interest, in atoms it is screened by atomic electrons; also, the 
atomic EDM induced by the electric octupole moment is small since it does 
not mix $s$ and $p$ electron orbitals.)

The mechanism generating collective $P,T$-odd moments is the following. 
In the ``frozen'' body frame collective moments 
can exist without any $P,T$-violation.
However, the nucleus rotates, and this makes the expectation value of 
these moments vanish in the laboratory frame 
if there is no $P,T$-violation.
(For example, the intrinsic Schiff moment is 
directed along the nuclear axis, ${\bf S}_{\rm intr}=S_{\rm intr}{\bf n}$,
and in the laboratory frame the only possible correlation 
$\langle {\bf n}\rangle \propto {\bf I}$ violates parity and time reversal invariance.) 
The $P,T$-odd nuclear forces mix rotational states of 
opposite parity and create an average orientation of the nuclear axis 
${\bf n}$ along the nuclear spin ${\bf I}$,
\begin{equation}
\label{nz}
\langle n_{z}\rangle =2\alpha \frac{KM}{I(I+1)} \ ,
\end{equation}
where 
\begin{equation}
\alpha =\frac{\langle \psi _{-}|\hat{W}|\psi _{+}\rangle}{E_{+}-E_{-}}
\end{equation}
is the mixing coefficient of the opposite parity states,
$K=|{\bf I}\cdot {\bf n}|$ is the absolute value of the 
projection of the nuclear spin ${\bf I}$ on the nuclear axis, 
$M=I_{z}$, and $\hat{W}$ is the effective single-particle potential 
(\ref{w}).
The Schiff moment in the laboratory frame is
\begin{equation}
S_{z}=S_{\rm intr}\langle n_{z}\rangle =S_{\rm intr}\frac{2\alpha KM}{I(I+1)} 
\ .
\end{equation}

In the ``frozen'' body frame the surface of an axially 
symmetric deformed nucleus is described by the following expression 
\begin{equation}
\label{defsurf}
R(\theta)=R_{N}\big( 1+\sum_{l=1}\beta_{l}Y_{l0}(\theta)\big) \ .
\end{equation}
To keep the center-of-mass at $r=0$ we have to fix $\beta _{1}$ 
\cite{bm}:
\begin{equation}
\label{beta1}
\beta_{1}=-3\sqrt{\frac{3}{4\pi}}\sum_{l=2}\frac{(l+1)\beta_{l}\beta_{l+1}}
{\sqrt{(2l+1)(2l+3)}} \ .
\end{equation}
Assuming that the distributions of the protons and neutrons are the same, 
the electric dipole moment $e\langle {\bf r}\rangle=0$
(since the center-of-mass of the 
charge distribution coincides with the center-of-mass)
and hence there is no screening contribution to the Schiff moment. 
We also assume constant density for $R<R(\theta)$.
The intrinsic Schiff moment $S_{\rm intr}$ is then \cite{AFS,SAF}
\begin{equation}
\label{intrschiff}
S_{\rm intr}=eZR_{N}^{3}\frac{3}{20\pi}\sum_{l=2}
\frac{(l+1)\beta _{l}\beta _{l+1}}{\sqrt{(2l+1)(2l+3)}}
\approx eZR_{N}^{3}\frac{9\beta _{2}\beta_{3}}{20\pi \sqrt{35}} \ ,
\end{equation}
where the major contribution comes from $\beta_{2}\beta_{3}$, the 
product of the quadrupole $\beta_{2}\sim 0.1$ and octupole $\beta _{3}\sim 0.1$ 
deformations. 
The estimate of the Schiff moment in the laboratory frame gives~\cite{SAF}
\begin{equation}
\label{snum}
S\sim \alpha S_{\rm intr}\sim 0.05 ~e \beta_{2}\beta_{3}^{2}ZA^{2/3}\eta 
r_{0}^{3}\frac{\rm eV}{E_{+}-E_{-}}\sim 700 \times 
10^{-8}~\eta e {\rm fm}^{3} \ ,
\end{equation}
where $r_{0}\approx 1.2~{\rm fm}$ is the internucleon distance, 
$E_{+}-E_{-}\sim 50~{\rm keV}$. This estimate (\ref{snum}) is about 
$500$ times larger than the Schiff moment of a spherical nucleus 
like Hg (see Eq. (\ref{eq:SHg})). 
%
%give values of \beta _{2} and \beta _{3}
%
%

See Table \ref{tab:deformedschiff} for calculations \cite{SAF} of Schiff moments 
in nuclei assuming static octupole deformation. 
An attractive candidate for $P,T$-odd studies is radium, and recently several 
laboratories around the world have considered performing EDM experiments with it. 
As well as the possibility for a large Schiff moment, the atomic 
EDM is large due to high Z. And if measurements can be performed for metastable atomic states, 
further enhancement can occur due to the presence of close opposite parity levels 
\cite{flambaum1999,DFG2000}. A Woods-Saxon calculation for the Schiff moment 
of $^{225}{\rm Ra}$ gives \cite{SAF}
\begin{equation}
\label{eq:SRa}
S(^{225}{\rm Ra})=300\times 10^{-8}~\eta _{np}~e~{\rm fm}^{3} \ .
\end{equation}
Recently, a self-consistent calculation of the nuclear Schiff moment of $^{225}$Ra was 
performed, with core polarization taken into account \cite{engel2003}, 
\begin{equation}
\label{eq:SRaE}
S(^{225}{\rm Ra})=-5.06~g\bar{g}_{0}+10.4~g\bar{g}_{1}-10.1~g\bar{g}_{2}~e~{\rm fm}^{3} \ .
\end{equation} 
This calculation was carried out in the approximation of the zero-range $P,T$-odd interaction. 
It was found that the Schiff moment in the rotating frame is up to 
twice as large as the value calculated in \cite{SAF} (see Table \ref{tab:deformedschiff}). 
However, the Schiff moment 
in the laboratory frame was found to be suppressed by between 1.5 and 3 times due 
to suppression of the matrix element of the $P,T$-odd interaction. 
[Comparison between Eqs. (\ref{eq:SRa},\ref{eq:SRaE}) can be made using the relation following 
Eq. (\ref{eq:SHgDS}); Eq. (\ref{eq:SRa}) then gives 
$S(^{225}{\rm Ra})\sim 20~g(\bar{g}_{0}+\bar{g}_{1}-2\bar{g}_{2})~e~{\rm fm}^{3}$.]
Improved calculations (taking into account the finite-range of the $P,T$-odd interaction) 
are in progress \cite{engel2003}.
It is seen by comparison with the calculations for $^{199}$Hg (\ref{eq:SHg},\ref{eq:SHgDS}) 
that the radium Schiff moment is several hundred times larger. 

Note that $S$ in Eq. (\ref{snum}) is proportional 
to the squared octupole deformation parameter $\beta_{3}^{2}$. 
In Ref. \cite{EFH00} it was pointed out that in nuclei with a soft octupole vibration 
mode $\langle \beta _{3}^{2}\rangle \sim (0.1)^{2}$, i.e., 
the Schiff moments induced in nuclei with a soft octupole vibration mode are of 
the same magnitude as those induced in nuclei with static octupole deformation.
In a recent work \cite{FZ02} Schiff moments of nuclei with soft octupole vibrations 
were calculated and found to have a similar enhancement as in the static case.
This means that a number of heavy nuclei can have large collective Schiff moments.  

The effect of static octupole deformation on the size of the single-particle 
MQM was considered in \cite{FMO}. It was found that generally there is no 
significant enhancement due to this mechanism. 
%However, it was pointed out that if there 
%is octupole deformation in the nucleus $^{229}$Pa then this mechanism would 
%produce a single-particle MQM of the same order of magnitude as the collective 
%MQM due to the spin-hedgehog mechanism. 

There are several experiments in preparation aimed to detect EDMs of heavy atoms 
with deformed nuclei; see Section \ref{section7iv}. 
%Rn, Ra

%
%
%\subsection{The size of atomic EDMs: comparison of Schiff and magnetic quadrupole moments}
%

\subsection{Calculations of atomic EDMs induced by $P,T$-violating nuclear moments; 
intepretation of the Hg measurement in terms of hadronic parameters} 
%\label{section7iii} 
\label{ss:nmatomic}

Results of atomic calculations of EDMs (of current interest) induced by 
nuclear Schiff moments are presented in Table \ref{tab:atomicEDMsnm}. 
Of particular interest is the 
calculation for $^{199}$Hg. A recent calculation for the atomic EDM induced 
by the Schiff moment yielded the result
\cite{atomicedms}
\begin{equation}
\label{eq:edmS}
d(^{199}{\rm Hg})=-2.8\times 10^{-17}
\Big( \frac{S(^{199}{\rm Hg})}{e~{\rm fm}^{3}}\Big)~e~{\rm cm} \ .
\end{equation}
This value was obtained using the new finite-size form for the Schiff potential 
(\ref{phigen}) and is the average of two calculations: 
one performed in the potential $\hat{V}^{N}$ with core 
polarization taken into account using the TDHF method and the other 
performed in the $\hat{V}^{N-2}$ potential using the combined 
MBPT+CI method. 
The error of the result (\ref{eq:edmS}) is about 20\%.
The previous value for $d(^{199}{\rm Hg})$ induced by the nuclear Schiff moment, 
$d(^{199}{\rm Hg})=-4$ (in the same units as Eq. (\ref{eq:edmS})), 
was estimated in Refs. \cite{flambaum85,FKS86} from an atomic calculation 
\cite{martpend85} of the EDM induced by the tensor electron-nucleon interaction. 

From (\ref{eq:hgedm},\ref{eq:edmS}), the best limit on the Schiff moment follows,
\begin{equation}
\label{eq:schifflim}
S(^{199}{\rm Hg})=(3.8\pm 1.8\pm 1.4)\times 10^{-12}~e~{\rm fm}^{3} \ .
\end{equation}
As we have discussed, the Schiff moment can be induced from a number of
$P,T$-violating mechanisms:
due to a $P,T$-violating nucleon-nucleon interaction 
or due to a static EDM of an unpaired nucleon. 
The limit on $\eta _{np}$ from Eqs. (\ref{eq:schifflim},\ref{eq:SHg}) is
\begin{equation}
\label{eq:etanplimit}
\eta _{np}=-(2.7\pm 1.3\pm 1.0)\times 10^{-4}\ .
\end{equation}
%
%what about many-body corrections?
%
The limits on neutron and proton EDMs are [Eqs. (\ref{eq:schifflim},\ref{eq:S(Hg,d)})]
\begin{eqnarray}
d_{n}&=&(1.7\pm 0.8 \pm 0.6)\times 10^{-25}~e~{\rm cm} \label{eq:neutronHg}\\
d_{p}&=&(17\pm 8 \pm 6)\times 10^{-25}~e~{\rm cm} \label{eq:protonHg}\ .
\end{eqnarray}
Using instead the recent calculation (\ref{eq:S(Hg,d)RPA}) will change the limits 
on $d_{n}$ and $d_{p}$ only slightly. 
The result (\ref{eq:SHgDS}) suggests that the limits on the 
isoscalar and isotensor $P,T$-odd couplings may be substantially weaker than 
those that would follow from Eq. (\ref{eq:etanplimit}) using Eq. (\ref{eq:SHg}).

%d_{n}= (17\pm 8\pm 6)\times 10^{-26}~e~{\rm cm}\ .
%\end{equation}
%The limit on the proton EDM from (\ref{eq:schifflim}) is
%\begin{equation}
%d_{p}=(1.7\pm 0.8\pm 0.6)\times 10^{-24}~e~{\rm cm} \ .
%\end{equation}

%Limits on the parameters of the electron-nucleon interactions 
%and on the electron EDM from the Hg experiment have already been 
%considered in Sections \ref{sss:e-n} and \ref{sss:electron}. 
%
%A summary of these limits and the corresponding limits at the more 
%fundamental level, with a comparison of the best limits from 
%other measurements, is presented in Table \ref{tab:limits}, Section 
%\ref{section7iv}.

\section{Current limits on fundamental $P,T$-violating parameters and prospects for 
improvement} 
\label{section7iv}

\subsection{Summary of limits}

EDM measurements have already excluded several models of $CP$-violation and 
the parameter space of currently popular models is strongly constrained. 
In this review we will not discuss the sensitivities of the various EDM measurements 
to different $CP$-violating models; for such an analysis, 
see, e.g., the reviews \cite{benreutherreview,barrreview,comminsreview} and the 
book \cite{cpoddbook}. 
The problem is that in new theories there are many free parameters, 
and this gives a whole range of possible values for $CP$-violating effects.
Here we compare the sensitivities of different measurements to fundamental 
$CP$-violating interactions by placing limits on phenomenological parameters.

In Tables \ref{tab:limitsI},\ref{tab:limitsII} we present the best limits on fundamental 
$P,T$-violating parameters extracted from EDM measurements in atoms, molecules, and neutrons. 
In previous sections we presented limits on hadronic and semi-leptonic $CP$-violating 
parameters at the nucleon scale 
[nucleon EDMs, Eqs. (\ref{eq:neutronlimits},\ref{eq:neutronHg},\ref{eq:protonHg}); 
nucleon-nucleon interaction, Eqs. (\ref{eq:etanplimit}); electron-nucleon interaction, 
Eqs. (\ref{eq:limitCsp},\ref{eq:limitCt},\ref{eq:limitCps})] and in the leptonic sector 
directly on the electron EDM [Eq. (\ref{eq:electronedmlimit})].
These are summarized in Table \ref{tab:limitsI}. 
Here we constrain $CP$-violation at the quark scale from limits at the nucleon scale. 
Fig. \ref{fig:flowdiagram} shows which of these parameters are related.

We will begin with a discussion of limits on $CP$-violating parameters in the 
hadronic sector. Currently, the best limits on the neutron EDM come from direct neutron 
measurements (\ref{eq:neutronlimits}), and the limit from mercury is not far behind 
(\ref{eq:neutronHg}). The best proton EDM limit comes from mercury (\ref{eq:protonHg}). 
Constraints on nucleon EDMs constrain the following parameters at the 
quark level: quark EDMs, chromoelectric dipole moments (CEDMs) 
[CEDMs are analogous to quark EDMs, though with the external electromagnetic field replaced 
by a gluonic field], 
$P,T$-odd quark-quark four-fermion interactions, 
the QCD phase defining the strength of the term $G\tilde{G}$, 
and another parameter defining the strength of the term $GG\tilde{G}$ 
[in all models of CP violation considered this 
parameter is not the dominating mechanism inducing neutron EDMs \cite{bigi1991}, 
so we will not discuss it further]. 
Limits on nucleon EDMs can also constrain $P,T$-odd lepton-quark four-fermion interactions 
which induce quark EDMs at the one-loop level \cite{HMcK1997}; however, the limits 
involving light quarks are weaker than those obtained at tree-level from 
atomic measurements.

We use the following calculations to relate the nucleon and quark parameters, with 
the corresponding limits presented in Table \ref{tab:limitsII}. 
The limit on the QCD $\theta$ term is arrived at using the relation obtained in 
Ref. \cite{pospelov99}, $d_{n}=1.2\times 10^{-16}\bar{\theta}e~{\rm cm}$. 
%This gives a limit also on the $P,T$-odd pion-nucleon coupling constant, 
%$\bar{g}_{\pi NN}\approx -0.027\bar{\theta}$. 
To relate $d_{n}$ to the quark CEDMs $\tilde{d}_{q}$ and quark EDMs $d_{q}$ 
we use the recent calculation \cite{pospelov01}, 
$d_{n}=(1\pm 0.5)[0.55e(\tilde{d}_{d}+0.5\tilde{d}_{u})+0.7(d_{d}-0.25d_{u})]$.  
For the $P,T$-odd quark-quark interactions we use the phenomenological parameters 
\cite{KKY1988,KK1992,cpoddbook}
%check rel.
defined by the Hamiltonian 
$\hat{h}_{qq}= 
\frac{G}{\sqrt{2}}[k_{s}(\bar{q}_{1}i\gamma _{5}q_{1})(\bar{q}_{2}q_{2}) + 
k_{s}^{c}(\bar{q}_{1}i\gamma _{5}t^{a}q_{1})(\bar{q}_{2}t^{a}q_{2}) + 
k_{t}\frac{1}{2}\epsilon _{\mu \nu \alpha \beta}(\bar{u}\sigma _{\mu \nu}u)
(\bar{d}\sigma _{\alpha \beta}d) + 
k_{t}^{c}\frac{1}{2}\epsilon _{\mu \nu \alpha \beta}(\bar{u}\sigma _{\mu \nu}t^{a}u)
(\bar{d}\sigma _{\alpha \beta}t^{a}d)]$, 
where $t^{a}$ are the $SU(3)$ generators and the quarks $q_{1},q_{2}=u,d$.

From the limit on the $P,T$-odd nucleon-nucleon interaction $\eta$ from mercury follows 
the best limit on the $P,T$-odd pion-nucleon coupling $\bar{g}_{\pi NN}$; they are 
related through $\eta =g_{\pi NN}\bar{g}_{\pi NN}\sqrt{2}/(Gm_{\pi}^{2})$, 
where the strong coupling $g_{\pi NN}=13.6$.
%\cite{}
The relation $|\bar{g}_{\pi NN}|\approx 0.027|\bar{\theta}|$ (Ref. \cite{crewther79}) 
is used to place a constraint 
on the QCD $\theta$ term, tighter than those obtained from direct neutron measurements. 
%what about $\bar{\theta}$ ????????!!!!!!!!!
Limits on CEDMs of quarks can be obtained from the limit on $\bar{g}_{\pi NN}$. 
We use the calculation of Ref. \cite{pospelov2002}, 
%$\bar{g}^{(1)}_{\pi NN}
$\bar{g}_{\pi NN}=2(\tilde{d}_{u}-\tilde{d}_{d})/(10^{-14}{\rm cm})$. 
Again, this limit is better than those from the neutron. 
Constraints on the phenomenological $P,T$-odd quark-quark interactions defined above 
are obtained from the constraints on $\bar{g}_{\pi NN}$, with the exception of 
$k_{s}^{c}$ with $q_{1}\neq q_{2}$ and $k^{c}_{t}$ for which 
the $P,T$-odd neutral pion-nucleon vertex is insensitive \cite{cpoddbook}.

Limits on the $P,T$-odd electron-nucleon parameters can be broken down into limits 
on phenomenological $P,T$-odd electron-quark parameters $k_{1q}$, $k_{2q}$, $k_{3q}$ defined 
according to $\hat{h}_{eq}=\frac{G}{\sqrt{2}}[k_{1q}\bar{q}q\bar{e}i\gamma_{5}e 
+k_{2q}\frac{1}{2}\epsilon _{\kappa \lambda \mu \nu} \bar{q}\sigma _{\kappa \lambda}q 
\bar{e}\sigma _{\mu \nu}e +k_{3q}\bar{q}i\gamma_{5}q\bar{e}e]$; see Ref. \cite{cpoddbook}. 

Until this point we have not mentioned the $P,T$-odd electron-electron interaction, 
$\hat{h}_{ee}=\frac{G}{\sqrt{2}}k_{s}^{e}(\bar{e}i\gamma _{5}e)(\bar{e}e)$.  
This interaction induces atomic EDMs smaller than those induced by the 
electron-nucleon interaction. It does not benefit from two enhancement factors 
present in the latter interaction: the relativistic factor arising from the outer 
electrons being in the nuclear vicinity; and the interaction with $A$ nucleons 
(rather than two $K$-shell electrons in the former interaction); 
see Ref. \cite{cpoddbook}.
  
The limit on the electron EDM can be reduced to limits on the $P,T$-odd electron-quark 
interaction \cite{HMcK1997} and on the electron-electron interaction \cite{cpoddbook} 
that are competitive with those obtained at tree level. However, we do not include 
these into our table. 

In relating the $P,T$-odd parameters at the nucleon and quark levels we used the most 
recent calculations available. For references to other calculations, please see, e.g., 
the review \cite{barrreview} and book \cite{cpoddbook}.

\subsection{Ongoing/future EDM experiments in atoms, solids, and diatomic molecules}

There is a new generation of experiments in preparation aimed to measure EDMs of ions 
in solids (gadolinium gallium garnet and gadolinium iron garnet) \cite{lam02,hunter01}. 
Here it is expected that the sensitivity to $T$-violating effects (in particular, the 
electron EDM) will be improved by several orders of magnitude. 
The solid-state nature of the problem complicates the 
calculations required for interpretation of the measurements. 
A series of calculations have been performed in \cite{sushkovsolids}. 

One of the primary goals of the TRI$\mu$P facility under construction at Groningen 
is to measure permanent EDMs of radioactive 
atoms and ions, in particular Ra (see, e.g., \cite{trimup}). Groups at Yale, Argonne, 
and Los Alamos are also considering performing EDM experiments with Ra and Rn. 
These atoms can have very large EDMs due to their high $Z$ and the presence of 
nuclear (static/vibrational) octupole deformation (Section \ref{sss:octdef}). 
Also, metastable atomic states of Ra have close levels of opposite parity, and this 
can be exploited to obtain an enhanced EDM effect \cite{flambaum1999,DFG2000}. 

EDM experiments with both paramagnetic and diamagnetic diatomic molecules are underway. 
The first EDM experiment with paramagnetic molecules (YbF) was performed recently 
at the University of Sussex \cite{hudson2002}, and while the result gave a looser bound 
on the electron EDM than the Tl experiment \cite{tledm02}, a substantial improvement in 
the result is expected. 
Experiments with PbO excited to the metastable $a(1)$ state have begun at 
Yale \cite{demille2000}.

\section{Concluding remarks}
\label{conclusion}

Exciting developments in violations of fundamental symmetries are 
expected in the next few years. Improved EDM measurements underway, 
including the new generation of experiments in solids, diatomic 
molecules, and radioactive atoms, are expected to yield limits on 
electric dipole moments that are several orders of magnitude better than 
the current ones. Or perhaps an EDM will be unambiguously detected? 
Popular models such as supersymmetry will be put to the test. 
Improved precision tests of parity violation in 
atoms in a single isotope and in a chain of isotopes will provide crucial 
tests of physics beyond the standard model complementary to each other and 
to other electroweak tests. 
New measurements of the nuclear anapole moment are anticipated, and 
they will have important consequences for the theory of 
parity violating nuclear forces.

\acknowledgments

We would like to thank V.A. Dzuba and M.Yu. Kuchiev for useful discussions. 
Some of this work was carried out at the National Institute for 
Nuclear Theory, University of Washington, Seattle; we thank them for 
support and kind hospitality. 
This work was supported by the Australian Research Council.

\label{bibliography}

%---------
%
% summary of PNC calculations
%

%\bibitem{NSK1976}
%
%V.N. Novikov, O.P. Sushkov, and I.B. Khriplovich, 
%Zh. Eksp. Teor. Fiz. {\bf 71}, 1665 (1976) 
%[Sov. Phys. JETP {\bf 44}, 872 (1976)].
%
%\bibitem{HW1976}
%
%E.M. Henley and L. Wilets, Phys. Rev. A {\bf 14}, 1411 (1976).
%
%\bibitem{HKW1977}
%
%E.M. Henley, M. Klapisch, and L. Wilets, Phys. Rev. Lett. {\bf 39}, 994 (1977).
%
%\bibitem{SFK1976}
%
%O.P. Sushkov, V.V. Flambaum, and I.B. Khriplovich, 
%Zh. Eksp. Teor. Fiz. Pis'ma {\bf 24}, 502 (1976) 
%[Sov. Phys. JETP Lett. {\bf 24}, 461 (1976)].
%
%\bibitem{NC1977}
%
%D.V. Neuffer and E.D. Commins, Phys. Rev. A {\bf 16}, 844 (1977).
%
%\bibitem{das1982}
%
%B.P. Das, J. Andriessen, M. Vajed-Samii, S.N. Ray, and T.P. Das, 
%Phys. Rev. Lett. {\bf 49}, 32 (1982).
%
%\bibitem{hartleytl}
%
%A.C. Hartley and P.G.H. Sandars, J. Phys. B {\bf 23}, 4197 (1990).

\begin{table}
\caption{Results of atomic PNC experiments measured to better than 5\%. 
Results of optical rotation experiments are given in terms of ${\rm Im}(E_{PNC}/M1)$; 
Stark-PNC experiments are given in terms of ${\rm Im}(E_{PNC}/\beta)$.}
\label{tab:pncexp<5}
\begin{tabular}{lllllll}
Atom & Transition & Group & Year & Ref. & \multicolumn{2}{c}{Measurement}\\ 
&&&&& ${\rm Im}\big( E_{PNC}/M1\big)$ & ${\rm Im}\big( E_{PNC}/\beta \big)$ \\
&&&&& [$10^{-8}$] & [mV/cm] \\
\hline
$^{209}$Bi & $^{4}S_{3/2}-{^{2}D}_{3/2}$ & Oxford & 1991 & \cite{macpherson91} & $-10.12(20)$ & \\
$^{208}$Pb & $^{3}P_{0}-{^{3}P}_{1}$ & Seattle & 1993 & \cite{meekhof93} & $-9.86(12)$ & \\
& & Oxford & 1996 & \cite{phipp96} & $-9.80(33)$ & \\
$^{205}$Tl & $6P_{1/2}-6P_{3/2}$ & Oxford & 1995 & \cite{edwards95} & $-15.68(45)$ & \\
& & Seattle & 1995 & \cite{vetter95} & $-14.68(17)$ & \\
$^{133}$Cs & $6S_{1/2}-7S_{1/2}$ & Boulder & 1988 & \cite{noecker88} & & $-1.576(34)$ \\
 & & Boulder & 1997 & \cite{wieman} & & $-1.5935(56)$ \\
\end{tabular}
\end{table}
%****************************************************************
\begin{table}
\caption{Most precise calculations of PNC amplitudes $E_{PNC}$ for atoms and 
transitions listed in Table \ref{tab:pncexp<5}. Units: $10^{-11}iea_{B}(-Q_{W}/N)$.}
\label{tab:pnccalcs}
\begin{tabular}{llll}
Atom & Transition & $E_{PNC}$\tablenotemark[1] & Ref. \\
\hline
$^{209}$Bi & ${^{4}S}_{3/2}-{^{2}D}_{3/2}$ & $28(3)$ & \cite{DFSS88,DFS1989pnc} \\ 
%   & ${^{4}S}_{3/2}-{^{2}D}_{5/2}$ & $4(3)$ & \cite{DFSS88,DFS1989pnc} \\
$^{208}$Pb & ${^{3}P}_{0}-{^{3}P}_{1}$ & $30(2)$ & \cite{DFSS88} \\
$^{205}$Tl 
%& $6P_{1/2}-7P_{1/2}$ & 7.9(5) & \cite{DFSS1987b} \\
   & $6P_{1/2}-6P_{3/2}$ & 27.0(8) & \cite{DFSS1987b} \\
   &                     & 27.2(7) & \cite{kozlovtl} \\
$^{133}$Cs & $6S-7S$ & $0.904(5)$ & \cite{DFG02} \\
\end{tabular}
\tablenotetext[1]{The values for the PNC amplitudes for Cs \cite{DFG02} 
and for Tl $6P_{1/2}-6P_{3/2}$ \cite{kozlovtl} include corrections beyond 
the other calculations. In particular, for Cs the contributions of the 
Breit interaction and vacuum polarization due to the strong nuclear 
Coulomb field are included. 
%, and the neutron distribution are included. 
For Tl the Breit interaction is also included. 
The remaining corrections for Cs and Tl are discussed in detail in Sections 
\ref{section3iib},\ref{section3iic}, respectively. 
These corrections would be inside the error bars for the other atoms and 
transitions in the table.}
\end{table}
%****************************************************************
\begin{table}
\caption{Summary of experimental results for PNC in cesium $6S-7S$, 
$-{\rm Im}(E_{PNC})/\beta$; units: ${\rm mV}/{\rm cm}$.}
\label{tab:csexps}
\begin{tabular}{llll}
Group & Year & Ref. & Value \\
\hline
%Paris & 1982 & \cite{bouchiat82} & $1.34(22)(11)$ \\
%Paris & 1984 & \cite{bouchiat84} & 1.78(26)(12) \\
Paris & 1982,1984 & \cite{bouchiat82,bouchiat84,bouchiat85} & 1.52(18) \\
Boulder & 1985 & \cite{gilbert85} & $1.65(13)$ \\
%Paris & 1985,1986 & \cite{bouchiat85} & $1.52(18)$ \\
Boulder & 1988 & \cite{noecker88} & $1.576(34)$ \\
Boulder & 1997 & \cite{wieman} & $1.5935(56)$ \\
Paris & 2003 & \cite{guena2003} & $1.752(147)$ \\
\end{tabular}
\end{table}
%*************************************************************************
\begin{table}
\caption{Summary of calculations of the PNC E1 amplitude for the 
cesium $6S-7S$ transition; units are $10^{-11}iea_{B}(-Q_{W}/N)$.}
\label{tab:cscalcs}
\begin{tabular}{llll}
Authors & Year & Ref. & Value \\
\hline
%
%CHECK THESE!!!
%
Bouchiat, Bouchiat & 1974,1975& \cite{bouchiats1,bouchiats2}\tablenotemark[1] & $1.33$ \\
Loving, Sandars & 1975 & \cite{calc2}\tablenotemark[1] & $1.15$ \\
Neuffer, Commins & 1977 & \cite{calc3}\tablenotemark[1] & $1.00$ \\
%
%check number above
%
Kuchiev, Sheinerman, Yahontov & 1981 & \cite{calc4}\tablenotemark[1] & $0.75$ \\
Das & 1981 & \cite{calc5}\tablenotemark[2] & $1.06$ \\
Bouchiat, Piketty, Pignon & 1983 & \cite{calc6}\tablenotemark[1] & $0.97(10)$ \\
Dzuba, Flambaum, Silvestrov, Sushkov & 1984,1985 & \cite{calc7,DFSS1985}\tablenotemark[2] & $0.88(3)$ \\
Sch\"afer, M\"uller, Greiner, Johnson & 1984 & \cite{calc8}\tablenotemark[2] & $0.74$ \\
M{\aa}rtensson-Pendrill & 1985 & \cite{M-P1985jp}\tablenotemark[2] & $0.886$ \\
Plummer, Grant & 1985 & \cite{calc10}\tablenotemark[2] & $0.64$ \\
%
%???????????????????????? or is the above number $0.71$, or........? \\
%
Sch\"afer, M\"uller, Greiner & 1985 & \cite{calc11}\tablenotemark[2] & $0.92$ \\
%cite Johnson '85 work with value below???
Johnson, Guo, Idrees, Sapirstein & 1985,1986 & \cite{johnson8586}\tablenotemark[3] & $0.754,0.876,0.856$\\
Johnson, Guo, Idrees, Sapirstein & 1985,1986 & \cite{johnson8586}\tablenotemark[2] & $0.890$\\
Bouchiat, Piketty & 1986 & \cite{calc13}\tablenotemark[1] & $0.935(20)(30)$ \\
Dzuba, Flambaum, Silvestrov, Sushkov & 1987 & \cite{DFSS1987b}\tablenotemark[2] & $0.90(2)$ \\
Johnson, Blundell, Liu, Sapirstein & 1988 & \cite{calc15}\tablenotemark[2] & $0.95(5)$ \\
Parpia, Perger, Das & 1988 & \cite{parpia1988}\tablenotemark[3] & $0.879$ \\
Dzuba, Flambaum, Sushkov & 1989 & \cite{DFS1989pnc}\tablenotemark[2] & $0.908(9)$\\
Hartley, Sandars & 1990 & \cite{hartley1990a}\tablenotemark[3] & $0.904(18)$\\
Hartley, Lindroth, M{\aa}rtensson-Pendrill & 1990 & \cite{hartley1990b}\tablenotemark[2] & $0.933(37)$\\
Blundell, Johnson, Sapirstein & 1990,1992 & \cite{BJS1990}\tablenotemark[2] & $0.905(9)$ \\
Safronova, Johnson & 2000 & \cite{safronova2000}\tablenotemark[2] & $0.909(11)$ \\
Kozlov, Porsev, Tupitsyn & 2001 & \cite{kozlovcs}\tablenotemark[2]\tablenotemark[4] 
& $0.901(9)$ \\
Dzuba, Flambaum, Ginges & 2002 & \cite{DFG02}\tablenotemark[2]\tablenotemark[4] 
& 0.904(5) \\
\end{tabular}
\tablenotetext[1]{Semi-empirical calculations.}
\tablenotetext[2]{Ab initio many-body calculations.}
\tablenotetext[3]{Combined many-body and semi-empirical calculations.}
\tablenotetext[4]{The difference between the values of \cite{kozlovcs,DFG02} and 
previous ones is due to the inclusion of the Breit interaction in \cite{kozlovcs} 
and the Breit and strong field vacuum polarization in \cite{DFG02}.}
\end{table}
%****************************************************************
\begin{table}
\caption{Removal energies for Cs in units cm$^{-1}$.}
\label{tab:energies} 
\begin{tabular}{lllll}
State & RHF & $\hat{\Sigma} ^{(2)}$ & $\hat{\Sigma}$ & 
Experiment \tablenotemark[1]\\
\hline
$6S$ & 27954 & 32415 & 31492 & 31407 \\
$7S$ & 12112 & 13070 & 12893 & 12871 \\
$6P_{1/2}$ & 18790 & 20539 & 20280 & 20228 \\
$7P_{1/2}$ & 9223 & 9731 & 9663 & 9641 \\
\end{tabular}
\tablenotetext[1]{Taken from \cite{moore}.}
\end{table}
%
%\begin{table}
%\caption{Calculated removal energies, in units ${\rm cm}^{-1}$, 
%for low-lying states of Cs and percentage deviations (in brackets) from experiment.}
%\label{tab:energies} 
%\begin{tabular}{lrdrdrdr}
%State & \multicolumn{2}{c}{RHF} & \multicolumn{2}{c}{$\hat{\Sigma}^{(2)}$} & 
%\multicolumn{2}{c}{$\hat{\Sigma}$} & Experiment\tablenotemark[1]\\
%\hline
%$6S$ & 27954 & (-11.0) & 32415 & (3.2) & 31492 & (0.3) & 31407 \\
%$7S$ & 12112 & (-5.9) & 13070 & (1.5) & 12893 & (0.2) & 12871 \\
%$6P_{1/2}$ & 18790 & (-7.1) & 20539 & (1.5) & 20280 & (0.3) & 20228 \\
%$7P_{1/2}$ & 9223 & (-4.3) & 9731 & (0.9) & 9663 & (0.2) & 9641 \\
%\end{tabular}
%\tablenotetext[1]{Taken from \cite{moore}.}
%\end{table}
%*************************************************************************
%
\begin{table}
\caption{Radial integrals of E1 transition amplitudes for Cs in 
different approximations. The experimental values are listed in 
the last column. (a.u.)}
\label{tab:e1}
\begin{tabular}{ldddddl}
Transition & RHF& TDHF & $\hat{\Sigma} ^{(2)}$ & 
$\hat{\Sigma}$ & $\hat{\Sigma}$ & Experiment \\ 
       &      &              & with fitting &           & with fitting & 
 \\ 
\hline
$6S-6P_{1/2}$ & 6.464 & 6.093 & 5.499 & 5.497 & 5.509 & 
       5.5232(91)\tablenotemark[1], 5.4979(80)\tablenotemark[2],\\ 
&&&&&& 5.5192(58)\tablenotemark[3], 5.512(2)\tablenotemark[4], \\
&&&&&& 5.524(5)(1)\tablenotemark[5] \\
$7S-6P_{1/2}$ & 5.405 & 5.450 & 5.198 & 5.190 & 5.204 & 
5.185(27)\tablenotemark[6]\\
$7S-7P_{1/2}$ & 13.483 & 13.376 & 12.602 & 12.601 & 12.612 & 
12.625(18)\tablenotemark[7] \\
\end{tabular}
\tablenotetext[1]{Ref. \cite{young94}.}
\tablenotetext[2]{Ref. \cite{rafac99}.}
\tablenotetext[3]{Ref. \cite{DP02}. Deduced from the van der Waals coefficient $C_{6}$.}
\tablenotetext[4]{Ref. \cite{amiot2002}. Deduced from photoassociation spectroscopy.}
\tablenotetext[5]{Ref. \cite{amini03}. Deduced from their measurement of the static dipole polarizability.}
\tablenotetext[6]{Ref. \cite{bouchiat84e1}.}
\tablenotetext[7]{Ref. \cite{bennett99}.}
\end{table}
%
%\begin{table}
%\caption{Radial integrals of $E1$ transition amplitudes, in units $a_{B}$, for low-lying levels of Cs 
%in different approximations. Percentage deviations from experiment are indicated in brackets. }
%\label{tab:e1}
%\begin{tabular}{lllllll}
%Transition & RHF& TDHF & $\hat{\Sigma} ^{(2)}$ & 
%$\hat{\Sigma}$ & $\hat{\Sigma}$ & Experiment \\ 
%       &      &              & with fitting &           & with fitting & 
% \\ 
%\hline
%$6S-6P_{1/2}$ 
%& 17.0 & 10.3 & -0.4 & -0.5 & -0.3 & 5.5232(91)\tablenotemark[1]\\ 
%& 17.6 & 10.8 & 0.02 & -0.02  & 0.2 & 5.4979(80)\tablenotemark[2]\\ 
%& 17.1 & 10.4 & -0.4 & -0.4 & -0.2 & 5.5192(58)\tablenotemark[3]\\
%& 17.3 & 10.5 & -0.3 & -0.3 & -0.1 & 5.513(2)\tablenotemark[4]\\
%$7S-6P_{1/2}$ 
%& 4.2 & 5.1 & 0.3 & 0.1 & 0.4 & 5.185(27)\tablenotemark[5]\\
%$7S-7P_{1/2}$ 
%& 6.8 & 5.9 & -0.2 & -0.2 & -0.1 & 12.625(18)\tablenotemark[6] \\
%\end{tabular}
%\tablenotetext[1]{Ref. \cite{young94}.}
%\tablenotetext[2]{Ref. \cite{rafac99}.}
%\tablenotetext[3]{Ref. \cite{DP02}.}
%\tablenotetext[4]{Ref. \cite{amiot2002}}
%\tablenotetext[5]{Ref. \cite{bouchiat84}.}
%\tablenotetext[6]{Ref. \cite{bennett99}.}
%\end{table}
%************************************************************
\begin{table}
\caption{Calculations of the hyperfine structure of Cs 
in different approximations. In the last column the experimental 
values are listed. Units: MHz.}
\label{tab:hfs}
\begin{tabular}{ldddddd}
State & RHF & TDHF & $\hat{\Sigma} ^{(2)}$ & 
$\hat{\Sigma}$ & $\hat{\Sigma}$ & Experiment \\ 
       &    &     & with fitting &      & with fitting & \\ 
\hline 
$6S$ & 1425.0 & 1717.5 & 2306.9 & 2315.0 & 2300.3 & 
2298.2 \tablenotemark[1] \\
$7S$ & 391.6 & 471.1 & 544.4 & 545.3 & 543.8 & 
545.90(9) \tablenotemark[2] \\
$6P_{1/2}$ & 160.9 & 200.3 & 291.5 & 293.6 & 290.5 &
291.89(8) \tablenotemark[3] \\
$7P_{1/2}$ & 57.6 & 71.2 & 94.3 & 94.8 & 94.1 & 
94.35 \tablenotemark[1] \\
\end{tabular}
\tablenotetext[1]{Ref. \cite{arimondo77}.}
\tablenotetext[2]{Ref. \cite{gilbert83}.}
\tablenotetext[3]{Ref. \cite{rafac97}.}
\end{table}
%
%\begin{table}
%\caption{Percentage deviation from experiment of calculated 
%hyperfine structure constants in different approximations. 
%The experimental value is listed in the last column (MHz).}
%\label{tab:hfs}
%\begin{tabular}{ldddddl}
%State & RHF & TDHF & $\hat{\Sigma}^{(2)}$ & $\hat{\Sigma}$ & $\hat{\Sigma}$ & Experiment \\
% & & & with fitting    &          & with fitting & \\
%\hline
%$6S$       & -38.0 & -25.3 &  0.4 &  0.7 & 0.09 & 2298.2\tablenotemark[1] \\
%$7S$       & -28.3 & -13.7 & -0.3 & -0.1 & -0.4 & 545.90(9)\tablenotemark[2] \\
%$6P_{1/2}$ & -44.9 & -31.4 & -0.1 &  0.6 & -0.5 & 291.89(8)\tablenotemark[3] \\
%$7P_{1/2}$ & -39.0 & -24.5 & -0.05 & 0.5 & -0.3 & 94.35\tablenotemark[1] \\
%\end{tabular}
%\tablenotetext[1]{Ref. \cite{arimondo77}.}
%\tablenotetext[2]{Ref. \cite{gilbert83}.}
%\tablenotetext[3]{Ref. \cite{rafac97}.}
%\end{table}
%****************************************************************
\begin{table}
\caption{Contributions to the $6S-7S$ $E_{PNC}$ amplitude 
for Cs in units $10^{-11}iea_{B}(-Q_{W}/N)$. 
($\hat{\Sigma}$ corresponds to the (unfitted) ``dressed'' 
self-energy operator.)}
\label{tab:pnci}
\begin{tabular}{ld}
TDHF & 0.8898 \\
Brueckner-type correlations & \\
\qquad $\langle \psi _{7s}|\hat \Sigma_s(\epsilon_{7s})|\delta X_{6s}\rangle$ 
& 0.0773 \\
\qquad $\langle \delta\psi _{7s}|\hat \Sigma_p(\epsilon_{7s})|X_{6s}\rangle$ 
& 0.1799 \\
\qquad $\langle \delta Y_{7s}|\hat \Sigma_s(\epsilon_{6s})|\psi_{6s}\rangle$ 
& -0.0810 \\
\qquad $\langle Y_{7s}|\hat \Sigma_p(\epsilon_{6s})|\delta \psi_{6s}\rangle$ 
& -0.1369 \\
Nonlinear in $\hat{\Sigma}$ correction & -0.0214 \\
Weak correlation potential & 0.0038 \\
Structural radiation & 0.0029 \\
Normalization & -0.0066 \\
& \\
Subtotal & 0.9078 \\
& \\
Breit & -0.0055 \\ 
Neutron distribution correction & -0.0018 \\
QED radiative corrections & \\
\qquad Vacuum polarization (Uehling) & 0.0036 \\
\qquad Self-energy and vertex & -0.0072  \\
 & \\
Total & 0.8969 \\
\end{tabular}
\end{table}
%****************************************************************
\begin{table}
\caption{Values for $E_{PNC}$ in different approximations; 
units $10^{-11}iea_{B}(-Q_{W}/N)$.}
\label{tab:pncii}
\begin{tabular}{cddd}
 & $\hat{\Sigma}^{(2)}$ with fitting & $\hat{\Sigma}$ & 
$\hat{\Sigma}$ with fitting \\
\hline
$E_{PNC}$ & 0.901 & 0.904 & 0.903 \\
\end{tabular}
\end{table}
%****************************************************************
%
%	anapole moment
%
%	Cs
%
\begin{table}
\caption{Calculations of the anapole moment $\kappa _{a}$ for $^{133}$Cs.  
In the last column the value for the anapole moment constant $\kappa _{a}$ is 
presented, using DDH best values of the meson-nucleon couplings (see Table \ref{tab:DDH}).}
\label{tab:kappa_a(Cs)}
\begin{tabular}{llld}
 & Ref. & $\kappa _{a}\times 10^{2}$ & $\kappa _{a}(DDH)$ \\
\hline
Single- &\cite{FKS1984}\tablenotemark[1] & $8g_{p}$ & 0.36 \\
particle &\cite{FKS1984}\tablenotemark[2] & $6g_{p}=
(34f_{\pi}-9h_{\rho}^{0}-2h_{\rho}^{1}+1h_{\rho}^{2}-5h_{\omega}^{0} 
-5h_{\omega}^{1})\times 10^{2}$ & 0.27 \\
&\cite{DKTnp1994,DT1997,DT2000}\tablenotemark[3] & $4.9g_{p}+0.65g_{pn}$ & 0.26 \\
&\cite{BPzpc1991,BPplb1991}\tablenotemark[4] & $5.5g_{p}$ & 0.25 \\
Many- &\cite{HHM1989}\tablenotemark[5] & $23 \times 10^{2}f_{\pi}$ &  0.11 \\
%0.11 ????????
%also - check more recent contributions cited in 2002 paper
body & \cite{DT1997,DT2000}\tablenotemark[6] & $4.4g_{p}+0.45g_{n}+0.45g_{pn}-0.03g_{np}$ & 0.23 \\
& \cite{DT2000}\tablenotemark[7] & $2.9g_{p}+0.18g_{n}+0.36g_{pn}-0.02g_{np}$ & 0.15 \\  
& \cite{HLR-M2001}\tablenotemark[8] &
$(26.98f_{\pi} -7.01h_{\rho}^{0} -1.72h_{\rho}^{1}+ 0.16h_{\rho}^{2} 
-4.48h_{\omega}^{0}-2.16h_{\omega}^{1})\times 10^{2}$ & 0.21 \\
\end{tabular}
\tablenotetext[1]{Formulae (\ref{eq:kappa_ag_p2}) and (\ref{eq:g_p}).}
\tablenotetext[2]{Woods-Saxon potential with spin-orbit interaction, 
formulae (\ref{eq:kappa_ag_pCs}) and (\ref{eq:g_p}).}
\tablenotetext[3]{Woods-Saxon potential with spin-orbit interaction; 
the contact and spin-orbit currents are included. This calculation 
includes some contributions beyond the single-particle approximation.}
\tablenotetext[4]{Harmonic oscillator potential with spin-orbit interaction. 
Configuration mixing is taken into account semi-empirically.}
\tablenotetext[5]{Large-basis shell-model calculations; 
just the pion contribution was calculated.}
\tablenotetext[6]{Many-body effects taken into account in RPA.}
\tablenotetext[7]{Many-body effects taken into account in RPA. More 
complete treatment than Ref. \cite{DT1997}.}
\tablenotetext[8]{Large-basis shell-model calculations.}
\end{table}
%
%---------
%
\begin{table}
\caption{Nuclear calculations of $\kappa _{2}$ and atomic calculations of $\kappa _{Q}$ 
for $^{133}$Cs and $^{203,205}$Tl.}
\label{tab:kappa_2,Q}
\begin{tabular}{ldddd}
Ref. & \multicolumn{2}{c}{$^{133}$Cs} & \multicolumn{2}{c}{$^{203,205}$Tl}\\
& $\kappa _{2}$ & $\kappa _{Q}$ & $\kappa _{2}$ & $\kappa _{Q}$ \\
\hline
Eq. (\ref{eq:smcoeff})\tablenotemark[1] & -0.05 & - & -0.05 & - \\ 
\cite{FK85}\tablenotemark[2] & - & 0.017 & - & 0.014 \\ 
\cite{BPzpc1991,BPplb1991} & -0.038\tablenotemark[3] & 0.027 & -0.027\tablenotemark[3] & 0.022 \\ 
\cite{HLR-M2001}\tablenotemark[4] & -0.063 & - & -0.064 & - \\ 
\cite{JSS2003}\tablenotemark[5] & - & 0.017 & - & - \\ 
\end{tabular}
\tablenotetext[1]{Single-particle, standard model value with $\sin^{2}\theta _{W}=0.23$.}
\tablenotetext[2]{Eq. (\ref{eq:qwa2/3}).}
\tablenotetext[3]{Nuclear configuration mixing was taken into account semi-empirically.}
\tablenotetext[4]{Large-basis nuclear shell model calculations.}
\tablenotetext[5]{Atomic many-body calculations in third-order perturbation theory. 
Core polarization is included in RPA.}
\end{table}
%**************************
%
\begin{table}
\caption{DDH ``best values'' ($f_{\pi}$, $h$) in units $10^{-7}$ and effective coupling 
constants ($g$) at DDH values for the meson-nucleon couplings.}
\label{tab:DDH}
\begin{tabular}{ccccccccccc}
 $f_{\pi}$ & $h_{\rho}^{0}$ & $h_{\rho}^{1}$ & $h_{\rho}^{2}$ & $h_{\omega}^{0}$ & $h_{\omega}^{1}$
& $g_{pn}$ & $g_{np}$ & $g_{pp}=g_{nn}$ & $g_{p}$ & $g_{n}$ \\ 
\hline
4.6 & -11.4 & -0.19 & -9.5 & -1.9 & -1.1 
& 6.5 & -2.2 & 1.5 & 4.5 & 0.2 \\
\end{tabular}
\end{table}
%
%********************************
%	Tl
%
\begin{table}
\caption{Calculations of the anapole moment $\kappa _{a}$ for $^{203,205}$Tl.  
In the last column the value for the anapole moment constant $\kappa _{a}$ is 
presented, using DDH best values of the meson-nucleon couplings (see Table \ref{tab:DDH}).}
\label{tab:kappa_a(Tl)}
\begin{tabular}{llld}
 & Ref. & $\kappa _{a}\times 10^{2}$ & $\kappa _{a}(DDH)$ \\
\hline
Single- &\cite{FKS1984}\tablenotemark[1] & $11g_{p}$ &  0.48 \\
particle &\cite{FKS1984}\tablenotemark[2] & $10g_{p}=(56f_{\pi}-16h_{\rho}^{0}- 
4h_{\rho}^{1}+1h_{\rho}^{2}-9h_{\omega}^{0}-9h_{\omega}^{1})\times 10^{2}$ & 0.43 \\
&\cite{DKTnp1994,DT1997,DT2000}\tablenotemark[3] & $7.8g_{p}+0.85g_{pn}$ & 0.41 \\
&\cite{BPzpc1991,BPplb1991}\tablenotemark[4] & $6.0g_{p}$ & 0.27 \\
Many- & \cite{DT1997,DT2000}\tablenotemark[5] & $7.1g_{p}+0.35g_{n}+0.64g_{pn}-0.06g_{np}$ & 0.36 \\
body & \cite{AB1999}\tablenotemark[6] & $5.3g_{p}+0.4g_{n}$ & 0.24 \\
& \cite{DT2000}\tablenotemark[7] & $4.3g_{p}+0.1g_{n}+0.64g_{pn}-0.06g_{np}$ & 0.24 \\  
& \cite{HLR-M2001}\tablenotemark[8] &
$(13.98f_{\pi}-2.92h_{\rho}^{0}-0.26h_{\rho}^{1}+0.23h_{\rho}^{2}- 
2.14h_{\omega}^{0}-0.98h_{\omega}^{1})\times 10^{2}$ & 0.10 \\
\end{tabular}
\tablenotetext[1]{Formulae (\ref{eq:kappa_ag_p}) and (\ref{eq:g_p}).}
\tablenotetext[2]{Woods-Saxon potential with spin-orbit interaction. Eq. (\ref{eq:g_p}) 
is used to express the effective coupling constant $g_{p}$ in terms of parity violating 
weak meson-nucleon couplings.}
\tablenotetext[3]{Woods-Saxon potential with spin-orbit interaction; 
the contact and spin-orbit currents are included. This calculation 
includes some contributions beyond the single-particle approximation.}
\tablenotetext[4]{Harmonic oscillator potential with spin-orbit interaction. 
Configuration mixing was taken into account semi-empirically.}
\tablenotetext[5]{Many-body effects taken into account in RPA.}
\tablenotetext[6]{Large-basis shell-model calculations. Just the dominant 
spin-current contribution was considered.}
\tablenotetext[7]{Many-body effects taken into account in RPA. More complete 
treatment than Ref. \cite{DT1997}.}
\tablenotetext[8]{Large-basis shell-model calculations.}
\end{table}

%***************************************
% atomic EDMs
%
%**************************
\begin{table}
\caption{Calculations of atomic EDMs induced by $P,T$-odd electron-nucleon interaction.}
\label{tab:e-n}
\begin{tabular}{llll}
Atom & $d_{\rm atom}/C^{SP}$ & $d_{\rm atom}/C^{T}$ & $d_{\rm atom}/C^{PS}$ \\
     &   ($10^{-18}~e~{\rm cm}$) & ($10^{-20}~e~{\rm cm}$) & ($10^{-23}~e~{\rm cm}$) \\ 
\hline
Cs & $0.70$\tablenotemark[1]
%should sign be + or - ?????
 & $0.92$\tablenotemark[2] & $2.2$\tablenotemark[3] \\
   & $0.71$\tablenotemark[2] & & \\
   & $0.72$\tablenotemark[4] & & \\
Tl
 & $-5.1$\tablenotemark[2] & $0.5$\tablenotemark[2] & $1.5$\tablenotemark[3] \\
 & $-(7\pm 2)$\tablenotemark[4] 
%should sign be + or - ????????? 
 & \\
Xe & $-5.6\times 10^{-5}$ \tablenotemark[5] & $0.52$\tablenotemark[6] & $1.2$\tablenotemark[5] \\ 
   & $-4.4\times 10^{-5}$ \tablenotemark[7] & $0.41$\tablenotemark[8] & $0.95$\tablenotemark[7] \\ 
   & & $0.6$\tablenotemark[2] & \\
Hg & $-5.9\times 10^{-4}$ \tablenotemark[5] & $2.0$\tablenotemark[6] & $6.0$\tablenotemark[5] \\ 
   & & $1.3$\tablenotemark[2] & \\
%TlF
%YbF
%other?
\end{tabular}
\tablenotetext[1]{Ref. \cite{bouchiat1975}, semi-empirical calculation.}
\tablenotetext[2]{Ref. \cite{khriplovichpnc,cpoddbook}, simple analytical calculations.}
\tablenotetext[3]{Using formula (\ref{eq:PS<->T}) and calculations for 
$d_{\rm atom}/C^{T}$ from \cite{cpoddbook}.}
\tablenotetext[4]{Ref.\cite{M-PL1991}, TDHF calculation with correlation corrections 
estimated from the size of the corrections for the corresponding electron EDM 
enhancement factors in \cite{hartley1990b}.}
\tablenotetext[5]{Using formulae (\ref{eq:SP<->T}), (\ref{eq:PS<->T}) and the calculation 
$d_{\rm atom}/C^{T}$ from Ref. \cite{martpend85}.}
\tablenotetext[6]{Ref. \cite{martpend85}, TDHF calculations.}
\tablenotetext[7]{Using formulae (\ref{eq:SP<->T}), (\ref{eq:PS<->T}) and the calculation 
$d_{\rm atom}/C^{T}$ from Ref. \cite{dzuba85}.}
\tablenotetext[8]{Ref. \cite{dzuba85}, TDHF
%?
%cite also 86 work?
calculation.}
\end{table}
%
%++++++++++++++++++++++++++++++++
%
%should I cite semi-empirical and HF values for Johnson8586 ?????
%cite unshielded result of sandars1965???
%
\begin{table}
\caption{Enhancement factors for atoms of interest.}
\label{tab:enhancement}
\begin{tabular}{llll}
 & Atom & \multicolumn{2}{l}{Enhancement factor $K$} \\
 & & Semi-empirical & Ab initio \\
\hline
Paramagnetic & Rb & 24\tablenotemark[1], 
                    24.6\tablenotemark[2], 
                    16.1\tablenotemark[3], 
                    23.7\tablenotemark[3], 
                    22.0\tablenotemark[3]	& 24.6\tablenotemark[3] \\
&              Cs & 119\tablenotemark[1], 
                    131\tablenotemark[4],
%
%ignatovich semi-empirical???!!!
% 
		    138\tablenotemark[2],  
                    138\tablenotemark[5],
	            80.3\tablenotemark[3],  
		    106.0\tablenotemark[3], 
		    100.4\tablenotemark[3]      & 114.9\tablenotemark[3], 
						  114\tablenotemark[6],
						  115\tablenotemark[7]\\
& Fr &              1150\tablenotemark[1]       & 910(50)\tablenotemark[8]\\ %not really ab initio!
& Tl &              $-716$\tablenotemark[9], 
%is this supposed to be + or - ????!!!!!
                    $-500$\tablenotemark[5], 
		    $-502$\tablenotemark[3],
		    $-607$\tablenotemark[3],
		    $-562$\tablenotemark[3]      & $-1041$\tablenotemark[3],
  					          $-301$\tablenotemark[10],
      						  $-179$\tablenotemark[6],
      						  $-585$\tablenotemark[11]\\
& Xe~$^{3}P_{2}$ & 130\tablenotemark[12], 
                   120\tablenotemark[5]		& \\
Diamagnetic & Xe &                              & $-0.0008$\tablenotemark[13],
						$-0.0008$\tablenotemark[7] \\
& Hg &                                          & $-0.014$\tablenotemark[13],
						  $0.0116$\tablenotemark[7]\\ 
%MOLECULES
\end{tabular}
\tablenotetext[1]{Ref. \cite{sandars1965}.}
\tablenotetext[2]{Ref. \cite{sternheimer1969}.}
\tablenotetext[3]{Ref. \cite{johnson8586}.}
\tablenotetext[4]{Ref. \cite{ignatovich1969}.}
\tablenotetext[5]{Ref. \cite{flambaum76}.}
\tablenotetext[6]{Ref. \cite{hartley1990b}.}
\tablenotetext[7]{Ref. \cite{M-P1987}.}
\tablenotetext[8]{Ref. \cite{BDFM1999}.}
\tablenotetext[9]{Ref. \cite{sandars1975}.}
\tablenotetext[10]{Ref. \cite{kraftmakher1988edm}.}
\tablenotetext[11]{Ref. \cite{liu92}.}
\tablenotetext[12]{Ref. \cite{player1970}.}
\tablenotetext[13]{Ref. \cite{FK85}. Extracted from the 
many-body calculation of $d_{\rm atom}(C^{T})$ \cite{martpend85}; 
see Section \ref{sss:electron}.}
\end{table}

%*****************************
\begin{table}
\caption{Calculations of $P,T$-violating moments in spherical nuclei 
%????? octupole vibrations?
induced by the $P,T$-odd nucleon-nucleon interaction.}
\label{tab:nuclearmoments}
\begin{tabular}{lll}
Atom & $S(e~{\rm fm}^{3}\big)\times 10^{8}$ & 
$M\big(\frac{e}{m_{p}}~{\rm fm}\big)\times 10^{7}$ \\
\hline
$^{129}$Xe & $1.75\eta _{np}$\tablenotemark[1] & \\
$^{199}$Hg & $-1.4\eta _{np}$\tablenotemark[1] & \\
$^{133}$Cs & $3.0\eta_{p}$\tablenotemark[2] & $1.7\eta_{p}$\tablenotemark[2] \\
           &                            & $1.6\eta_{p}$, $2.6\eta_{p}$, $1.8\eta_{p}$\tablenotemark[3]
%check these numbers!!!
\\  
           && $-0.3\eta _{pp}-0.2\eta_{pn}-2.5\eta^{\prime}_{pp}+1.7\eta^{\prime}_{pn}$\tablenotemark[4] 
%check these numbers!!!!!!!
\\   
$^{203,205}$Tl & $-2\eta_{p}$\tablenotemark[2] & \\
               & $1.2\eta _{pp}-1.4\eta _{pn}$\tablenotemark[1]\tablenotemark[5] & \\
\end{tabular}
\tablenotetext[1]{Calculated using wave functions and Green's functions 
found in the Woods-Saxon potential with spin-orbit interaction, 
Refs. \cite{flambaum85,FKS86}.}
\tablenotetext[2]{Woods-Saxon potential with spin-orbit interaction, Ref. \cite{SFK84}.}
\tablenotetext[3]{Contributions of the valence nucleon carried 
out in different potentials: Eq. (\ref{eq:rhoU}), harmonic oscillator, Woods-Saxon, 
respectively. Ref. \cite{DKT1994}.}
\tablenotetext[4]{Core contribution, Ref. \cite{DTFD1996}.}
\tablenotetext[5]{This value is comprised of a valence nucleon and core contribution. 
The valence proton gives a contribution of $-0.04\eta _{pp}-1.4\eta _{pn}$ while the 
core contribution is $1.23\eta _{pp}$, in the units in the table. 
The difference between the valence contributions of the works  
\cite{flambaum85,FKS86} and \cite{SFK84} is due to a difference in the potentials used; 
remember that in $^{203,205}$Tl the Schiff moment is very sensitive to the 
potential, see Eq. (\ref{eq:S1/2}).}  
\end{table}
%********************************
\begin{table}
\caption{Nuclear Schiff moments $S_{\rm intr}$ and $S$ 
(in rotating and laboratory frames, respectively) 
%static???? 
calculated in the Woods-Saxon potential. 
Static octupole deformation is assumed.}
\label{tab:deformedschiff}
\begin{tabular}{llllll}
 & $^{223}$Ra & $^{225}$Ra & $^{223}$Rn & $^{221}$Fr & $^{223}$Fr \\
%& $^{225}$Ac & $^{229}$Pa \\
\hline
$S_{\rm intr}(e~{\rm fm}^{3})$\tablenotemark[1] & 24 & 24 & 15 & 21 & 20 \\ 
%& 28 & 25 \\
$S(\eta ~e ~{\rm fm}^{3})\times 10^{8}$\tablenotemark[1] & 
400 & 300 & 1000 & 43 & 500 \\
%& 900 & $1.2\times 10^{4}$ \\
\end{tabular}
\tablenotetext[1]{Numbers are from Ref. \cite{SAF}. See \cite{SAF} for details.}
\end{table}
%****************************************************************
\begin{table}
\caption{Ground state EDMs of diamagnetic atoms induced by nuclear Schiff moments. 
Units: $10^{-17}\Big(S/(e~{\rm fm}^{3})\Big)e~{\rm cm}$.}
\label{tab:atomicEDMsnm}
\begin{tabular}{llll}
$^{129}$Xe & $^{223}$Rn & $^{199}$Hg & $^{225}$Ra \\ % & $^{239}$Pu \\
\hline
 0.27\tablenotemark[1], 0.38\tablenotemark[2] &
 2.0\tablenotemark[3], 3.3\tablenotemark[2] &
 $-4.0$\tablenotemark[4], $-2.8$\tablenotemark[2] &
 $-7.0$\tablenotemark[3], $-8.5$\tablenotemark[2] \\ 
%& $-10$\tablenotemark[5], $-11$\tablenotemark[2] \\
\end{tabular}
\tablenotetext[1]{Ref. \cite{dzuba85}. Relativistic Hartree-Fock calculation.}
\tablenotetext[2]{Ref. \cite{atomicedms}. Average of two ab initio many-body calculations; 
core polarization and correlation corrections included.}
\tablenotetext[3]{Ref. \cite{SAF}. Estimate found by scaling (with $Z$) calculations for lighter 
analogous atoms. Radon result scaled from xenon calculation \cite{dzuba85}; 
radium result scaled from mercury calculation \cite{flambaum85,FKS86}.}
\tablenotetext[4]{Ref. \cite{flambaum85,FKS86}. Estimated from the calculation of 
$d_{\rm atom}(C^{T})$ for mercury performed in Ref. \cite{martpend85}.}
%\tablenotetext[5]{Ref. \cite{fgschiff}.}
%
%
% ACCURACY
%
%
\end{table}
%****************************************************************
%table of limits I
%
\begin{table}
\caption{Best limits on $P,T$-violating parameters at the nucleon level. 
Signs of the central points are omitted. Errors are experimental. 
Some relevant theoretical works are presented in the last column.}
\label{tab:limitsI}
\begin{tabular}{llrll}

$P,T$-violating term & Value & System & Exp. & Theory \\
\hline

%& & & & \\
{\bf HADRONIC} & & & &\\
& & & &\\

neutron EDM $d_{\rm n}$ & $(17\pm 8\pm 6)\times 10^{-26}~e~{\rm cm}$ 
& $^{199}{\rm Hg}$ & \cite{hgedm} & \cite{khriplovich76,cpoddbook} \\
%check sandars reference  - what is it doing here????!!!!

 & $(1.9\pm 5.4)\times 10^{-26}~e~{\rm cm}$ & n & \cite{ILL} & \\

 & $(2.6\pm 4.0\pm 1.6)\times 10^{-26}~e~{\rm cm}$ & n & 
\cite{PNPI} & \\

& & & & \\

proton EDM $d_{\rm p}$ & $(1.7\pm 0.8\pm 0.6)\times 10^{-24}~e~{\rm cm}$ 
& $^{199}{\rm Hg}$ & \cite{hgedm} & 
\cite{khriplovich76,cpoddbook,atomicedms} \\

& $(17\pm 28)\times 10^{-24}~e~{\rm cm}$ & TlF & 
\cite{tlfedm} & \cite{sandars1967,petrov02} \\

& & & & \\

$\frac{G}{\sqrt{2}}\frac{1}{2m}\eta \mbox{\boldmath$\sigma$}\cdot \mbox{\boldmath$\nabla$}\rho$ 
&
$\eta _{\rm np}=(2.7\pm 1.3\pm 1.0)\times 10^{-4}$ & 
$^{199}{\rm Hg}$ & \cite{hgedm} & 
\cite{flambaum85,FKS86} \\

& & & & \\

$\bar{g}_{\pi NN}$ & $\bar{g}_{\pi NN}=(3.0\pm 1.4\pm 1.1)\times 10^{-12}$ & 
$^{199}{\rm Hg}$ & \cite{hgedm} & \cite{crewther79,cpoddbook} \\

& & & & \\
{\bf SEMI-LEPTONIC} & & & & \\
& & & & \\

$\frac{G}{\sqrt{2}}C^{SP}\bar{N}N\bar{e}i\gamma_{5}e$ &
$(0.40C^{SP}_{\rm p}+0.60C^{SP}_{\rm n})=(18\pm 8\pm 7)\times 10^{-8}$ 
& $^{199}{\rm Hg}$ & \cite{hgedm} & \cite{FK85,cpoddbook} \\

& $(0.40C^{SP}_{\rm p}+0.60C^{SP}_{\rm n})=(6\pm 6)\times 10^{-8}$
& $^{205}{\rm Tl} $& \cite{tledm02} & \cite{M-PL1991} \\

& & & & \\

$\frac{G}{\sqrt{2}}C^{PS}\bar{N}i\gamma_{5}N\bar{e}e$ & 
$C^{PS}_{\rm n}=(1.8\pm 0.8\pm 0.7)\times 10^{-6}$ & 
$^{199}{\rm Hg}$ & \cite{hgedm} & \cite{FK85,cpoddbook} \\

& & & & \\

$\frac{G}{\sqrt{2}}C^{T}\bar{N}i\gamma_{5}\gamma_{\mu \nu}N\bar{e}
\sigma _{\mu \nu}e$ &
$C^{T}_{\rm n}=(5.3\pm 2.5\pm 2.0)\times 10^{-9}$
& $^{199}{\rm Hg}$ & \cite{hgedm} & \cite{martpend85} \\

& & & & \\
{\bf LEPTONIC} & & & & \\
& & & & \\

electron EDM $d_{e}$ & $(6.9\pm 7.4)\times 10^{-28}~e~{\rm cm}$ & 
$^{205}{\rm Tl}$ & \cite{tledm02} & \cite{liu92} \\

& $(9.1\pm 4.2\pm 3.4)\times 10^{-27}~e~{\rm cm}$ 
& $^{199}{\rm Hg}$ & \cite{hgedm} & \cite{FK85}

\end{tabular}
\end{table}

%****************************************************************
%table of limits II
%
\begin{table}
\caption{Best limits on $P,T$-violating parameters at the quark level. 
Signs of the central points are omitted.
Errors are experimental. 
Some relevant theoretical works are presented in the last column.}
\label{tab:limitsII}
\begin{tabular}{llrll}

$P,T$-violating term & Value & System & Exp. & Theory \\
\hline

%& & & & \\
{\bf HADRONIC} & & & &\\
& & & &\\

$\frac{G}{\sqrt{2}}k_{s}\bar{q}_{1}i\gamma_{5}q_{1}\bar{q}_{2}q_{2}$ 
& $k_{s}=(3\pm 1\pm 1)\times 10^{-7}$ & $^{199}{\rm Hg}$ & \cite{hgedm} & \cite{KKY1988,cpoddbook} \\
& $k_{s}=(2\pm 5)\times 10^{-5}$ & n & \cite{ILL} & \cite{KKY1988,KK1992,cpoddbook} \\
& $k_{s}=(3\pm 4\pm 2)\times 10^{-5}$ & n & \cite{PNPI} & \cite{KKY1988,KK1992,cpoddbook} \\

& & & & \\

$\frac{G}{\sqrt{2}}k_{s}^{c}\bar{q}_{1}i\gamma_{5}t^{a}q_{1}\bar{q}_{2}t^{a}q_{2}$ 
& & & & \\
$q_{1}=q_{2}$ 
& $k_{s}^{c}=(1\pm 1\pm 1)\times 10^{-6}$ & $^{199}{\rm Hg}$ & \cite{hgedm} & \cite{KKY1988,cpoddbook} \\
& $k_{s}^{c}=(1\pm 2)\times 10^{-4}$ & n & \cite{ILL} & \cite{KKY1988,KK1992,cpoddbook} \\
& $k_{s}^{c}=(1\pm 1\pm 1)\times 10^{-4}$ & n & \cite{PNPI} & \cite{KKY1988,KK1992,cpoddbook} \\

& & & & \\

$q_{1}\neq q_{2}$ 
& $k_{s}^{c}=(5\pm 2\pm 2)\times 10^{-4}$ & $^{199}{\rm Hg}$ & \cite{hgedm} & \cite{KKY1988,cpoddbook} \\
& $k_{s}^{c}=(1\pm 2)\times 10^{-4}$ & n & \cite{ILL} & \cite{KKY1988,KK1992,cpoddbook} \\
& $k_{s}^{c}=(1\pm 1\pm 1)\times 10^{-4}$ & n & \cite{PNPI} & \cite{KKY1988,KK1992,cpoddbook} \\

& & & & \\

$\frac{G}{\sqrt{2}}\frac{k_{t}}{2}\epsilon_{\mu \nu \alpha \beta}
\bar{u}i\sigma _{\mu \nu}u\bar{d}\sigma _{\alpha \beta}d$ 
& $k_{t}=(1\pm 1\pm 1)\times 10^{-5}$ & $^{199}{\rm Hg}$ & \cite{hgedm} & \cite{KKY1988,cpoddbook} \\
& $k_{t}=(1\pm 2)\times 10^{-5}$ & n & \cite{ILL} & \cite{KKY1988,KK1992,cpoddbook} \\
& $k_{t}=(1\pm 1\pm 1)\times 10^{-5}$ & n & \cite{PNPI} & \cite{KKY1988,KK1992,cpoddbook} \\

& & & & \\

$\frac{G}{\sqrt{2}}\frac{k_{t}^{c}}{2}\epsilon_{\mu \nu \alpha \beta}
\bar{u}i\sigma _{\mu \nu}t^{a}u\bar{d}\sigma _{\alpha \beta}t^{a}d$ 
& $k_{t}^{c}=(5\pm 3\pm 2)\times 10^{-5}$ & $^{199}{\rm Hg}$ & \cite{hgedm} & \cite{KKY1988,cpoddbook} \\
& $k_{t}^{c}=(1\pm 2)\times 10^{-5}$ & n & \cite{ILL} & \cite{KKY1988,KK1992,cpoddbook} \\
& $k_{t}^{c}=(1\pm 1\pm 1)\times 10^{-5}$ & n & \cite{PNPI} & \cite{KKY1988,KK1992,cpoddbook} \\

& & & & \\

CEDMs $\tilde{d}$ and & 

$e(\tilde{d}_{\rm d}-\tilde{d}_{\rm u})
=(1.5\pm 0.7\pm 0.6)\times 10^{-26}e{\rm cm}$ 
& $^{199}{\rm Hg}$ & \cite{hgedm} & \cite{pospelov2002} \\ 

EDMs $d$ of quarks &
$e(\tilde{d}_{\rm d}+0.5\tilde{d}_{\rm u})+1.3d_{\rm d}-0.3d_{\rm u}$ 
& & & \\

& $\qquad 
=(3.5\pm 9.8)\times 10^{-26}e{\rm cm}$ & n & 
\cite{ILL} & \cite{pospelov01} \\

& $\qquad
=(4.7\pm 7.3\pm 2.9)\times 10^{-26}e{\rm cm}$ & n & 
\cite{PNPI} & \cite{pospelov01}\\

& & & & \\

QCD phase $\bar{\theta}$ & 
$\bar{\theta}=(1.1\pm 0.5\pm 0.4)\times 10^{-10}$ & 
$^{199}{\rm Hg}$ & \cite{hgedm} & \cite{crewther79,cpoddbook} \\ 

& $\bar{\theta}=(1.6\pm 4.5)\times 10^{-10}$ & n & \cite{ILL} & 
\cite{pospelov99} \\

& $\bar{\theta}=(2.2\pm 3.3\pm 1.3)\times 10^{-10}$ & n & \cite{PNPI} & 
\cite{pospelov99} \\

& & & & \\
{\bf SEMI-LEPTONIC} & & & & \\
& & & & \\

$\frac{G}{\sqrt{2}}k_{1q}\bar{q}q\bar{e}i\gamma_{5}e$ & 
$k_{1u}+k_{1d}=(3\pm 1 \pm 1)\times 10^{-8}$ & $^{199}{\rm Hg}$ & \cite{hgedm} & \cite{cpoddbook} \\

& $k_{1u}+k_{1d}=(1\pm 1)\times 10^{-8}$ & $^{205}{\rm Tl}$ & \cite{tledm02} & \cite{cpoddbook} \\

& & & & \\

$\frac{G}{\sqrt{2}}\bar{q}i\gamma_{5}q\bar{e}e$ &
$k_{3q}=(2\pm 1\pm 1)\times 10^{-8}$ & $^{199}{\rm Hg}$ & \cite{hgedm} & \cite{cpoddbook} \\

& & & & \\

$\frac{G}{\sqrt{2}}\bar{q}i\gamma_{5}\sigma_{\mu \nu}q\bar{e}
\sigma _{\mu \nu}e$ &
$k_{2q}=(5\pm 3\pm 2)\times 10^{-9}$ & $^{199}{\rm Hg}$ & \cite{hgedm} & \cite{cpoddbook}
\end{tabular}
\end{table}

\center
\widetext
\input psfig
\psfull

\begin{figure}[b]
\centerline{\psfig{file=feyncorr.eps, clip=}}
\caption{Correlation corrections to energy in the Feynman diagram technique.}
\label{feyncorr}
\end{figure}

\begin{figure}[b]
\centerline{\psfig{file=hpchain.eps, clip=}}
\caption{Hole-particle interaction in the polarization operator.}
\label{hpchain}
\end{figure}

%\begin{figure}
%\psfig{figure=../../papers/review/section/diagrams/anfig3.ps,height=2in}
%\caption{A cross-section (in the $x$-$z$ plane) of the current
%distribution due to the spin helix
%(produced by the parity violating nucleon-nucleus interaction).
%The anapole moment points along the $z$ direction.}
%\label{f3}
%\end{figure}
%
%\begin{figure}
%\psfig{figure=../../papers/review/section/diagrams/anfig4.ps,height=2in}
%\caption{A cross-section (in the $x$-$y$ plane) of the magnetic field
%due to the spin helix (produced by the parity violating
%nucleon-nucleus interaction).
%The anapole moment points along the $z$
%direction.}
%\label{f4}
%\end{figure}

\end{document}